\newif\iffigs\figstrue

\iffigs
   \documentstyle[12pt,epsf]{article}
\else
   \documentstyle[12pt]{article}
\fi

\def\thefootnote{\alph{footnote}}
\def\thefootnote{\fnsymbol{footnote}}
\newcommand{\eq}{\begin{equation}}
\newcommand{\en}{\end{equation}}
\newcommand{\eqa}{\begin{eqnarray}}
\newcommand{\ena}{\end{eqnarray}}

\newcommand{\lbl}{\label}
\newcommand{\bm}{\boldmath}
\newcommand{\ubm}{\unboldmath}

\newcommand{\bhk}{J}

\newcommand{\ca}{{\cal A}}
\def\dop#1{{\rm d}\hskip -1pt #1}
\def\real{{\rm Re}\hskip 1pt}
\def\ee#1{{\rm e}^{#1}}
\def\trace{{\rm Tr}\hskip 1pt}
\def\ii{{\rm i}}
\def\diag{{\rm diag}}
\def\Vdag{V^\dagger}
\def\Udag{U^\dagger}

\def\mycaptionl#1#2{%
\begin{center}
\hskip 1pt\vskip -1cm
\begin{minipage}{14cm}
\small {\bf #1}: {\sl #2}
\end{minipage}
\null\hskip 1pt\vskip -0.2cm
\end{center}}
\newcommand{\NP}[1]{Nucl.\ Phys.\ {\bf #1}}
\newcommand{\PL}[1]{Phys.\ Lett.\ {\bf #1}}

\newcommand{\PR}[1]{Phys.\ Rev.\ {\bf #1}}
\newcommand{\PRL}[1]{Phys.\ Rev.\ Lett.\ {\bf #1}}
\newcommand{\MPL}[1]{Mod.\ Phys.\ Lett.\ {\bf #1}}
\newcommand{\IJMP}[1]{Int.\ J.\ Mod.\ Phys.\ {\bf #1}}

\begin{document}
\hskip 10cm \vbox{\hbox{DFTT 43/96}
\hbox{NORDITA 96/66P}
\hbox{October 1996}}
\vskip 0.7cm
\centerline{\Large\bf Finite Temperature Lattice QCD}
\centerline{\Large\bf in the Large $N$ Limit}
\vskip 0.4cm
\centerline{\large M. Bill\'{o}\footnote{E--mail: {\tt billo@alf.nbi.dk};
supported by I.N.F.N., Italy}}
\vskip .2cm
\centerline{\small\it Nordita, Blegdamsvej 17, Copenhagen \O, Denmark}
\vskip .4cm
\centerline{\large M. Caselle\footnote{E--mail: {\tt caselle@to.infn.it}},
A. D'Adda}
\vskip .2cm
\centerline{\small\it Istituto Nazionale di Fisica Nucleare, Sezione di Torino}
\centerline{\small\it  Dipartimento di Fisica
Teorica dell'Universit\`a di Torino}
\centerline{\small\it  via P.Giuria 1, I-10125 Turin,Italy}
\vskip .4cm
\centerline{\large S. Panzeri\footnote{Present address: {\it Department
of Experimental Psychology, University of Oxford, Oxford OX1 3UD, UK.}
E--mail: {\tt stefano.panzeri@psy.ox.ac.uk}}}
\vskip .2cm
\centerline{\small\it  SISSA,Via Beirut 2-4, I-34013, Trieste, Italy}
\vskip 0.5cm
\begin{center}
{\bf Abstract}\\
\vskip 0.3cm
\begin{minipage}{14cm}
\small 
Our aim is to give a self-contained review of recent advances in the 
analytic description of the deconfinement transition and determination 
of the deconfinement temperature in lattice QCD at large $N$. 
We also include some new results, as for instance in the comparison of 
the analytic results with Montecarlo simulations. 
We first review the general set-up 
of finite temperature lattice gauge theories, using asymmetric lattices, 
and develop a consistent perturbative expansion in the coupling $\beta_s$ 
of the space-like plaquettes. We study in detail the effective models for 
the Polyakov loop obtained, in the zeroth order approximation in $\beta_s$,  
both from the Wilson action (symmetric lattice) and from the heat kernel 
action (completely asymmetric lattice). The distinctive feature of the heat 
kernel model is its relation with two-dimensional QCD on a cylinder; the 
Wilson model, on the other hand, can be exactly reduced to a twisted 
one-plaquette model via a procedure of the Eguchi--Kawai type. In the weak 
coupling regime both models can be related to exactly solvable Kazakov--Migdal 
matrix models. The instability of the weak coupling solution is due in both 
cases to a condensation of instantons; in the heat kernel case, this is directly 
related to the Douglas--Kazakov transition of QCD2. A detailed analysis of 
these results provides rather accurate predictions of the deconfinement 
temperature. In spite of the zeroth order approximation they are in good 
agreement with the Montecarlo simulations in $2+1$ dimensions, while in $3+1$ 
dimensions  they only agree with  the Montecarlo results away from the 
continuum limit.
\end{minipage}
\end{center}
\noindent
\vfill
\eject
\newpage
\setcounter{footnote}{0}
\def\thefootnote{\arabic{footnote}}
\tableofcontents
\vfill\eject
\setcounter{footnote}{0}
\def\thefootnote{\arabic{footnote}}
\section{Introduction}
\label{introsec}
In the last twenty years the lattice regularization has proved to be a very
powerful tool to understand and describe the non-perturbative features of
non-abelian gauge theories. However, while impressive results have been
obtained  by means of Montecarlo simulations, very few
progresses have been achieved with analytic techniques. Lattice Gauge Theories
(LGT in the following) can be solved exactly in two dimensions for any gauge
group, but become unaffordably complex in more than two dimensions, even in
absence of quarks. Moreover, most of the approximation techniques which are
usually successful in dealing with simpler statistical mechanical systems, like
(suitably improved) mean field methods or strong coupling expansions 
turn out to be less useful in the case of LGT.
A remarkable exception to this state of art is
represented by the large $N$ approximation~\cite{ln} which is able to keep the 
whole complexity of the finite $N$ models. 
Unfortunately, even in
the large $N$ limit (despite the fact that, as we shall discuss below, some 
major simplifications occur) it is not possible to give exact solutions 
(the so called ``Master Field'') to the Lattice SU$(N)$ models. 
Notwithstanding this,  several interesting results have 
been obtained in the past years even without the explicit knowledge of the 
Master Field.  

In this review we shall deal with one of the most interesting features of non
abelian gauge theories: the presence of a deconfinement transition at finite
temperature. We shall apply large $N$ techniques  to finite temperature LGT,
with the aim of constructing the phase diagram of the model, 
locate the critical points and identify their order. 
We shall deal for most part of the review with the pure gauge
theory (namely without quarks); only in section \ref{adjsec}
we shall comment on the phase diagram in presence of quarks. 
We shall keep $d$, the number of space-like dimensions as a free parameter, 
and shall study in particular the
cases $d=2$ and $d=3$ which are the most interesting from a phenomenological 
point of view and for which there exist Montecarlo simulations  to compare with
our predictions. 

This paper is an update rather than a complete review on 
this topic. We shall mainly concentrate on the most recent results 
(say, of the last five years).    
There are some very good reviews both on lattice gauge theories and on 
large $N$ models which cover most of the results obtained up to the 
second part of the eighties. In particular, for the large $N$  approach we 
suggest the reviews of S. Coleman~\cite{col}, S.R. Das~\cite{das} and the
recent contribution of Campostrini, Rossi and Vicari~\cite{crv}. 
For a throughout introduction to Lattice Gauge Theories, 
with a good discussion  both of the large $N$ limit and of the finite 
temperature regularization, we refer to~\cite{dz}.

During the eighties most of the efforts in the study of finite temperature 
large $N$ LGT where
devoted in the following two directions:
{\em a)}\, Eguchi--Kawai (EK) type \cite{ek} models; 
{\em b)}\, dimensionally reduced model.
Both these approaches have advantages and drawbacks.
Twisted or Quenched EK models maintain the whole complexity of the
theory and  reduce it to  models of just $(d+1)$ independent matrices. However
these models could not be solved exactly, and the only
fruitful approach has been so far to extract numerical results by using 
Montecarlo simulations. 
We shall discuss below some of these results. Let us mention here
that the reliability of Montecarlo simulations based on the TEK model 
is strongly affected by the presence of a first order
bulk phase transition which shadows the true deconfinement  transition. Finding
a precise characterization of the TEK or QEK  phase diagram,
from the results of the Montecarlo simulations only, 
is a difficult and open problem.

On the other side, by doing some rather crude approximation, it has been 
possible to construct dimensionally reduced matrix models which are exactly 
solvable. 
The price to pay in this case is that the simplifications 
needed to reach the exact
solvability are so strong that the resulting models show sometimes only 
a very faint similarity with the original models. 

In the last few years  it has been realized that some insight into the
large $N$ structure of LGT can be obtained by using the results that
have been accumulating in the meantime on the large $N$ solution of a variety
of matrix models. These models have found important applications in
various contexts. Let us mention among the others: the random matrix 
description \cite{mehtabook} of various physical systems
(ranging from quantum wires to chaotic systems),
the exact solution of two dimensional LGT's 
\cite{contqcd2,latqcd2,grmat1,grmat2} 
and the related string like behaviour \cite{grosstaylor}, the large $N$ matrix
approach to two dimensional quantum gravity \cite{qg} and the 
``induced QCD'' models proposal by Kazakov and Migdal \cite{kazmig}. 

The present review is mostly concerned with the results  
\cite{bcdmp,zar,semzar,bilda} obtained in the last few
years in finite temperature LGT by using the matrix models techniques mentioned
above. 
We shall show that by using
both some recent results of two dimensional QCD and  the exact solution
\cite{grosskm} of the induced QCD models in any dimension, one can solve 
dimensionally reduced models (of the type {\em b)} listed above)  
that involve far less crude approximations than those discussed in  the 
eighties. 
By using these exact results we can also solve, within a very good 
approximation, a simplified TEK model and understand its phase diagram. 
Our results can be compared with those obtained with Montecarlo 
simulations in the $N=2$ and $N=3$ cases. 
There are by now many rather precise numerical estimates of the
deconfinement temperature for these models both in $d=2$ and $d=3$. We shall
show that our results are rather good if compared with the simulations at $d=2$,
while they are in general poor (except for the lowest values of $n_t$, the
lattice size in the time direction) in $d=3$. We shall give some arguments to
explain this failure and indicate how our results could be improved. 
Let us stress however that our aim is not to compete with the Montecarlo
simulations, which certainly remain the most powerful tool to obtain
quantitative result in LGT. Our idea is rather that any new analytic result in 
this context can teach us quite a
lot on  the non-perturbative regime of non abelian gauge theories. 
Moreover, even if our results  are often only crude approximations 
of the exact ones, the pattern of the phase diagram that we obtain has 
good chances to be the correct one. 
Finally let us stress that there are situations in which
the methods discussed here could even become competitive with the Montecarlo
simulations. This is the case, for instance, when the model contains
more that one coupling constant (like the mixed
fundamental-adjoint action), and the phase diagram is so complex that it is
difficult to study it numerically. 

The paper is organised as follows.
For the sake of being  as self-contained as possible, we devote a
rather large introductory section (sec. \ref{gensetsec}) to describe 
the general setting of finite temperature LGT in a form which is suitable
for our purposes. In sec. \ref{effacsec} we review in some detail the 
derivation of refined  effective action for the Polyakov loops, 
both in the case of a LGT described by the heat kernel action and 
by the Wilson action. 
In sec. \ref{hksec} the
heat kernel effective model is discussed; its phase diagram is
described,  and a numerical estimate for the deconfinement 
temperature is obtained.
In sec. \ref{extsec} the phase diagram is re-discussed as a function of the
space dimensionality and in the case in which
the gauge fields are coupled to certain
external static sources or to an external ``magnetic'' field.
Sec. \ref{wilssec} is devoted to the analysis
of the effective model obtained from the Wilson action; in particular we 
discuss its twisted reduction \'a la Eguchi--Kawai, that
gives rise to a very interesting one-plaquette matrix model. 
In sec. \ref{mcsec} the results for the deconfinement temperature 
are discussed and compared with the available Montecarlo simulations.  
\section{General Setting}
\label{gensetsec}
\subsection{Asymmetric Lattices}
Let us consider a finite temperature lattice gauge theory  (LGT) with 
gauge  group SU$(N)$, defined
on a $d+1$ dimensional cubic lattice.
In order to describe a finite temperature theory  we require that one 
dimension (which we shall call ``time" from now on) is compactified with
periodic boundary conditions and compactification length ${1 \over T}$,
if $T$ is the temperature. We shall assume that in the other ``space" 
dimensions the extension of the lattice is infinite, or anyway much 
greater than ${1 \over T}$.
If we denote by $N_t$ the number of lattice spacings in the time 
direction, and $a_t$ the corresponding lattice spacing, then we have 
${1 \over T}=N_t a_t$. 
It is convenient to work with an asymmetric lattice, namely with 
different lattice spacing in the time and space directions. 
Let us  denote by  $a_s$ the lattice spacing in the space directions. 
The ratio between $a_t$ and $a_s$ defines the asymmetry parameter $\rho$:
\eq
\rho = { a_{s} \over a_t}= T N_t a_s~.
\label{rho}
\en
We shall require that different values of $\rho$ correspond to different,
but equivalent, lattice regularization of the same model.
In order to implement such requirement we introduce different bare 
couplings in the time and space directions. Let us call them $\beta_t$ 
and $\beta_s$  respectively.
The Wilson action is then
\eq
S_{\rm W}= N^2 \sum_{\vec{x},t}~  \left( \sum_i~   \beta_t \hat{G}_{0i}
(\vec{x},t)+\sum_{i<j}~  \beta_s  \hat{G}_{ij}(\vec{x},t) ~~~\right),
\label{wilson}
\en
where here and in what follows we denote by $G_{0i}$ and $G_{ij}$  
the time-like and space-like plaquette variables:
\begin{eqnarray}
G_{0i}(\vec{x},t)& = &   V(\vec{x},t)U_i
(\vec{x},t+1) V^{\dagger}(\vec{x}+\hat{\i},t)U^{\dagger}_i(\vec{x},t)~,
 \nonumber \\ G_{ij}(\vec{x},t)& = &   U_i(\vec{x},t)U_j(\vec{x}+
\hat{\i},t)U^{\dagger}_i(\vec{x}+ \hat{\j},t)U^{\dagger}_j(\vec{x},t)
\label{plaq}
\end{eqnarray}
and by $\hat{G}_{0i}(\vec{x},t)$ and $\hat{G}_{ij}(\vec{x},t)$ the real 
part of the suitably normalized traces:
\eq
\label{trplaq}
\hat{G}_{0i}(\vec{x},t)= {1 \over N} \real \trace G_{0i}(\vec{x},t)~,~~~
~~~~~~~~~~\hat{G}_{ij}(\vec{x},t)={1 \over N}\real\trace 
G_{ij}(\vec{x},t)~.
\en

In eq. (\ref{plaq}) we have denoted by $U_i(\vec{x},t)$ the link 
variables in the space directions ($i \in \{1,2,...,d \} $) and by 
$V(\vec{x},t)$ the link variables in the time direction. 
The components of $\vec{x}$ and $t$ are integers, with $t$ periodic 
modulo $N_t$.
The normalization of the couplings $\beta_s$ and $\beta_t$ has been 
chosen, by extracting a $N^2$ factor in front of $S_{\rm W}$, in such a way to 
have a smooth large $N$ limit.

For a given $\rho$ the relation between the bare couplings $\beta_t$ and 
$\beta_s$ can be obtained by requiring that the Wilson action 
(\ref{wilson}) reproduces in the naive continuum limit a gauge theory 
with the same coupling constant $g$ for all components of the field 
strength.
This leads to the following equations, which relate $\rho$ and the bare 
gauge coupling $g$ to $\beta_t$ and $\beta_s$:
\eq
\frac{2}{Ng^2}=a_s^{3-d}\sqrt{\beta_s\beta_t}~,
\hskip 1cm
\rho = \sqrt{\frac{\beta_t}{\beta_s}}~.
\label{couplings}
\en

It is clear from eqs. (\ref{rho}) and (\ref{couplings}) that equivalent 
regularizations with different values of $\rho$ require different
values of $N_t$. Hence, to maintain the  equivalence, $N_t$ must be
a function of $\rho$ according to eq. (\ref{rho}). 

Among all these equivalent regularizations a particular role is played 
by the symmetric one, which is defined by:
\eq
\beta\equiv\frac{2}{Ng^2} a^{d-3}
\label{coupsym}
\en
(from now on we shall distinguish the symmetric regularization from the
asymmetric ones by eliminating the subscripts $t$ and $s$ in $\beta$ and
$a$).
By comparing eq.s (\ref{rho},\ref{couplings},\ref{coupsym}) we see that 
all regularizations are equivalent to a symmetric one provided
\eq
\beta=\rho\beta_s=\frac{\beta_t}{\rho}~,
\label{rel1}
\en
\eq
N_t(\rho)=\rho ~n_t~,
\label{rel2}
\en
where $n_t$ is the number of links in the time direction with a symmetric
regularization: $n_t = N_t(\rho=1)$.

Notice however that these equivalence relations have been derived in the 
{\em naive} or ``classical'' continuum limit, and quantum corrections 
are in general present. These corrections were studied and calculated in
the (3+1) dimensional case  by F.Karsch in~\cite{k81}.  
They lead to the following expressions:
\eq
\beta_t=\rho(\beta+c_\tau(\rho))~,
\label{rel3}
\en
\eq
\beta_s=\frac{\beta+c_\sigma(\rho)}{\rho}~.
\label{rel4}
\en
The quantum effects are encoded in the two functions $c_\sigma(\rho)$ 
and $c_\tau(\rho)$\footnote{Let us notice, to avoid confusion, 
that our functions $c_\sigma$ and $c_\tau$ correspond 
to those of~\cite{k81} multiplied by the
factor $2/N$, which  ensures a smooth limit as $N\to\infty$.}.
We shall be
interested in their behaviour at  large $\rho$, that can be extracted 
from ~\cite{k81} and is given by the asymptotic expansion:
\eq
c_{\sigma,\tau}\equiv \alpha^{0}_{\sigma,\tau}
+\frac{\alpha^{1}_{\sigma,\tau}}{\rho}+\ldots~.
\label{rhoex}
\en

The numerical values of the $\alpha$'s in the large $N$ limit can be 
obtained from ~\cite{k81}, and they are given by:
$\alpha^{0}_{\tau}=-0.2629$; $\alpha^{1}_{\tau}=1/4$;
$\alpha^{0}_{\sigma}=1/4$; $\alpha^{1}_{\sigma}=1/4$. 
\subsection{Center symmetry and the Polyakov loop}
\label{centersec}
The major consequence of the periodic boundary conditions in the time direction
is the appearance of a new global symmetry of the action, with symmetry  
group the center $C$ of the gauge group (in our case ${\bf Z}_N$).

This symmetry can be realized as follows. Let us transform 
all the timelike links 
of a given space-like slice with the same element $W_0$ belonging to the center
of the gauge group.
\eq
V(\vec{x},t) \to W_0V(\vec{x},t)  \hskip 1cm \forall\, \vec{x},~~~t~
\mbox{fixed}~.
\en
The space-like plaquettes are not affected by the transformation, while in each
timelike plaquette two contribution appear: $W_0$ and $W_0^{-1}$. Since they
belong to the center, they commute with all other matrices in the plaquette and
can be moved so as to cancel each other. 
So the Wilson action is invariant under such transformation. 
The important point is that, due to the periodic boundary 
conditions in the time direction, it is impossible to
re-absorb this global twist by means of local gauge transformations..  

A second consequence of the periodic boundary conditions is that
it is possible to define
gauge invariant observables which are topologically non-trivial.
The simplest choice is the Polyakov loop, 
defined  in terms of link variables as:
\eq
\hat P(\vec{ x})\equiv {\rm Tr}
 \prod_{t=1}^{N_t} V(\vec x,t)~.
\label{polya}
\en
In the following we shall often use the untraced quantity $P({\vec x})$, 
defined as:
\eq
P(\vec{ x})\equiv
\prod_{t=1}^{N_t} V(\vec x,t)~,
\label{polya2}
\en
which will be referred to as ``Polyakov line''.

The relevant feature of the Polyakov loop is that it transforms
 under the above discussed symmetry as follows:
\eq
\hat P(\vec{ x}) \to W_0 \hat P(\vec{ x})~;
\label{polya3}
\en
 thus it is a natural order parameter for this symmetry.
It will acquire a non zero expectation value if the center symmetry is
spontaneously broken. 
In a pure LGT  the Polyakov loop has 
a deep physical interpretation, since its  expectation value is related to the
free energy of a single
isolated quark. 

Hence the fact that the Polyakov loop acquires a non-zero expectation 
value can be considered as a signature of deconfinement and the 
phase transition which separates the regime in which the center 
symmetry is unbroken from the broken
symmetric phase will be the deconfinement transition.

General arguments show that,
in $d>1$, finite temperature gauge theories admit such a
deconfinement transition for some critical value of the temperature $T=T_c$, 
separating the
high temperature, deconfined, phase ($T>T_c$) from the low
temperature, confining domain ($T<T_c$).
In the following we shall be interested in the phase diagram of the
model as a function of $T$, and we shall make some attempt to locate the
critical point $T_c$. 

A peculiar feature of the behaviour of the Polyakov loop in the high
temperature phase is that its (non-zero) expectation value is {\em not}
a generic element of SU$(N)$ but tends to fluctuate around one of 
the elements of the center. 
These fluctuations become smaller and smaller as the temperature
increases and finally in the infinite temperature 
limit the expectation value of
the Polyakov loop {\em exactly} becomes an element of $C$
(see for instance~\cite{sy} for a discussion of this point).

This feature will be  relevant in the following, when we shall describe 
these small fluctuations of the Polyakov loop in the high temperature phase 
by using a suitable generalization of the Kazakov--Migdal model. 
\subsection{Svetitsky--Yaffe conjecture}
\label{sysec}
The peculiar role played by the Polyakov loops in the above discussion,
suggests to use some kind  of effective action for the Polyakov loops 
to study the deconfinement
transition and, more generally, the physics of finite temperature LGT.

Let us make this statement more precise. Constructing an {\em exact} effective
action for the Polyakov loops, equivalent to the original LGT is clearly
impossible (even in the simplest possible, non-trivial, LGT, namely the (2+1)
dimensional Ising gauge model)
since it would require to 
integrate out exactly  all the space-like
gauge degrees of freedom of the original model. However one can try to 
approximate somehow this integration and, at the same time, treat exactly the
timelike degrees of freedom which are related to the Polyakov loops.
The discussion of sec. 2.2 then tells us that
this type of approximation is the one which better preserves the finite
temperature behaviour of the original model. 
The resulting {\em approximate} effective action will keep {\em all} 
the symmetries of the original model relevant to the deconfinement dynamics 
and as a consequence it will give a faithful {\em qualitative}
description of this transition. We can also hope to have a {\em quantitative}
agreement between the results obtained with the effective action and those of
the gauge model. This agreement will become better and better as
we improve our approximation in the space-like degrees of freedom.

This approach was proposed and discussed in~\cite{sy} and accordingly we shall 
refer to it, in the following, as the ``Svetitsky--Yaffe program''.
The most appealing feature of this program is that, by resorting only to some
general results of statistical mechanics, before doing any calculation,
one can obtain several interesting
predictions on the phase diagram of the model. Let us see some of them: 
\begin{itemize}
\item[{\em a)}]
If the original gauge theory lives in ($d$+1) dimension, then
the effective theory for the Polyakov loops is a $d$-dimensional spin
system with symmetry group the center $C$ of the original gauge group. 
\item[{\em b)}]
The deconfinement transition of the
original gauge model becomes the order--disorder transition of the
effective spin system. The ordered phase of the spin model corresponds to the 
deconfined phase in the original gauge theory. In this phase the Polyakov loop
acquires a non-zero expectation value.  
\item[{\em c)}]
This effective theory would obviously have very complicate
interactions, but Svetitsky and Yaffe were able to argue that all these
interactions should be short ranged. As a consequence, if the
deconfinement transition of the original gauge model 
(and hence of the effective spin model)
is continuous, near this critical point,
where the correlation length becomes infinite, the precise form of
the short ranged
interactions should not be important. 
\item[{\em d)}]
Let us consider now the much simpler spin
model with only nearest neighbour
interactions and the same global symmetry group. 
If also the order-disorder phase transition  of this model is 
continuous then, due to point {\em (c)} above, its
universality  class should coincide with that of the
deconfinement transition.
\end{itemize}
This last result is usually known as the ``Svetitsky--Yaffe conjecture'', 
and has been confirmed in the last years by several Montecarlo simulations.
It is certainly the most important consequence of this whole approach. 

Let us stress again that it applies only to the case in which $both$ the 
deconfinement transition of the gauge
model and the order-disorder transition of the nearest neighbour
effective spin model are
continuous. So it does not apply in our case since in the effective 
actions that we shall study in the following the order-disorder transition turns
out to be of the first order.

There are however, in our large $N$ limit framework
 two other intriguing questions which are raised by the SY analysis. The first
 one refers to the difference between the deconfinement transition in
  ${\bf Z}_N$ and SU$(N)$ LGT's. Both have the same center  ${\bf Z}_N$
 so they should
be described by the same effective spin model. 
However we know very well that the
two LGT's have rather different features. This differences becomes particularly
evident in the large $N$ limit where in one case we have the U(1) LGT which,
for instance, in (2+1) dimensions is always confining, while in the other case
one finds the large $N$ deconfinement transition which will be the subject of
the forthcoming sections.

A second, related, problem is that, according to the SY point of view, the
deconfined phase should correspond to the ordered phase of the effective spin
model, but in two dimensions the Mermin--Wegner theorem~\cite{mw} tells us that
the fluctuation are always strong enough to restore the U(1) symmetry thus
forbidding the existence of an ordered phase and consequently of a deconfined
phase in the original (2+1) LGT. This prediction is in clear contradiction with
the analysis of the present paper and also with the commonly accepted scenario
for non-abelian LGT's in (2+1) dimensions.
\bm
\subsection{${\bf Z}_N$ versus SU$(N)$}
\label{mwsec}
\ubm
In this subsection we shall try to answer  these questions and to better
understand the difference between  ${\bf Z}_N$
and SU$(N)$ LGT's.  Let us start our discussion with the analysis of 
the  ${\bf Z}_N$ models, which are simpler. 
In this case one easily realizes
that the nearest neighbour effective spin model which describes the Polyakov
loops dynamics is  the ${\bf Z}_N$ 
symmetric ``clock model''. Its  phase diagram in $d=2$ is well 
known~\cite{eps}. It is rather non trivial and admits,
for finite $N$  
and large enough  $\beta$, an ordered phase in which the 
${\bf Z}_N$ symmetry is spontaneously broken.
In fact the phase diagram of the ${\bf Z}_N$ models (if $N > 4$) 
is composed by three phases: 
the usual (low $\beta$) disordered  and (high $\beta$) ordered phases and a new
intermediate phase which is critical and has the same features of the high 
$\beta$ phase of the U(1) model. 
This intermediate phase is separated from the high and low $\beta$ phases
by two phase transitions of the Kosterlitz--Thouless~\cite{kt} (KT) type. 
The ${\bf Z}_N$ model in two dimensions is self dual, and this greatly 
simplifies the study of the phase diagram. 
Ordered and disordered phases are related by duality
(just like in the well known 2d Ising model) and also the two 
KT transitions are dual to each other~\cite{eps}. 
The coupling of the dual theory is related to the inverse of
the original one by  a factor $N^2$. As $N$ increases, the location of the 
first KT transition (the one which divides the disordered from the critical
phases) remains more or less unchanged, and as $N\to\infty$ it becomes the KT
transition of the U(1) model, whose critical coupling $\beta_{KT}$ is 
of order unity.
The other KT transition and the ordered phase,  which begins at this  KT 
transition, due to the $N^2$ factor mentioned above, are pushed to
infinity as $N$ increases, and finally, in the $N\to\infty$ limit, disappear.
This agrees with the U(1) phase diagram discussed above, which in fact does not
admit an ordered phase. 
This discussion suggests that in the
large $N$ limit the (2+1) dimensional ${\bf Z}_N$ LGT can never have a truly 
deconfined phase, and that the
fluctuations of the Polyakov loops are always strong enough to restore the U(1)
symmetry. 

\iffigs
\begin{figure}
\begin{center}
\null\hskip 1pt
\epsfxsize 13cm
\epsffile{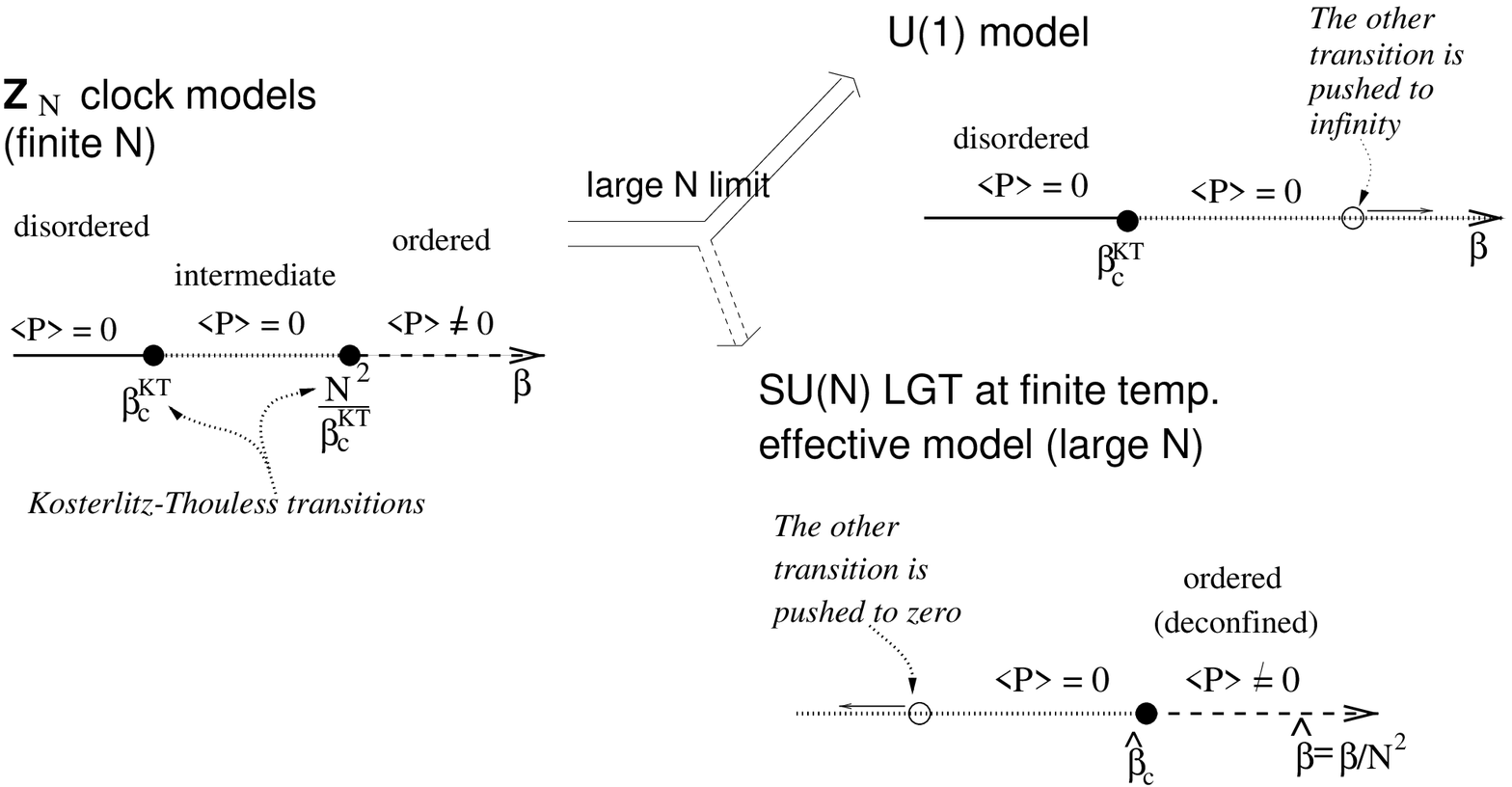}
\vskip 0.15cm\noindent
\end{center}
\mycaptionl{Fig. 1}{Relation between the phase 
diagrams of the ${\bf Z}_N$ ``Clock
models'', the U(1) model and the effective model for the Polyakov loops in
large $N$ LGT at finite temperature. When taking the large $N$ limit of the
${\bf Z}_N$ models, depending if the coupling constant is rescaled with $1/N^2$
(LGT case) or not (U(1) case), one of the two Kosterlitz--Thouless
transitions of the ${\bf Z}_N$ models is pushed to infinity or to zero,
respectively, and disappears for $N=\infty$.}
\end{figure}
\fi
In the SU$(N)$ case the global symmetry of the effective spin model is again
${\bf Z}_N$, but the model is much less trivial. In each site instead of a
single spin which can take $N$ values we have a collection of $N$  spins
(the eigenvalues of the SU$(N)$ matrix) and the nearest neighbour effective 
action  between these collections of spins is highly non trivial. If we label
the eigenvalues with an index $i$ then the action, which, according to the SY
conjecture, is local in the real
space, turns out to be non local in the index space. It is exactly this
last feature which makes the difference and allows 
 the existence
of a high $\beta$ phase in which the Polyakov loop expectation value is
 different from zero. 
Once the large $N$ limit has been taken and a master field
 configuration for the eigenvalues has been assumed, it turns out to be
 very difficult to understand  how the non locality in the index space could
 allow to circumvent the Mermin--Wegner theorem.
However, the above discussion on the phase diagram of the 
${\bf Z}_N$ models allows us to gain some intuition of this phenomenon
from a different point of view and, in particular, 
 to see that there is no contradiction between the SY conjecture and the
presence of a deconfined phase in the large $N$ limit of SU$(N)$ LGT's.

In fact, let us take  
a value of $N$ large, but finite and let us assume that for large $\beta$ 
the SU$(N)$ LGT admits a deconfined phase. This means that all the Polyakov
loops are ``frozen'' in the same direction (one of the ${\bf Z}_N$ roots of
unity). If $\beta$ is large enough all the eigenvalues (namely all the spins in
each given site) take the same value and the non-locality in the index space
becomes trivial. The effective action becomes $N^2$ times the action of the 
${\bf Z}_N$ clock model\footnote{Let us stress that this is a 
simplified picture
and that fluctuations in the eigenvalues distribution are present for any
finite value of $\beta$. These fluctuations are very important in the discussion
of the phase diagram of the theory and we shall deal with them in sec. 4. 
But with respect to the present discussion they only represent a higher order
correction and do not affect the validity of the argument.} .  
This means that the critical
coupling which separates the deconfined (i.e. ordered, in the language of
the ${\bf Z}_N$ model) phase from the intermediate phase scales\footnote{
 This is already apparent in the
definition of the coupling (see eq. (\ref{wilson}) ) 
where we have factorized a $N^2$
factor to ensure a smooth $N\to\infty$ limit.} with $N^2$.
Thus we are exactly in the region where the  ${\bf Z}_N$ clock model admits a
broken symmetric phase and we see explicitly that there is no contradiction
between our assumption of the presence of a deconfined phase and the SY
dimensional reduction. Moreover we see that we can take smoothly in both models
(LGT and  spin model) the large $N$
limit and  keep the agreement between the two phase diagrams.

This allows us to better understand the relation between  
SU$(N)$  and ${\bf Z}_N$ LGT's in (2+1) dimensions. For suitable values of
$\beta$ both have the same two-dimensional
${\bf Z}_N$ spin model as effective action, but they are described by two very
different regimes in the coupling space of the spin model.. Low $\beta$ (of order
unity) for the ${\bf Z}_N$ LGT, high $\beta$ (of order $N^2$) for the SU$(N)$
theories. This consideration shares some more light on the SY conjecture and
explains how is it possible that gauge models, which have a very different 
dynamics (like the ${\bf Z}_N$ and SU$(N)$ ones) could be described by the same
effective action. 
\bm
\subsection{Character expansion in the large $N$ limit}
\ubm
An important role in the following analysis will be played by the
character expansion.
Let us briefly summarize few results (for more details see Ref.~\cite{dz}).
We shall particularize them to the SU$(N)$ case, 
but most of them hold for any Lie
group $G$ and with minor modifications also for discrete groups.
The irreducible characters $\chi_r(U)$ are the traces of the irreducible 
representations (labelled by $r$)  of
the group. They form a complete orthonormal basis for the class functions  on 
the group. A function $f(U)$ on the group is called a ``class function'' if it
satisfies the relation:
\eq
f(U)=f(VUV^{\dagger}) \hskip 2cm \forall\, V\in {\rm SU}(N)~.
\en
In particular, the characters themselves are class functions.
The pure gauge action, eq. (\ref{wilson}), is a class function.

The following orthogonality relations between characters hold:
\eq
\int DU\, \chi_r(U)~\chi^*_s(U)=\delta_{r,s}~,
\label{c2}
\en
\eq
\sum_{r} d_r \chi_r(U~V^{\dagger})= \delta(U,V)~,
\label{c3}
\en
where $DU$ denotes the Haar measure (normalized to unity) on SU$(N)$ and
$d_r$ denotes the dimension of the $r^{\rm th}$ representation.

Besides the above orthogonality relations we shall make use
of  two important integration formulas of the characters, namely:
\eq
\int DU\, \chi_r(V_1 U)~\chi_s(U^{\dagger}V_2)=\delta_{r,s}\frac{\chi_r(V_1 
V_2)}{d_r}~;
\label{c4}
\en
\eq
\int DU\, \chi_r(UV_1U^{\dagger}V_2)=\frac{1}{d_r}\chi_r(V_1)\chi_r(V_2)~.
\label{c5}
\en
Any class function can be expanded in the basis of the characters:
\eq
f(U)=\sum_r \chi_r(U) f_r~,
\en
where the sum is over the set of all irreducible representations of the group,
and the coefficients $f_r$ are given by
\eq
f_r\equiv \int DU\, \chi^*_r(U) f(U)~.
\en
Let us construct now the character expansion 
for the Wilson action.

The Boltzmann factor associated to each plaquette in
the Wilson action is (we neglect the index ``$t$'' or ``$s$'' of $\beta$, 
which is irrelevant for the following analysis) :
\eq
{\rm e}^{N\beta {\rm Re} {\rm Tr} G} = \sum_r F_r(\beta) \chi_r(G)~,
\label{chexp}
\en
where $G$, given by eq. (\ref{plaq}) above, denotes the ordered product of the
link variables around the plaquette and the coefficients $F_r$ are given by:
\eq
F_r(\beta)\equiv
\int DU\, {\rm e}^{N\beta {\rm Re} {\rm Tr} U}\chi^*_r(U)
= \sum_{n=-\infty}^{\infty} {\rm det}\,I_{r_j-j+i+n}(N\beta)~.
\en
The $r_j$'s are a set of integers labelling the representation 
$r$ and they are constrained by: $r_1\geq\cdots\geq r_N=0$.
The indices $1\leq i,j \leq N$ label the entries of the $N\times N$ matrix of 
which the determinant is taken and
$I_n(\beta)$ denotes the modified Bessel function of order $n$.

As a consequence of the factor $d_r$ at the denominator in eq.s  (\ref{c4},
\ref{c5}) the relevant coefficients in the character expansion (\ref{chexp}),
namely the ones that will appear in the strong coupling expansions, are not the
$F_r$ themselves, but the following normalized coefficients:

\eq
\label{dierre}
D_r(\beta)=\frac{F_r(\beta)}{d_r F_0(\beta)}~.
\en
These coefficients have two remarkable properties, which will be important in
the following:
\begin{itemize}
\item[{\em a)}]
In the large $N$ limit they have a very simple form, if the limit is taken
keeping $\beta< 1$ fixed: 
\eqa
F_0(\beta)&\sim & {\rm e}^{\left(\frac{N\beta}{2}\right)^2}~, \nonumber \\
F_f(\beta)&\sim &\frac{N\beta}{2}\, {\rm e}^{\left(\frac{N\beta}{2}\right)^2}~, 
\nonumber
\ena
where the index $f$ denotes the 
fundamental representation (whose dimension is $N$).  
The above relations imply that in the large $N$ limit 
\eq
D_f(\beta)=\frac{\beta}{2}~.
\label{241}
\en
Similar simplified relations hold also for higher representations.
\item[{\em b)}]
For any fixed value of $N$, in the large $\beta$ limit the coefficients $D_r$
become equivalent to the heat kernel coefficients: 
\eq
\lim_{\beta\to\infty} D_r(\beta)={\rm e}^{-\frac{C_r}{2N\beta}}~,
\label{242}
\en
where $C_r$ denotes the quadratic Casimir in the $r^{\rm th}$ representation.
As it is easy to see, this exponential form greatly simplifies the 
construction of strong coupling expansions. Moreover, this limit
is particularly relevant
for us, since this is exactly the situation that we have for the timelike part
of the action when the asymmetry parameter $\rho$ is sent to infinity.
\end{itemize}
\subsection{Scaling behaviour}
\label{scalsec}
The ultimate test of the correctness of any lattice regularization
is that, as the continuum limit is approached, the various dimensional
quantities  follow the correct scaling behaviour.
This scaling behaviour can be easily obtained by writing explicitly the
dependence on the lattice spacing $a$ of the relevant (dimensional) observables.
The form of these scaling laws depends on the number of spacetime dimensions
 of the lattice. 
So we shall discuss separately the two cases $d=2$ and $d=3$ which are the most
relevant ones for the physical applications.
We shall concentrate in particular on the scaling  behaviour of the 
critical temperature $T_c$ (which has the dimension of a mass) since
this is the simplest physical observable which we can study with our techniques.
We shall be interested in the scaling laws as  functions of the coupling $\beta$
and of the lattice size in the time direction $N_t$.
We shall study for simplicity the case of a symmetric lattice
$\beta_s=\beta_t\equiv\beta\equiv\frac{4}{g^2}$.  If $\rho\neq 1$, the 
change in the scaling laws due to the asymmetry is completely encoded in the
equations (\ref{rel3}) and (\ref{rel4}) and in the 
functions $c_\sigma(\rho)$ and $c_\tau(\rho)$ discussed in sec. 2.1.
\bm
\subsubsection{$d$=2}
\ubm
In the $d=2$ case the coupling constant $g^2$ has the dimensions of a mass
(see eq. (\ref{coupsym})). This simplifies the analysis since in this case the
coupling constant itself sets the overall mass scale. So near the continuum
limit a physical observable, like $T_c$, with the dimensions of a mass can be
written, according to the renormalization group equations, as a series in 
powers of $g^2$. Hence in terms of the coupling $\beta$ we have:
\eq
aT_c=\frac{a_1}{\beta}+\frac{a_2}{\beta^2}+\cdots~.
\label{251}
\en
The critical temperature is obtained by looking at the critical coupling
$\beta_c$ at which the deconfinement transition occurs; hence, 
if the lattice size in the time direction is $N_t\equiv\frac{1}{Ta_t}$, 
we can rephrase eq. (\ref{251}) as a scaling law for the behaviour of the 
critical coupling $\beta_c$ as a function of $N_t$. 
Keeping only the first term in eq. (\ref{251}) we find:
\eq
\beta_c(N_t)=a_1 N_t~.
\label{252}
\en
Thus  from the simple observation that in (2+1) dimensions $g^2$ has 
dimensions of a mass we immediately deduce that near the continuum limit
$\beta_c$ must be a {\em linear} function of $N_t$.
If the lattice is asymmetric we can use eq.s (\ref{rel3}) and (\ref{rel4})
to reconstruct the equivalent symmetric coupling and then use again eq.
(\ref{252}).
\bm
\subsubsection{$d$=3}
\ubm
This case is less trivial, since by
looking at eq.s (\ref{couplings}, \ref{coupsym}) we see 
that  for $d=3$ the coupling constant $g^2$ is dimensionless and the theory
dynamically generates a dimensional scale $\Lambda_L$ in units of
which we must measure any dimensional quantity on the lattice.
The dependence of the lattice spacing on $\beta$  can be reconstructed in the
continuum limit by using  the renormalization group equations. The well known
two loop result is:
\eq
a\Lambda_L=\left({b_0 g^2}\right)^{-\frac{b_1}{2b_0^2}}
\exp\left(-\frac{1}{2b_0g^2}\right)~,
\label{lambda}
\en
where  $b_0, b_1$
are the first two coefficients of the Callan--Symanzik equation which for 
SU$(N)$  are:
\eq
b_0=\frac{11 N}{48\pi^2}~,\hskip 1cm
b_1=\frac{34}{3}\left(\frac{ N}{16\pi^2}\right)^2~.
\en
Plugging eq. (\ref{lambda}) into the definition of the critical temperature:
\eq
T_c=\frac{1}{aN_t}~,
\label{tcsym}
\en
we find:
\eq
\frac{T_c}{\Lambda_L}=\frac{1}{N_t}
\left(\frac{24\pi^2\beta}{11}\right)^{-\frac{51}{121}}
\exp\left(\frac{12\pi^2\beta}{11}\right) .
\label{tccont}
\en
If the continuum limit is correctly taken then the ratio $T_c/\Lambda_L$
should approach for large enough values of $\beta$ 
(hence, in our case, also for
large values of $N_t$) a constant value. Inserting this constant into eq.
(\ref{tccont}) and keeping only the first perturbative contribution 
in eq. (\ref{lambda}) we immediately see that, for $d$=3,
$\beta_c$ is  a {\em logarithmic} function of $N_t$.

\section{Construction of the Effective Action}
\label{effacsec}
The aim of this section is to construct an effective action for the 
finite temperature LGT in terms of Polyakov loops only.
In agreement with the Svetitsky--Yaffe program outlined above, we shall
try to integrate  over the space-like 
variables $U_i(\vec{x},t)$ in the Wilson action (\ref{wilson}) so that the only 
remaining degrees of freedom will be  at the end the Polyakov loops. 
The resulting effective action will live in $d$ dimensions, one dimension 
less than the starting model.
The integration over the space-like variables can be done in principle 
in two distinct steps. First one integrates over all the space link variables
except for the ones on an arbitrarily chosen time layer. 
The result is a lattice theory with $N_t=1$ in which the time-like 
links are the open Polyakov loops. 
This first step will be denoted for obvious reasons as 
``dimensional reduction''. The second and last step consists in the 
integration over the remaining space-like link variable, leading to an 
effective action with the Polyakov loops as the only dynamical variables.  
As already remarked in the introduction the integration over
the space-like variables cannot be performed exactly and some approximation is 
needed. The approach that we shall follow  
consists in treating the time-like part $S_t$ of 
the Wilson action as a Born term and the space-like part $S_s$ as a 
perturbation.
This means making a strong coupling expansion in $\beta_s$ while 
treating the time-like part of the action exactly.
The first term of this expansion, which we shall call in the
following ``zeroth order approximation''  simply corresponds to neglecting the 
space-like plaquettes. 
In this case both steps in the integration over the space-like links can be 
performed exactly, at least as a character expansion, but the result is  
in some respect unsatisfactory. 
In this limit in fact the result is exactly the same that one would obtain with
a standard Migdal--Kadanoff \cite{migkad} 
bond-moving approximation and it has the same drawbacks.
In fact, in $3+1$ dimensions, it gives a good approximation of the
whole theory only for $N_t=1$ while for larger $N_t$'s,
although it  still  gives a good {\em qualitative} 
description of the phase diagram, 
it fails to predict the critical properties of the
deconfinement transition, namely the scaling behaviour of the critical coupling
$\beta_c$ as a function of $N_t$. This is a major problem, since it is
only by following the correct scaling behaviour that one can finally take the
continuum limit and extract, for instance, a reliable estimate for the
deconfinement temperature.
The situation is different in $2+1$ dimensions where, at least to the 
leading order $N_t$, the scaling behaviour of the ``zeroth order approximation''
and of the full theory coincide. As a consequence we may expect that
the ``zeroth order approximation'' gives in this case reliable results
even at large values of $N_t$. 
Qualitatively the picture is the following: in $3+1$ dimensions
the statistical weight of the space-like plaquettes is large enough to
affect the scaling properties; hence they can be neglected only for small
values of  $N_t$, 
where the critical value of $\beta$ is small and the correlations between 
Polyakov loops induced by the space-like plaquettes negligible.
On the contrary in $2+1$ dimensions the weight of the space-like plaquettes 
is too small to affect the scaling, and there are a good indications that they
are not important not only for small $N_t$ but also near the continuum limit.

The effect of the space-like plaquettes 
can be taken into account perturbatively,
order by order in $\beta_s$. 
In~\cite{ourpaper} we constructed the first non trivial order in
$\beta_s$ for the SU(2) model and indeed found a better agreement
with the expected scaling law in $d=3$. Unfortunately the effective action at
this order becomes very cumbersome, no exact solution can be found and one has
to rely on mean field estimates of the critical coupling. Moreover, it is easy
to see that the
complexity of the calculations further increases as higher orders 
are taken into account or if larger values of $N$ are considered.
 
In the large $N$ limit   a
completely different approach, which is  more elegant and powerful, 
is available. By using  suitable modifications of the 
Eguchi--Kawai techniques it is 
possible to perform an exact dimensional reduction, and obtain an action 
with $N_t=1$ {\em  which is completely equivalent to the original 
Wilson action}. Although the last step, namely the integration over the 
remaining space-like link 
variables, still proves to be too difficult, this is a major improvement 
with respect to previous approaches. In particular, as discussed in detail in 
sec. 6, the solution of the dimensionally reduced action in the $\beta_s=0$
limit gives the next to leading order in the scaling behaviour of $\beta_c$,
which turns out to be in excellent agreement, in $2+1$ dimensions, with the 
available Montecarlo simulations.

This section is  divided  into two parts.
In the first one (sec. 3.1) we review the construction of  the ``naive'' 
effective actions, namely the ones obtained with the ``zeroth order 
approximation''.
In sec.s 3.1.1 and 3.1.2 we shall construct respectively  
the zeroth order approximation of the  heat kernel action (which can
be thought of as the $\rho\to\infty$ limit of the standard Wilson action) and
of the Wilson action itself. 
Then we shall obtain the corresponding scaling behaviours (sec. 3.1.3), 
and we shall discuss the reasons why they are  unsatisfactory. 
Finally we shall describe a simplified version of the effective action 
(sec. 3.1.4), which is simple enough to be solved exactly 
and at the same time accurate enough to give 
a good {\it qualitative} description
of the phase diagrams of the deconfinement transition.  
All the material collected in this first part is rather old. 
It mainly refers to results already obtained at the beginning of the
eighties, even if they are discussed here in the new framework of our strong
coupling expansion.

On the contrary, in the second part we deal with some completely new results.
First we shall discuss how to construct higher order terms in the
$\beta_s$ expansion (sec. 3.2.1).
We shall also show that, in spite of the 
$N^2$ factor in front of the space-like 
coupling, the expansion in powers of $\beta_s$  is convergent order by order 
in the large $N$ limit. This is a non-trivial and important check of the 
consistency of the strong coupling expansion in $\beta_s$.
Finally in sec. 3.2.2 we shall derive an exact dimensional reduction
by using techniques typical of the Eguchi--Kawai models. 
\bm
\subsection{Naive dimensional reduction ($\beta_s=0$) 
and related effective actions.}
\ubm
The starting point for our considerations is the complete action $S_{\rm W}$
defined in (\ref{wilson}) on a lattice with arbitrary asymmetry parameter
$\rho$. More precisely we shall consider
\eq
\label{fullac}
e^{S_{\rm W}} = \prod_{\vec{x},t,i} \left\{\sum_r d_r D_r(\beta_t) 
\chi_r (G_{0i}(\vec{x},t))  
\right\} e^{N^2 \sum_{i<j}~  \beta_s \hat{G}_{ij}(\vec{x},t)}~,
\en
where a character expansion of the contributions of the time-like
plaquettes has been performed according to eq.s  (\ref{chexp}) and
(\ref{dierre}).
The naive dimensional reduction  can be achieved by simply setting
$\beta_s=0$ in (\ref{fullac}). In this case  it is easy to integrate out  
 the space link variables $U_i(\vec{x},t)$. 
This can be done exactly because each link variable $U_i(\vec{x},t)$ 
belongs only to two timelike plaquettes.
In fact, consider the ladder of plaquettes obtained from any given 
timelike plaquette by moving in the time direction. Its contribution to the
action is:
\eq
\label{a1}
A(\vec{x},i)\equiv
 \prod_{t} \left\{\sum_r d_r D_r(\beta_t) \chi_r (G_{0i}(\vec{x}
,t)) \right\}~. 
\en
The integration over the
first $N_t -1$ space-like link variables on the ladder can be done by 
using eq. (\ref{c4}), leading to the following result:
\eq
\label{a2}
\sum_r d_r \left[ D_r(\beta_t)\right]^{N_t} \chi_r \left(U_i(\vec{x},t=0) 
P(\vec{x}+\hat{\i}) U^\dagger_i(\vec{x},t=N_t)P^\dagger(\vec{x})\right). 
\en
The effect of this integration is to reduce the original action (which had
$N_t$ space-like slices in the time direction) 
to an effective action with only one link in the
timelike direction and one space-like slice. 
The timelike plaquettes of this reduced action are very peculiar: the two
space-like links coincide due to the periodic boundary conditions, while 
the two timelike links exactly correspond to the Polyakov lines defined 
in eq. (\ref{polya2}), which are indeed the  degrees of
freedom  which we are interested in  to construct our effective action. 

The last step is now to integrate the remaining space-like links.
Because of the periodicity in the time direction this integration
is of the type given in (\ref{c5}).

By repeating the same procedure to all ladders the integration over all 
space-like link variables can be explicitly performed, leading to the 
following effective action for the Polyakov loops at  $\beta_s=0$:       
\begin{eqnarray}
e^{S_{\rm Pol}(\beta_s=0)} &=& \int \prod_{\vec{x},t,i} DU_i(\vec{x},t) 
\left\{ \sum_r d_r \chi_r \left[ D_r(\beta_t) \right]^{N_t} 
(G_{0i}(\vec{x},t) )  \right\} \nonumber \\ 
& = &\prod_{\vec{x},i}
\sum_r  \left[ D_r(\beta_t) \right]^{N_t} \chi_r(P(\vec{x}+\hat\i)) 
\chi_r(P^\dagger(\vec{x})) ~.  
\label{zeta1}
\end{eqnarray}
Notice that up to eq. (\ref{a2}) we were still dealing 
with an ordinary (even if asymmetric) lattice gauge theory. 
With the last integration, the gauge theory
disappears and we end up with a spin model in one dimension less than the
original model. However it is still possible to recognize some remnant of
 the original gauge
symmetry in the invariance of the characters under transformations of the type
$V\to W^{-1}VW$. 
Everything that has been done so far,  in particular eq. (\ref{zeta1}),
is valid for any value of the asymmetry parameter $\rho$. In the next
subsections we shall consider the two extreme situations, namely 
$\rho=\infty$ and $\rho=1$. It should be stressed that the effective actions   
obtained from (\ref{zeta1}) by setting $\rho=\infty$ and $\rho=1$ are equivalent 
regularizations of the effective $\beta_s=0$ theory in the continuum,
but they differ away from the continuum limit.
\bm
\subsubsection{$\rho\to\infty$ limit: heat kernel action}
\ubm
The large $\rho$ limit of eq. (\ref{zeta1}) can be easily obtained
from the asymptotic behaviour of the $D_r$ coefficients given in
(\ref{242}), namely\footnote{$\beta_t$ is a free parameter of the
theory and as such is not related to $N_t$. It becomes a function
of $N_t$ if we require that as $N_t$ changes the effective model
(\ref{zeta1}) describes the same physics. This implies for instance  
that the expectation value of the Polyakov loop and of other observables
does not depend on the value of $N_t$.}
\eq
\label{largero}
\left[ D_r(\beta_t) \right]^{N_t} 
\stackrel{\rho \rightarrow\infty}{\longrightarrow} 
\exp \left(- {C_r n_t \rho \over 2 N \beta_t(n_t \rho)}\right) = 
\exp \left(- {C_r n_t  \over 2 N \beta_{\rm hk}}(n_t)\right)~, 
\en
where we have used the fact that $N_t = \rho n_t$, and defined a new
coupling $\beta_{\rm hk}$ as
\eq
\label{betahk}
\beta_{\rm hk}(n_t) = \lim_{\rho \to \infty} \frac{\beta_t(n_t \rho)}{\rho}
\en
This new ``heat kernel'' coupling is related in the continuum 
limit, that is for large $n_t$, to the coupling $\beta$ of the 
symmetric lattice by the relation (\ref{rel3})\footnote{Notice however 
that, since the space-like plaquettes have been neglected, the quantum 
corrections and hence the numerical values of $\alpha_{\tau}^{0,1}$ may be 
different from the ones given in sec. 2.1.} :
\eq
\label{bew}
\beta_{\rm hk}(n_t)  \stackrel{n_t \rightarrow \infty}{\sim} \beta(n_t) 
+ \alpha^0_{\tau}
\en
Away from the continuum limit, the effective action with $\rho=\infty$
and the one defined on a symmetric lattice are not analytically related,
and $\beta_{\rm hk}$ should be regarded as an independent coupling.
By inserting (\ref{largero}) into (\ref{zeta1}) we find
\eq
\label{zetaheat}
e^{S_{\rm Pol}^{\rm hk}(\beta_s=0)}  = \prod_{\vec{x},i}
\sum_r  \chi_r(P(\vec{x}+\hat\i)) \chi_r(P^\dagger(\vec{x})) 
\exp \left(- {C_r n_t \over 2 N \beta_{\rm hk}}\right)~,
\en
which is the final form of the heat kernel ($\rho \to \infty$) effective
action for the Polyakov loop.

\subsubsection{Symmetric lattice: Wilson action}
Let us consider now the case of a symmetric lattice ($\rho=1$),
which is particularly important if one wants to compare the
analytic results with the results of Montecarlo simulations.
The action is obtained from (\ref{zeta1}) by replacing 
$\beta_t$ with the symmetric coupling $\beta$ and $N_t$
with $n_t$:
\eq
e^{S^{\rm W}_{\rm Pol}(\beta_s=0)} = \prod_{\vec{x},i}
\sum_r  [D_r(\beta^)]^{n_t}
\chi_r(P(\vec{x}+\hat\i)) \chi_r(P^\dagger(\vec{x}))~, 
\label{zetawil}
\en
As a consequence of eq. (\ref{largero}) the heat kernel 
and the Wilson effective actions, given by eq.s (\ref{zetaheat})
and (\ref{zetawil}), coincide in the large $\beta_{\rm hk}$ and $\beta$ limit
provided the ratio of the two couplings goes to $1$ in the continuum
limit ($n_t \to \infty$).  This is guaranteed by eq. (\ref{bew}) as the
constant term $ \alpha^0_{\tau}$ is negligible in that limit. 
Hence, as expected, the heat kernel
and the Wilson effective actions have the same continuum limit.
For finite $n_t$ the two model are different and the coupling
constants $\beta_{\rm hk}$ and $\beta$ are in principle unrelated.
However, in the strong coupling limit (small 
$\beta$), $S^{\rm W}_{\rm Pol}$ and $S^{\rm hk}_{\rm Pol}$  coincide again
provided a suitable relation between the couplings of the two actions
is established. In fact in that limit one retains in the character expansion
of eq.s (\ref{zetaheat}) and (\ref{zetawil}) only the contributions of 
the fundamental representation and
its conjugate, which are the leading contributions in the strong coupling
limit.
A direct comparison of the formulas, taking into account eq. (\ref{241})
and the relation $C^{(2)}_f=N$, gives:
\eq
\beta=2\,{\rm e}^{-\frac{1}{2\beta_{\rm hk}}}+\ldots \hskip 2cm 
(\beta,\beta_{\rm hk}\to 0)~,
\label{323}
\en
where the dots denote higher order corrections which vanish in the
$\beta\to 0$ limit.
\subsubsection{Scaling behaviours}
This analysis and the knowledge of the explicit form of the two actions,
allows us to obtain some general information on the scaling behaviour that we
may expect for $T_c$. 
In the heat kernel case the result is very
simple. It can be read directly from eq. (\ref{zetaheat}) by requiring that the
physics does not change if we simultaneously change $\beta_{\rm hk}$ and $n_t$.
This implies a linear scaling of $\beta_{\rm hk}$ with $n_t$. 
\eq
\beta_{\rm hk}(n_t)=J n_t~.
\label{331}
\en
The ``renormalized coupling'' $J\equiv\beta_{\rm hk}(n_t=1)$ 
is the relevant parameter in the continuum
limit; it will play a major role in the following.

In the Wilson case, due to the complicated form of the coefficients
$D_r(\beta)$ the situation is less simple. However we can all the same
deduce {\em asymptotic} scaling relations from the {\em exact}
scaling of the heat kernel action (eq. (\ref{331})) and the relations between
$\beta$ and $\beta_{hk}$ (eq.s (\ref{bew},\ref{323})).
Let us define also in the Wilson case a ``renormalized coupling'' $J_{\rm W}$  
by setting $J_{\rm W} = \beta(n_t=1)$.
We must distinguish then the  weak coupling from the strong coupling regime. 
In the weak coupling  the two actions coincide and also
for the Wilson action we find a linear scaling. In fact by simply replacing 
eq. (\ref{331}) into (\ref{bew}) we find:
\begin{equation}
\beta(n_t)= J \,n_t~- \alpha_{\tau}^0 = 
(J_{\rm W} + \alpha_{\tau}^0 )\,n_t~- \alpha_{\tau}^0~,
\label{331bis}
\end{equation}
where in the last step we have anticipated the relation between 
$J$ and $J_{\rm W}$ discussed below (eq. (\ref{334})).
As already mentioned the numerical value of $\alpha_{\tau}^0$ may not coincide
with the one found by Karsch in ~\cite{k81} and reported in sec. 2.1, as the 
contributions of the space-like plaquettes have not been taken into account
here. The value of $\alpha_{\tau}^0$ is calculated in this context in
sec. 6.1 and it turns out to be close to the one calculated in  ~\cite{k81}
including the space-like plaquettes.
In the strong coupling regime by using eq. (\ref{241}) 
we find 
\eq
J_{\rm W}  = 2 \left( \frac{\beta(n_t)}{ 2} \right)^{n_t}~.
\label{332}
\en
Let us stress again that both eq. (\ref{332}) and the linear scaling in 
the weak coupling
regime must be considered as asymptotic behaviours, while eq. (\ref{331})
for the heat kernel action is exact.
It is useful to write explicitly the relation between the normalized 
couplings $J_{\rm W}$ and $J$. They can be found immediately by setting $n_t=1$
in (\ref{bew}) in the weak coupling and from  (\ref{323}) 
in the strong coupling. In the weak coupling regime we have
\eq
J_{\rm W} = J -  \alpha_{\tau}^0~,
\label{334}
\en
and in the strong coupling 
\eq
J_{\rm W} = 2\,{\rm e}^{-\frac{1}{2J}}+\ldots ~.
\label{335}
\en

The important lesson that we learn from this analysis is that in the continuum
limit all these actions obey a {\em linear} scaling law.
Notice that  this scaling behaviour does not depend on the number of spatial
dimensions $d$ of the model. This is clearly an artifact of our
approximation, and a quite disappointing one, since we know very 
well that the scaling laws do indeed depend on $d$. 
The situation is somehow reminiscent of what
happens in the case of the mean field approximation. However, 
while in the case of
mean field we know that the approximation becomes better and better as $d$
increases, here the situation is exactly reverted, and the approximation that we
make by neglecting the space-like plaquettes 
becomes worse and worse as $d$
increases. As a matter of fact this linear scaling
agrees with what we expect for the (2+1) dimensional gauge models, but 
disagrees with our expectations in the (3+1) dimensional case. 

\subsubsection{A simplified effective action}
\label{simp}
It is instructive, before analysing the rather complex model given in 
(\ref{zetaheat}), to study the simplified effective action which is obtained
from the Wilson one, eq. (\ref{zetawil}), by truncating the character expansion
to the lowest order term:
\eq
S_{\rm eff}(J_{\rm W})=\sum_{\vec x}~{\rm Re}\,\left\{J_{\rm W}\sum_{i}
~({\hat P}(\vec x))({\hat P}^\dagger(\vec{x}+\hat\i))\right\}~.
\label{seff}
\en
\iffigs
\begin{figure}
\begin{center}
\null\hskip 1pt
\epsfxsize 10cm
\epsffile{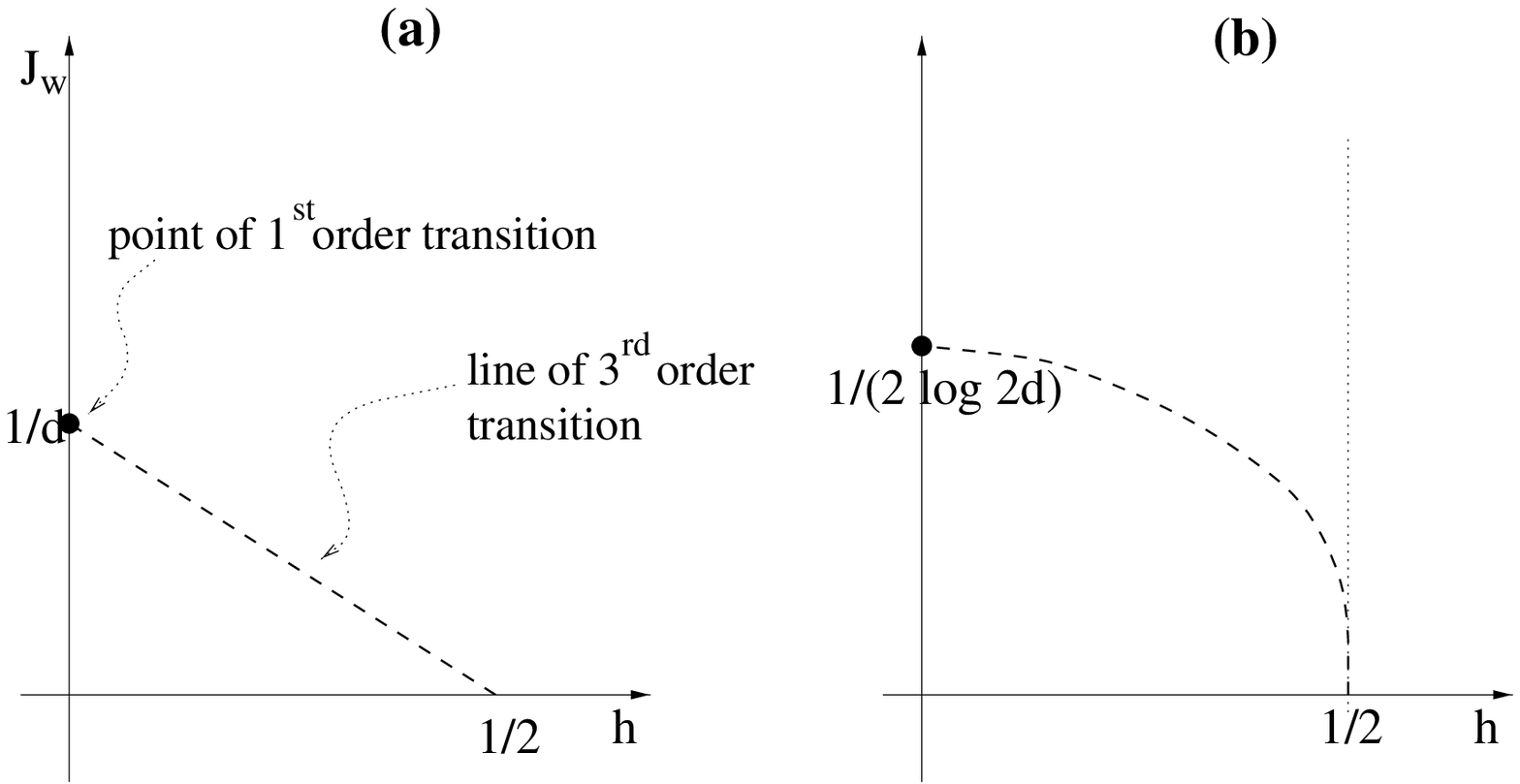}
\end{center}
\vskip 0.2cm\noindent
\mycaptionl{Fig. 2}{{\bf (a)} Phase diagram resulting
from the large $N$ limit of the
effective model eq.(\ref{sjh}); 
{\bf (b)} The same diagram expressed in terms of the
heat kernel coupling $J$ (related to the Wilson one by $J_{\rm W} = 2
\exp \{- 1/(2J)\}$, see sec. 3.1.3, eq. (\ref{335})).}
\label{phdiagfig}
\end{figure}
\fi
This model is a rather crude approximation of the original action.
Due to the truncation, it does not even treat exactly the timelike part of the
action, so it cannot be trusted in the weak coupling regime or in the
$\rho\to\infty$ limit where all the terms in the character expansion become
important. However this model is very simple to study, it can be easily
solved exactly and for this reason it has been very popular in the past years.
It was  extensively studied in the literature, 
both at finite $N$~\cite{og,djk} and in the large $N$ 
limit~\cite{dp,gnr}, with strong coupling~\cite{ps,gk} and 
mean field~\cite{djk} approximations,
and with Montecarlo simulations~\cite{djk,dh}.
Notwithstanding its simplicity it turned out to be 
a valuable tool to understand the general features
of the phase diagram.  Its exact solution in the large $N$ limit was
derived in~\cite{dp,gnr}, for any value of the space dimensions $d$,
leading to a phase diagram with a first order deconfinement transition located 
at $J_{\rm W}=1/d$. 
It is even possible to solve exactly the more general model in which the
Polyakov loop is also coupled to an external ``magnetic'' field $h$,
\eqa
S_{\rm eff}(J_{\rm W},h) & = & \sum_{\vec x}~{\rm Re}\, 
\left\{J_{\rm W}\sum_{i}
~{\rm Tr}(P(\vec x)){\rm Tr}(P^\dagger(\vec{x}+\hat\i))\right\}
~~+ \nonumber \\
& + & hN\sum_{\vec x}\left[
{\rm Tr}(P(\vec x) + P^\dagger(\vec{x}))\right]~.
\label{sjh}
\ena
In this framework it becomes apparent that the first order
phase transition is the end point of a line of third order phase
transitions of the Gross--Witten \cite{grwit} type, located along the line
\eq
J_{\rm W}=\frac{1-2h}{d}~,~~~h,J_{\rm W}\geq 0~.
\en

As we shall see below, keeping the whole complexity of 
the heat kernel (or Wilson) effective action, the phase diagram turns out to 
be more rich and complex and also the location of the deconfinement transition 
changes. 
In particular in sec. 5.3 we shall discuss the phase diagram of the Wilson
action  coupled to  a ``magnetic'' field $h$ as in eq. (\ref{sjh})  
up to the second order in the strong coupling expansion. 

\subsection{Toward an exact dimensional reduction}
\bm
\subsubsection{Higher orders in $\beta_s$}
\ubm
We have seen in the previous subsections that the naive (zeroth order) 
effective actions
fails to catch the correct scaling behaviour of the theory.
Our hope is that we can overcome this problem  by taking into account
higher order contributions in $\beta_s$.
In~\cite{ourpaper} we tested
this expectation in the  case of the SU(2) model in (3+1) dimensions and
found a remarkable improvement in the scaling behaviour of the model by adding 
the terms of order $\beta_s^2$. This is well summarized in Fig. 3 where the two
scaling behaviours of the zeroth order and of the $\beta_s^2$ order actions are
compared with the results of the Montecarlo simulations. 
\iffigs
\begin{figure}
\null\vskip -1cm\hskip 2cm
\epsfxsize = 10.8 truecm
\epsffile{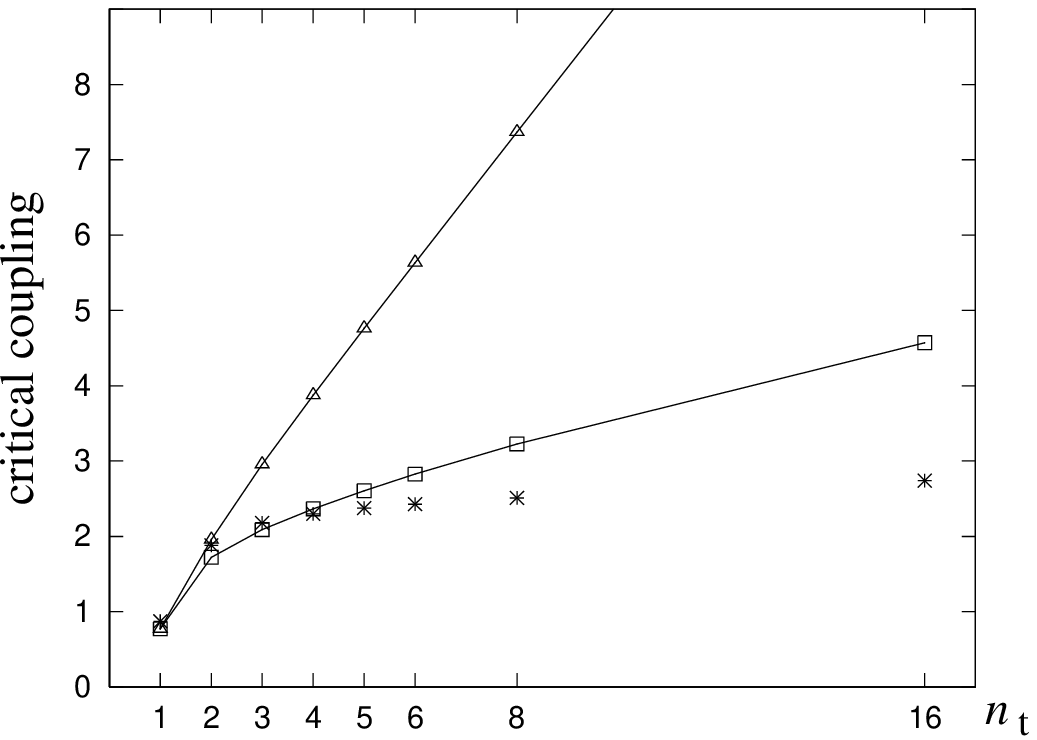}
\vskip 0.3cm
\mycaptionl{\bf Fig. 3}{Values of the critical coupling 
$\beta_c$ in the SU(2) theory are plotted for
different values of the number of time-like links $n_t$. Results obtained
with Montecarlo simulations, which are denoted by *,
are compared with those obtained in~\cite{ourpaper}:
$\triangle$ represents the data for $\beta_{c}|_{0}$,
the critical coupling in the zeroth order approximation and $\Box$  
the data for $\beta_c|_{1}$, 
the critical coupling including the effect of
the space-like plaquettes al the lowest non trivial order.}
\label{data}
\end{figure}
\fi

A similar analysis in the large
$N$ limit is still missing. In this  subsection we shall give some
preliminary result in this direction and in particular we shall show that, at
least, the problem is well defined and
higher order contributions do not diverge as $N\to\infty$.

The expansion in powers of $\beta_s$ of the full partition function
(\ref{fullac}) can be formally written as:
\begin{eqnarray}
e^{S_{\rm Pol}} & = & e^{S_{\rm Pol}(\beta_s=0)} \{ 1 + N^2 \beta_s 
\sum_{\vec{x},t} \sum_{i<j} <G_{ij}(\vec{x},t)>  \nonumber 
\\  & +&  {1 \over 2} N^4 \beta_s^2 \sum_
{\vec{x}_1,t_1} \sum_{\vec{x}_2,t_2} \sum_{i_1<j_1} \sum_{i_2<j_2} 
<G_{i_1j_1}(\vec{x}_1,t_1) G_{i_2j_2}(\vec{x}_2,t_2)> + \ldots 
\}
\label{zetab}
\end{eqnarray}
where the expectation values are taken with respect to the unperturbed 
action, namely with $\beta_s=0$.
The series in brackets at the r.h.s. of eq. (\ref{zetab}) 
can be re-exponentiated, leading to the following expansion for $S_{\rm Pol}$:
\begin{eqnarray} 
S_{\rm Pol} &=&S_{\rm Pol} (\beta_s=0)+  N^2 \beta_s \sum_{\vec{x},t} 
\sum_{i<j} <G_{ij}(\vec{x},t)>_c  \nonumber \\ 
 &+& {1 \over 2} N^4 
\beta_s^2 \sum_{\vec{x}_1,t_1} \sum_{\vec{x}_2,t_2} \sum_{i_1<j_1} 
\sum_{i_2<j_2} <G_{i_1j_1}(\vec{x}_1,t_1)
G_{i_2j_2}(\vec{x}_2,t_2)>_c + \ldots~,  
\label{zetab2}
\end{eqnarray}
where the subscript $c$ denotes  the connected part, for instance 
$$<G_1 G_2>_c = <G_1 G_2> - <G_1>< G_2> .$$
A few remarks about the general structure of this perturbative expansion
are in 
order here: first it should be noticed that according to a general
counting~\cite{das} 
of the powers of $N$ in the large $N$ limit we have $<G_1 G_2 ...G_k>_c 
= O({1\over N^{2k - 2}}) $, so that all the terms in the exponent at 
the l.h.s of (\ref{zetab2}) are of the same order $N^2$. Secondly, 
each term of order $\beta_s^k$ involves the expectation value of $k$ 
space-like plaquettes, and a sum over  all their possible space-time 
positions. 
If we compare different regularizations, corresponding to different 
values of $\rho$, the sum over the time positions of the plaquettes is of 
order $\rho^k$ while $\beta_s^k$ rescales like $\rho^{-k}$.
So, although $\beta_s \rightarrow 0$ as $\rho \rightarrow \infty$, the 
effective coupling of our expansion does not vanish and it coincides
with $\beta + \alpha^0_\sigma$ (see eq. (\ref{rel4})). 
Indeed, as $\beta$ is the coupling of the symmetric 
regularization, it is known from asymptotic freedom to grow 
logarithmically with $n_t$ as the continuum limit is approached. 
 
It is easy to see that the terms of order $\beta_s$ in (\ref{zetab2})
vanish identically  and that at the order $\beta_s^2$ the only 
surviving contributions come from two space-like plaquettes in the 
same space position and separated by an arbitrary time interval.
These type of contributions were explicitly calculated in the case 
of a gauge group SU$(2)$ in 
Ref.~\cite{ourpaper}; for an arbitrary $N$ the
calculation up to order $\beta_s^2$ has not been done yet and it would
involve the explicit evaluation of an integral of the type
\eq
\int DU\, D\tilde{U}\, U_{\alpha \beta}\, \tilde{U}_{\gamma \delta}^{\dagger}
\, \chi_r(U \tilde{U}^{\dagger})\, \chi_s(\tilde{U} \Omega_1 U^{\dagger}
\Omega_2^{\dagger})~.
\label{diffint}
\en
This could be done by resorting to the same techniques (use of suitable
Schwinger-Dyson equation) used in the SU(2) case~\cite{ourpaper}.


\subsubsection{Eguchi--Kawai approach}
\label{egu}
In the present section we shall derive an exact dimensional reduction to
$N_t=1$. This can be achieved by using ideas similar to the ones used to reduce in
the large $N$ limit lattice gauge theories to twisted\footnote{The twist 
consists in a suitable phase factor belonging to the center of SU$(N)$ 
that multiplies each plaquette variable in the action.} one plaquette models. 
This idea was first proposed  by  Eguchi and Kawai \cite{ek} and subsequently 
perfected by several authors and it is based on the observation that in
the large $N$ limit a suitably twisted lattice gauge theory  on a lattice 
consisting of just one site and one link variable for each
space-time direction generates the same set of loop equations as a theory 
defined on a large lattice, typically consisting of $N^{(d+1)/2}$ sites.
Hence twisted one plaquette models
can be used to describe lattice gauge theories on large lattices, 
by essentially mapping space-time degrees of freedom into internal 
degrees of freedom. A general review of the Eguchi--Kawai model, 
including applications to finite temperature 
lattice gauge theories, can be found in  ~\cite{das}.
Following~\cite{bilda}, rather than applying directly the
Eguchi--Kawai method, we shall use a similar technique to reduce 
to one the size of the lattice only in the compactified time dimension. 
The fact that such exact dimensional
reduction is possible is in itself an interesting result. In sec.
\ref{egusec} we shall solve the model in the zeroth order approximation 
($\beta_s=0$) by  assuming as usual that the master
field is  invariant under translations in space. 
In this limit the final result is the same that one would obtain with a
standard  hot twisted Eguchi--Kawai model, where the dimensional
reduction is done at the same time in all space-time direction. 
The present approach 
however has the advantage of being consistent in any space-time dimension, 
while the Eguchi--Kawai reduction only works for even dimensions. 
This will allow us to apply it in the case of (2+1) dimensions.

Let us consider again the Wilson action given in eq. (\ref{wilson}).
A naive prescription for the reduction of the degrees of freedom in
the time direction would be
\begin{eqnarray}
\label{ttekred}
V(\vec x,t) & \rightarrow\hskip 10pt V(\vec x)~, \nonumber\\
U_i(\vec x,t) & \rightarrow\hskip 10pt U_i(\vec x)~.
\end{eqnarray}
Let us denote with $S_{\rm W}(N_t=1)$ the action resulting from (\ref{wilson})
with the substitution (\ref{ttekred}).
It is easy to show by  standard methods (see for instance~\cite{das} )
that $S_{\rm W}(N_t=1)$ leads to the same set of loop equations in the large 
$N$ limit as the full $S_{\rm W}$ theory, {\it provided all 
loops which are closed in the reduced
lattice ($N_t=1$)  but correspond to open loops in the original lattice
have vanishing expectation value}. This would be granted by the fact that,
in addition to local gauge invariance, $S_{\rm W}(N_t=1)$ is endowed with the 
symmetry
\begin{equation}
\label{symm}
V(\vec x) \rightarrow \ee{i{2 \pi n \over N}} V(\vec x)
\end{equation}
The trace along ``open" lines is not
invariant under the symmetry (\ref{symm}) as they do not contain the
same number of $V(\vec x)$ and $\Vdag(\vec x)$ fields. So these
contributions vanish  unless the symmetry is broken.
The symmetry however is actually broken in the  weak
coupling regime, that is also in the continuum limit, where $V(\vec x)$
is close to ${\bf 1}$ (more generically to an element of ${\bf Z}_N$)
and the traces of open lines do not vanish.
Consequently the reduction prescription must be 
modified, as in the twisted Eguchi--Kawai model, according to the
formula:
\begin{eqnarray}
\label{tekred}
V(\vec x,t) &\rightarrow& D(\vec x,t)V(\vec x) D^{\dagger}(\vec x,t)~,
\nonumber\\ U_i(\vec x,t) &\rightarrow & D(\vec x,t)U_i(\vec x) 
D^{\dagger}(\vec x,t)~, 
\end{eqnarray}
where $D(\vec x,t)$ is given by
\begin{equation}
\label{red}
D(\vec x,t) = (\Gamma)^{\sum x_i} (\Gamma_0)^t~,
\end{equation}
with $\Gamma$ and $\Gamma_0$ traceless SU$(N)$ matrices 
satisfying the 't Hooft algebra
\begin{equation}
\label{hooft}
\Gamma \Gamma_0 = \ee{\ii {2 \pi \over N} m} \Gamma_0  \Gamma ~.
\end{equation} 
In the last equation $m$ in an integer number to be determined.
By performing in the Wilson action (\ref{wilson}) the  replacement 
(\ref{tekred}) and redefining the variables according  
to the substitution $U_i(\vec x) \rightarrow
U_i(\vec x) \Gamma$ and $V(\vec x) \rightarrow V(\vec x) \Gamma_0$
we obtain the ``reduced'' partition function
\begin{eqnarray}
\label{ttekaction}
Z_{\rm red} & = & \int\prod_{\vec x}\bigl[DV(\vec x)\prod_{i=1}^d
DU_i(\vec x)\bigr]\, \exp (S_R)~,
\nonumber\\
S_{\rm red} &=& \beta_t N \sum_{\vec x} \sum_{i=1}^d
\real~ \ee{\ii{2\pi m\over N}}\trace
\bigl[U_i(\vec x) V(\vec x + \hat\i)
\Udag_i(\vec x) \Vdag(\vec x)\bigr] +
\nonumber\\
&&\beta_s N \sum_{\vec x} \sum_{i > j} \real\trace \bigl[
U_{i}(\vec x) U_{j}(\vec x+ \hat\i) \Udag_{i}(\vec x+ \hat\j)
\Udag_{j}(\vec x)\bigr]~.
\end{eqnarray}
Notice that, unlike the twisted Eguchi--Kawai model, the twists are 
present in (\ref{ttekaction}) only in the contributions from time-like 
plaquettes, as the  reduction has been done only in the time direction.

Consider now the loop equations for the reduced theory (\ref{ttekaction}).
We already remarked that as long as the symmetry (\ref{symm}) is unbroken 
the loop equations of the reduced theory coincide with the ones of 
(\ref{wilson}). We will show now that in the twisted theory this is the
case {\it also in the weak coupling regime}. 
Indeed in the extremely weak coupling the fields tend to their vacuum
configurations,  which in our twisted reduced theory are
\begin{eqnarray}
\label{ttekvacuum}
U_i(\vec x) &\rightarrow  &P_{{N \over m}}\otimes {\bf 1}_{m}~,\nonumber\\
V(\vec x) &\rightarrow &Q_{{N \over m}}\otimes {\bf 1}_{m}~,
\end{eqnarray}
where ${\bf 1}_{m}$ is the $m\times m$ unit matrix and $P_{{N \over m}}$ 
and $Q_{{N \over m}}$ are the usual building blocks for the twist eating 
configurations:
\begin{equation}
\label{PandQ}
(P)_{ab}=\delta_{a+1,b} \hskip 0.5cm ; \hskip 0.5cm
(Q)_{ab}=\delta_{ab}\ee{\ii{2\pi  m \over N} a} \hskip 0.5cm
,\hskip 0.5cm  a,b=1,\ldots N/m
\end{equation}
with periodicity in the the indices $a $ and $b$, namely   
$a={N \over m}+1$ means  $a=1$.
 It is clear from the context that we have to restrict the values of $N$
so that $N/m$ is an integer.
With these vacuum configurations the trace of open lines is proportional
to $\trace (Q_{{N \over m}})^t$, where $t$ is the difference between the
number of $V$'s and $\Vdag$'s in the trace, namely the difference 
between the time coordinate of the initial and the final point of the 
path in original unreduced lattice.
It is elementary to see from eq. (\ref{PandQ}) that 
\begin{equation}
\label{traceqt}
\trace (Q_{{N \over m}})^t = 0  \hskip 10pt \mbox{unless} \hskip 10pt t=k 
{N \over m}
\end{equation}
for integer $k$. In the unreduced lattice  closed loops correspond to 
$t = k N_t$, so the comparison of the two equations determines $m$ :
\begin{equation}
\label{emme}
m = {N \over N_t}~.
\end{equation}
With the above replacement eq. (\ref{ttekaction}) is, in the large $N$ limit,
an exact dimensional reduction of (\ref{wilson}) on a
$d$-dimensional lattice.

\section{Phase diagram of the Effective Model}
\label{hksec}
While the exact solution of the effective model for the Polyakov loops
given in eq. (\ref{zetab}) is well
beyond our computational capabilities,
its zeroth order approximation in the $\beta_s$ expansion 
can be studied and solved analytically, within 
reliable approximation schemes, in both the weak coupling and the
strong coupling regime.
In particular, we consider in this section the  effective action 
for the Polyakov loop at $\beta_s=0$, eq. (\ref{zetaheat}), that is obtained
in the $\rho\to\infty$ limit (heat kernel action), 
with the aim of describing its phase diagram. 

The solution of the model (\ref{zetaheat}) will rely, 
in both the weak and the strong coupling
regime, on the assumption that a translational invariant master 
field describes the eigenvalue distribution of the Polyakov loop in
the large $N$ limit.
In the weak coupling region, namely for large $J=\beta_{\rm hk}(n_t=1)$,
we know that the model is in a deconfined phase where 
${\hat P}(\vec x)\not= 0$, and the invariant angles of the Polyakov loop
are distributed around $\theta=0$. The solution of model can then be obtained
by retaining only quadratic fluctuations of the eigenvalues around $\theta=0$.
This is done in subsection \ref{wkesec}, where it is shown that the 
model obtained from this quadratic approximation
is a solvable model of the type known as ``Kazakov--Migdal model''
\cite{kazmig}.
The solution is a semicircular distribution of  eigenvalues, centered around 
$\theta=0$, with a radius that increases as $J$ decreases and acquires an 
imaginary part at a critical value of $J$. As first noticed by Zarembo
in ~\cite{zar}, this denotes an instability of the weak coupling
solution and puts a lower bound for a phase transition.
A deeper insight into the meaning of this phase transition, as well as 
a more accurate determination of the critical value of $J$, is 
obtained in subsection \ref{dksec} following the remark that 
the  contribution to the action (\ref{zetaheat}) from any pair of
nearest neighbour
 Polyakov loops is exactly the action of QCD2 on a cylinder,
of area ${1 \over J}$ and with the holonomies at the ends of the cylinder
coinciding with the open Polyakov loops.
When the holonomies at the boundaries are given by semicircular
distributions of eigenvalues, as in the quadratic approximation
mentioned above, QCD2 on a cylinder can be solved exactly and it 
is known to have a third order phase transition ~\cite{cdmp2}, as Douglas and 
Kazakov \cite{dougkaz} first discovered in the case of a sphere, due to the 
condensation of instantons. In the present context this not 
only clarifies the nature of the phase transition but, 
as already mentioned, increases by a few percent the lower bound
already established by Zarembo's argument.
The strong coupling regime  is studied in subsec. \ref{scesec}. 
In this  phase (small $J$) the vacuum is  symmetric, 
${\hat P}({\vec x}) = 0$, namely it is characterised by a uniform 
distribution of eigenvalues. 
Two types of results can be obtained in this context.
A strong-coupling expansion of the effective model
(that was pushed up to the $4^{\rm th}$ order in \cite{bcdmp}) 
shows that a new maximum, other than the symmetric one, 
appears as $J$  increases, it becomes competitive with the
symmetric one at a critical coupling $J_c$ and it is energetically favoured
for $J>J_c$. We have therefore, at $J=J_c$, a $1^{\rm st}$ order transition. 
It is also possible \cite{zar} to calculate exactly, 
namely at all orders in $J$, the mass of the lowest excitation in the 
fluctuations around the symmetric vacuum,
and hence to calculate exactly for which value of $J$ the symmetric vacuum
becomes unstable.
This argument determines an upper bound for the critical value of $J$,
which, at least for not too high values of the number $d$ of space dimensions,
is 
higher
than the lower bound determined by the weak coupling analysis, thus 
restricting to a rather narrow band the region in which 
the deconfinement transition must take place. 
\subsection{Weak coupling expansion}
\label{wkesec}
The building block of the partition function (\ref{zetaheat}),
that is the contribution of  two nearest neighbour Polyakov loops,
is given by a kernel of the form: 
\begin{equation}
{\cal K}_2(g_1,g_2;\ca) = 
\sum_r \,\chi_r(\phi)\,\chi_r(\theta){\rm e}^{-{C_r\, {\cal A}\over 2 N}}~,
\label{expla2}
\end{equation}
where $\phi$ and $\theta$ denote the invariant angles respectively of
$g_1$ and $g_2$. It is well known (see for instance ~\cite{latqcd2,contqcd2})
that ${\cal K}_2(g_1,g_2;\ca)$ is  the partition 
function of QCD2 on a cylinder of area $\ca$ and boundary holonomies 
$g_1$ and $g_2$. With this notation the partition function (\ref{zetaheat})
can be rewritten as
\begin{equation}
Z=\int\prod_{\vec x} DP({\vec x})\, \prod_{{\vec x},i} 
{\cal K}_2\bigl(P({\vec x}),P({\vec x} + i);1/J\bigr),
\label{zhk0}
\end{equation}
where $J$ is the ``renormalized'' coupling $J = \beta_t(N_t=1)$ introduced
in eq. (\ref{331}).

In the large $N$ limit we can
assume that the saddle point solution is translational invariant and, as
a consequence, the action reduces to a one plaquette integral.
The partition function is then simply given by
\eq
Z = \int DP \left[ {\cal K}_2(P,P^{\dagger};1/{J}) \right]^d~,
\label{zetahk}
\en
where $d$ is the number of space dimensions.
Solving the model amounts to finding the eigenvalue distribution $\rho(\theta)$
for the eigenvalues of the Polyakov loop $P={\rm e}^{\ii \theta}$ 
that is an extreme of the free energy associated with eq. (\ref{zetahk}).
In order to do that let us first 
rewrite eq. (\ref{zetahk}) explicitly as an integral over
the invariant angles of $P$. This involves the explicit expression for the
characters $\chi(\theta)$ of SU$(N)$:
\begin{equation}
\chi_r(\theta) = { {\rm det}\,\{{\rm e}^{\ii r_i \theta_j} \}  
\over {\cal J}(\theta) } (-\ii)^{N(N-1)/2}
\label{carattere}
\en
where ${\cal J}(\theta)$ is the Vandermonde determinant for a unitary matrix,
\eq
{\cal J}(\theta) = \prod_{i<j} 2 \sin \frac{\theta_i - \theta_j}{2}~,
\lbl{vandermonde}
\en
and the set of integers $r_i$ label the representation $r$ of SU$(N)$.
By using eq. (\ref{carattere}), the explicit expression of the Casimir 
$C_r$, and the Poisson summation formula
\begin{equation}
\label{poissonsum}
\sum_{l=-\infty}^{\infty} \exp\left(-{(\theta + l)^2\over 4 t}\right)
= (4\pi t)^{1/2} \,\sum_{n=-\infty}^{\infty} \exp\left(-4\pi^2 n^2 t\right)
\,\exp\left(2\pi {\rm i} n\theta\right)~,
\end{equation}
we can rewrite (\ref{zetahk}) in the following way\footnote{The calculation is
almost straightforward in the case of $U(N)$ where the sum over the integers 
$r_i$ is unrestricted and the Casimir is simply given by $C_r = \sum_i r_i^2$.
The details of the calculation in the SU$(N)$ case can be found 
in ~\cite{cdmp1}.}:   
\eq
Z = \int \prod {\rm d}\theta_i \left[ {\cal J}^2(\theta) \right]^{(1-d)}
\left[\sum_{l_i} \sum_P (-1)^{\sigma(P)} {\rm e}^{-\frac{1}{2} N
J \sum_i (\theta_i - \theta_{P(i)} + 2 \pi l_i)^2}
\right]^d~.
\label{zetahk2}
\en
where $P$ denotes a permutation (of signature $\sigma(P)$) of the indices.
The r.h.s. of (\ref{zetahk2}) depends upon a new set of integers $l_i$,
which are the winding numbers of the eigenvalues 
on the unit circle\footnote{Similarly
the integers $r_i$ labelling the representations can be interpreted as 
discretized momenta, and the Poisson summation formula as the corresponding
discrete Fourier transform. This will be further clarified in the next
section \ref{dksec}.}.
In the weak coupling (large $J$) regime, we can assume that the invariant angles
$\theta_i$ are small and that 
the contributions of the winding (i.e. $l_i \not = 0$)
configurations, which are exponentially depressed, can be 
neglected. We shall actually prove in the next section that in the large $N$
limit and above a critical value of $J$ the vanishing of the contributions
of the winding modes is an exact result.
Hence we set $l_i=0$ in eq. (\ref{zetahk2}).
Moreover, for small $\theta$ we have
\eq
{\cal J}^2(\theta) = \Delta^2(\theta) {\rm e}^{-
\frac{1}{12} N \sum_i \theta_i^2 + O(\theta^4)}~,
\label{Jdelta}
\en
where $\Delta(\theta) = \prod_{i<j}(\theta_i - \theta_j)$ is the usual
Cauchy--Vandermonde determinant.
Inserting these results in eq. (\ref{zetahk2}) we obtain
\eq
\int \prod_i {\rm d}\theta_i \left[ \Delta^2(\theta) \right]^{(1 - d)}
{\rm e}^{-N [d J - \frac{d-1}{12}] \sum_i \theta_i^2 }
\left[\sum_P (-1)^{\sigma(P)} e^{N J
\theta_i \theta_{P(i)}} \right]^d~.
\label{zetahk3}
\en
Notice that, provided $J$ is larger than its critical value, the only
approximation needed to go from eq. (\ref{zetahk}) to eq. (\ref{zetahk3})
consists in neglecting the $O(\theta^4)$ terms in (\ref{Jdelta}).
\subsubsection{Solution via  the Kazakov--Migdal model}
\label{kmsec}
The model given in (\ref{zetahk3}) coincides
with a Kazakov--Migdal (KM) matrix model with quadratic potential,
which is exactly solvable.

The KM model ~\cite{kazmig} can be defined, 
in its discretized version, by the  following partition function on a 
$d$-dimensional hypercubic lattice:
\begin{eqnarray}
Z_{KM} & = &  \int\, \prod_{\vec x} D\phi({\vec x}) \, \prod_{{\vec x},i}
DU_i({\vec x}) \,\times
\nonumber\\
&& \exp \left\{ \sum_{\vec x} N\trace \left(
-V\left(\phi({\vec x})\right) + \sum_{i=1}^d \phi({\vec x}) U_i({\vec x})
\phi({\vec x + \hat\i}) U_i^\dagger({\vec x}) \right)\right\} \,\, ,
\label{km1}
\end{eqnarray}
where $\phi({\vec x})$ is an Hermitian $N\times N$ matrix 
and  the link matrices $U_i({\vec x})$ belong to SU$(N)$.
$V\left(\phi\right)$ is a potential term.

While the model (\ref{km1}) contains no self-interaction between the ``gauge'' 
fields $U_i({\vec x})$, if one integrates out the $\phi({\vec x})$ fields
the resulting effective model for the gauge fields (known as ``induced gauge
model'') includes peculiar gauge self-interactions. The original hope that the
induced gauge model could describe directly QCD in $d$ dimensions was 
however shown to fail \cite{kmsw} due to the 
super-confining behaviour of the former.

Another way to deal with eq. (\ref{km1}) is to carry out first the integration
over the link matrices, 
to obtain an effective model for the $\phi({\vec x})$'s.
To do so, one utilizes typical matrix model techniques. First the gauge is 
fixed so that the $\phi$'s are diagonal:
\begin{eqnarray}
\phi({\vec x})^a_{\hskip 3pt b} & \rightarrow & \delta^a_{\hskip 3pt b}\,
\lambda_a({\vec x}) \nonumber \\
D\phi({\vec x}) & \rightarrow & 
\left[\prod_a {\rm d}\lambda_a({\vec x})\right] \, \Delta^2(\lambda)\,\, .
\end{eqnarray}
Then the integration over the links is carried out utilizing the 
Harish--Chandra--Itzykson--Zuber--Mehta integral \cite{izhcm}:
\begin{equation}
\int DU_i({\vec x})\, \ee{N\trace U_i({\vec x}) \lambda({\vec x}+\hat\i)
U_i^\dagger({\vec x}) \lambda({\vec x})} 
= \sum_P (-1)^P {\ee{N \sum_a \lambda_a({\vec x})\, 
\lambda_{Pa}({\vec x} +\hat\i)} 
\over \Delta\left(\lambda({\vec x})\right)\,\Delta\left(\lambda({\vec x}+i)
\right) } \,\, .
\end{equation}
For a quadratic potential $V\left(\phi({\vec x})\right) 
= {1\over 2} m^2 \phi^2({\vec x})$,
and in the large $N$ limit, in which one looks for a translationally invariant
master field $\lambda_a({\vec x}) \rightarrow \lambda_a$, the resulting
model is:
\begin{equation}
Z_{KM} = \int\,\prod_a{\rm d}\lambda \,\left[\Delta^2(\lambda)\right]^{1-d}
\, \ee{-N {m^2\over 2} \sum_a\lambda_a^2}\,
\left[\sum_P (-1)^{\sigma(P)} \ee{N \sum_a \lambda_a
\lambda_{Pa}}\right]^d~.
\label{quadkm}
\end{equation}
Gross \cite{grosskm} found the master field that {\em exactly} solves, 
for any $d$, this model to be given by a semicircular Wigner eigenvalue
distribution:  
\eq
\rho(\lambda) = \frac{2}{\pi R^2} \sqrt{R^2 - \lambda^2}~,
\label{Wigner1}
\en
where the radius $R$ is given by
\eq
R^2 = \frac{4(2 d-1)}{(m^2 (d-1) + d \sqrt{ m^4 -
 4(2d-1)})}~.
\label{erre2}
\en
This exact solution provides us, as we shall see in a moment, with the
solution for the small-$\theta$ expression eq. (\ref{zetahk3}) of our effective 
model.

Let us still remark that, although unsuitable for the description of QCD in
$d$ dimensions, the $d$-dimensional KM model was soon argued to be related
with finite-temperature QCD in $d+1$ dimensions \cite{cdp}.
Indeed the presence of ``matter'' fields $\phi({\vec x})$ in the adjoint
representation of the gauge group led to conjecture that they could be the 
remnants of the components of the gauge fields in an extra compactified 
direction. This extra dimension is naturally interpreted as the time-like 
direction in a finite-temperature model. 

Going back to our effective model for the Polyakov loops, given, 
in the approximation of small eigenvalues, by eq. (\ref{zetahk3}),
and comparing it with eq. (\ref{quadkm}), we see that indeed it is a 
quadratic KM model\footnote{The overall factor of $J$ in the exponent in
eq. (\ref{zetahk3}) amounts just to a different normalization of the
$\theta$'s.} with mass 
\eq
m^2 = 2d-\frac{d-1}{6 J}~.
\label{emme2}
\en
The solution of the our model, which is exact as long as eq. (\ref{zetahk3})
is exact, i.e. as long as quartic and higher terms in the $\theta$'s are
negligible, is therefore given by a Wigner distribution $\rho(\theta)$
of radius 
\eq
r^2 = \frac{4(2 d-1)}{J\,(m^2 (d-1) + d \sqrt{ m^4 -
 4(2d-1)})}~.
\label{erretheta}
\en  
The radius of the distribution {depends on the coupling $J$} (a part from the 
overall factor of ${1\over J}$) 
through the expression (\ref{emme2}) of the mass.

Because of the dependence of $m^2$ from the coupling $J$, it is
easy to see that the argument of the square root in (\ref{erretheta})
decreases as $J$ decreases, and it eventually becomes negative.
Correspondingly the radius $r^2$ of the distribution acquires an imaginary
part, thus making the solution of our model inconsistent. 
It was argued in \cite{zar} that this is a signal that 
the weak coupling solution becomes instable and that we are in presence of 
a phase transition.
The corresponding values of $J$, for various dimensions $d$, 
are reported in the last column of Tab. I  at the end of the 
following section.

A deeper insight of this phase transition will be obtained in
the next section, following the identification of  ${\cal K}_2(g_1,g_2;\ca)$
with the action of QCD2 on a cylinder.
We just remark here that in the present calculation the instability of 
the weak coupling solution stems from the $J$-dependence of $m^2$, namely from the fact
that the Vandermonde determinant ${\cal J}^2(\theta)$ for unitary
matrices appears in eq. (\ref{zetahk2}) instead of the usual
Vandermonde determinant $\Delta^2(\theta)$. 
So, in spite of neglecting the winding modes $l_i$, our solution 
``knows'' that the eigenvalues live on a circle rather than on a line
and that the excitation of the winding modes becomes eventually
favourite, thus leading to a phase transition. 
\subsubsection{Douglas--Kazakov phase transition on a cylinder}
\label{dksec}

We have already remarked that the kernel ${\cal K}_2(g_1,g_2;\ca)$
defined in (\ref{expla2}) is the partition function
of QCD2 on a cylinder of fixed area and boundary holonomies.
The fact that it appears as the basic ingredient in our effective
action at $\beta_s=0$ is not really a surprise, as the space-like plaquettes
are absent in $d=1$ and $\beta_s=0$ is exact in that case. 
So, in order to compute the partition function of 
QCD2 on a cylinder we can  use a heat kernel regularization  and start from eq. 
(\ref{fullac}) with $\beta_s=0$.
The different steps of the calculation are depicted in Fig. 4, and correspond
the the steps leading from eq. (\ref{fullac}) with $\beta_s=0$ 
to eq. (\ref{zetaheat}) but with the additional integration 
over all intermediate timelike links.
The final integration over the remaining space-like link leads to 
the r.h.s of (\ref{zetaheat}), 
but with the space lattice consisting  just of the 
two end points of the cylinder, and with the area appearing in the exponent.
This is exactly ${\cal K}_2(g_1,g_2;\ca)$.
The partition function of QCD2, not just on a cylinder but on a generic 
space-time manifold is known exactly, and it has been extensively studied.
It was soon recognised by Douglas and Kazakov ~\cite{dougkaz} that in 
QCD2 on a sphere a third order phase transition occurs at a critical
value of the area. This result was later generalized to the case of
a cylinder ~\cite{cdmp2,bcdmp,grmat1,grmat2} and the corresponding phase 
transition will therefore be called in the following ``Douglas--Kazakov (DK)
phase transition on a cylinder''.
\iffigs
\begin{figure}
\begin{center}
\null\hskip 1pt
\epsfxsize 12cm
\epsffile{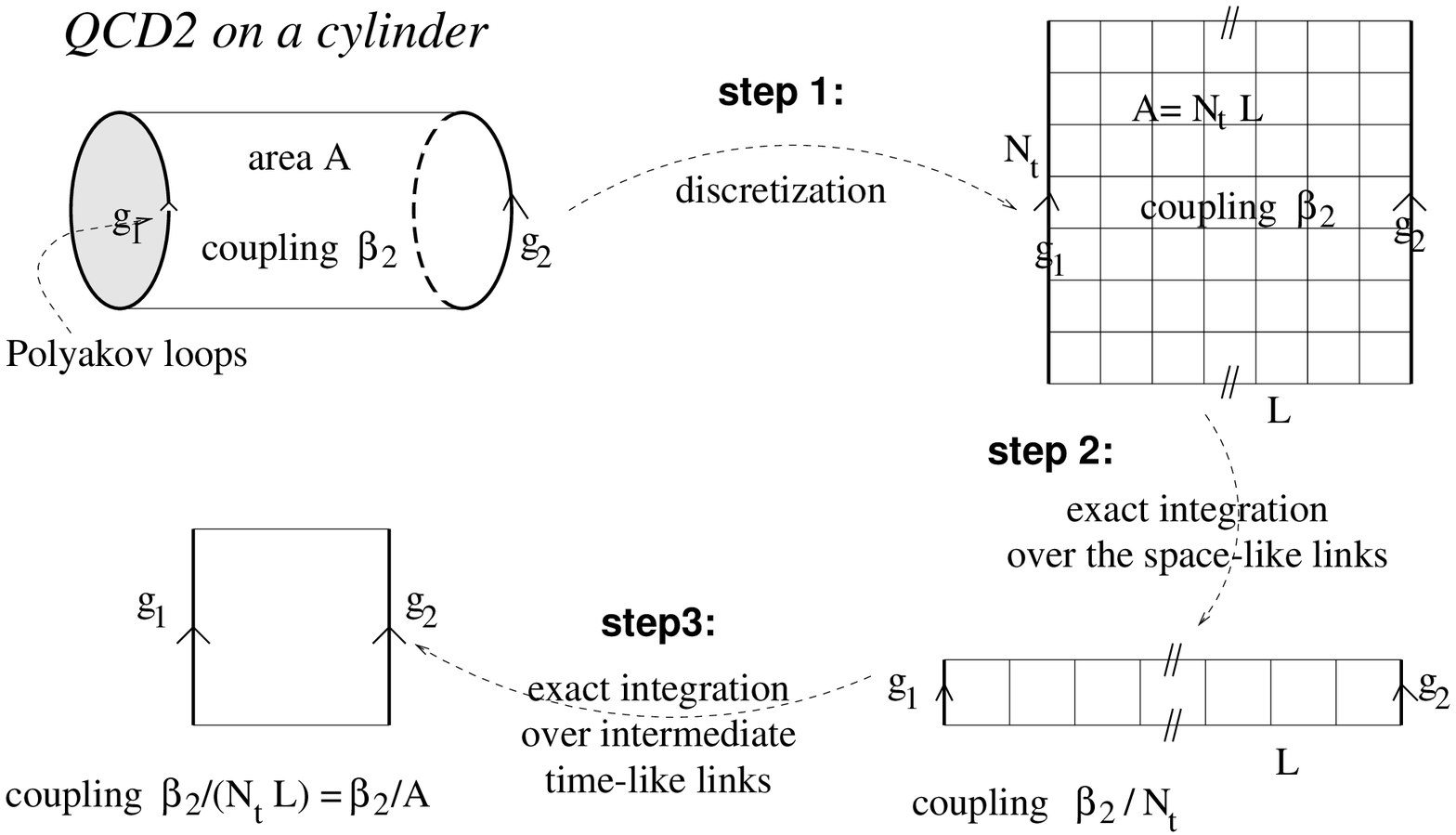}
\end{center}
\vskip 0.2cm\noindent
\mycaptionl{Fig. 4}{The partition function of the 
QCD2 on the cylinder can be derived
exactly via discretization. See the text for explanation.}
\label{qcd2fig}
\end{figure}
\fi
In order to study this transition let us first rewrite 
${\cal K}_2(g_1,g_2;\ca)$ by using the Poisson summation formula 
(\ref{poissonsum}). The result is  \cite{aiz,cdmp1}:
\eqa
\label{poisson}
{\cal K}_2 ( g_1,g_2;\ca) &=& \left( \frac{N}{4 \pi} \right)^{1/2}
 \exp \left(\frac{\ca}{24} (N^2 - 1) \right)\sum_P
\frac{(\frac{\ca}{N})^{(1-N)/2}}{{\cal J}(\theta)
{\cal J}(\phi)} \\
& & (-1)^{\sigma(P)  }
\sum_{\{l_{i}\}}
\exp \left[ - \frac{N}{2\ca} \sum_{i = 1}^N
\left( \phi_i - \theta_{P(i)} + 2 \pi l_i \right)^2 \right]~,
\nonumber
\ena
where as usual $P$ denotes a permutation of the indices.
It is easy to check directly on  (\ref{poisson}) that 
${\cal K}_2 ( g_1,g_2;\ca) $ is the solution of the heat kernel
equation on the SU$(N)$ group manifold
\eq
\left( N\frac{\partial}{\partial \ca} - \frac{1}{2}  {\cal J}^{-1}(\phi)
\sum_i \frac{\partial^2}{\partial \phi_{i}^{2}}  {\cal J}(\phi)
- \frac{1}{24} N(N^2-1)
\right) {\cal K}_2 (\phi,\theta; \ca)  =  0
\label{heateq}
\en
uniquely determined by the condition
\eq
\lim_{\ca \rightarrow 0} {\cal K}_2(g_1,g_2;\ca) =
\hat{\delta}(g_1,g_2)~,
\label{deltain}
\en
where, in the notations of Ref. \cite{cdmp1}, 
$\hat{\delta}$ is the invariant delta function on the group manifold.
As pointed out in \cite{cdmp1,panz}, redefining the Kernel by
${\cal K}_2(\phi,\theta;\ca)$ $\rightarrow$ $\hat{{\cal K}}_2 =
{\cal J}(\phi) {\cal J}(\theta){\cal K}_2$, eq. (\ref{heateq}) becomes
the (Euclidean) free Schroedinger equation for $N$ fermions on a circle,
where $\ca$ plays the role of (Euclidean) time.
This means that, because of the condition (\ref{deltain}),
we can interpret
${\cal K}_2(\phi,\theta;\ca)$ as the Euclidean transition amplitude
for this system of fermions, from the configuration $\{\phi_i\}$ at
zero time to the configuration $\{\theta_i\}$ at the time $\ca$.
The modular transformation eq. (\ref{poissonsum}), relating eq. (\ref{poisson})
to eq. \ref{expla2}, admits a straightforward interpretation in the
fermionic language: the integers labelling the unitary representations
in the character expansion (\ref{expla2}) correspond to discretized
momenta of the fermions on the circle, while eq. (\ref{poisson}) gives
the corresponding coordinate representation, and the integers in the
co-root lattice are the fermion winding numbers.

The expressions (\ref{expla2},\ref{poisson}) of the kernel depend upon two
sets of eigenvalues, $\{\phi_i\}$ and $\{\theta_i\}$.The specific
eigenvalues do not tend to any limit as $N$ goes to infinity; however, the
corresponding eigenvalue distributions\footnote{If $\{\phi_i\}$ is a set of
invariant angles, the corresponding distribution is defined as $\rho(x) =$
${1\over N} \sum_i \delta(x - \phi_i)$. Notice that distributions
corresponding to set of angles are periodic functions of $x$, with period
$2\pi$.} do have a large $N$ limit, and contain all the information needed
to evaluate the large $N$ asymptotics of ${\cal K}_2$.

In particular, it is possible to write, in the large $N$ limit, the time
evolution equation (\ref{heateq}) as a functional differential equation for
the smooth functional of the eigenvalue densities $F[\rho_0,\rho_1;{\cal
A}]$, defined as ${\cal K}_2(\phi,\theta;{\cal A}) = \exp(N^2
F[\rho_0,\rho_1;{\cal A}])$, $\rho_0$ and $\rho_1$ being the densities
corresponding to $\{\phi\}$ and $\{\theta\}$ respectively. This goal is
achieved by replacing partial derivatives in (\ref{heateq}) by derivatives
with respect to the eigenvalue densities,
\begin{equation}
\label{funcder}
N\, {\partial F\over\partial\phi} = {\partial\over \partial x}\left.
{\delta F\over \delta\rho(x)}\right|_{x = \phi_i}~,
\end{equation}
and all sums by integrals (see Ref. \cite{maty} for the details of
the calculation). The final result \cite{maty,bcdmp,zar} is that 
the time (area) evolution of the eigenvalue distribution $\rho(x)$ is
governed, at the leading order in $1/N$,
by a Das--Jevicki Hamiltonian ~\cite{dasjev},
\eq
H\left[ \rho(x),\Pi(x)\right] = {1\over 2}\int {\rm d}x\, 
\rho(x) \left\{ \left(
{\partial \Pi(x) \over \partial x} \right)^2 - {\pi^2 \over 3} \rho^2(x)
\right\}~,
\label{dasjev} 
\en
where $\Pi(x)$ is the canonical momentum conjugate to $\rho(x)$.
In terms of the complex quantity
\eq
f(x,t) = {\partial \Pi(x,t) \over \partial x} + {\rm i} \pi \rho(x,t)
\label{mah} \en
the equation of motion becomes the Hopf equation for an ideal fluid,
\eq
\frac{\partial f}{\partial t} +
f \frac{\partial}{\partial x} f = 0~,
\label{hopf}
\en
with the boundary conditions
\eq
\rho(x,t=0) = \pi \rho_0(x)~,~~~~~\rho(x,t=\ca) = \pi \rho_1(x)~.
\label{boundary}
\en

The solutions to the equations (\ref{hopf},\ref{boundary}) 
have been studied in detail by Gross and Matytsin~\cite{grmat2}.
The knowledge of these solutions allows to write the {\em exact} expression,
in the large $N$ limit, 
of the free energy corresponding to the partition function
(\ref{expla2}). The free energy exhibits a $3^{\rm rd}$ order phase transition, 
for a critical value ${\cal A}_c$ of the area [i.e. of the coupling] 
that we shall derive in a moment. 
This transition is the exact analogue of the $3^{\rm rd}$ order
transition of QCD2 on the sphere, discovered by Douglas and Kazakov 
\cite{dougkaz}; actually
the case of the sphere is retrieved when the boundary distributions
both degenerate to delta functions. 
 
For our purposes it is sufficient to follow the derivation first
given in \cite{cdmp2} of the critical value ${\cal A}_c$. 
In order to find a solution to the eq.s  (\ref{hopf},\ref{boundary})
consider the  ansatz corresponding to a semicircular Wigner 
distribution of variable radius,
\eqa
\rho(x, t)  & = &   \frac{2}{\pi r^2(t)} \sqrt{r^2(t) - x^2}~,~~~~~~
|x|< r^2(t) \nonumber \\
\rho(x, t)  & = & 0~,~~~~~r^2(t)<|x|<\pi~.
\label{Wigner}
\ena
By inserting this ansatz into the Hopf  equation we find that it 
gives a solution of the equation provided
the time dependence of the radius $r$ of the distribution is of
the form
\eq
r(t)  =  2 \sqrt{\frac{(t+\alpha)(\beta - t)}{\alpha + \beta}}~.
\label{gensol}
\en
The arbitrary  constants $\alpha,\beta$ are determined by the boundary
conditions (\ref{boundary}). For consistency, the initial and final
distributions in (\ref{boundary}) must be semicircular.
with radii given by $r(0) = r_0 , r(\ca) = r_1$.
Given $r_0$ and $r_1$, the radius of the distribution is determined at
any time along the cylinder by eq. (\ref{gensol}).
However, because of the periodicity condition on the eigenvalue
distribution, the density (\ref{Wigner},\ref{gensol}) is a solution of
the saddle-point equations (\ref{hopf},\ref{boundary}) only if
$r(t) < \pi$ for any $t$ on the trajectory.
For any given value of $r_0$ and $r_1$, the maximum value of $r(t)$
increases as the area ${\cal A}$ increases, so the solution
(\ref{hopf},\ref{gensol}) is valid only if the area ${\cal A}$ is
smaller than a critical value ${\cal A}_c$, where the maximum radius
equals $\pi$ and the eigenvalues fill up the whole circle.
The critical value of $\ca$ at which the transition occurs can be
easily calculated and is given in \cite{bcdmp}. It will be relevant in the
following only the case in which $r_0 = r_1 \equiv r$; then 
\begin{equation}
\label{4n4}
\left({\cal A}_c\right)^2  = \pi^4 - \pi^2 r^2~. 
\end{equation}

Notice that for $r=0$ the  partition function of QCD on a sphere is
retrieved. Consistently, the critical value (\ref{4n4}) becomes
${\cal A}_c = \pi^2$, which is just the value found by Douglas and
Kazakov \cite{dougkaz}.

{}From the previous discussion one can deduce that below the critical
area ${\cal A}_c$ the eigenvalue distribution is confined in an interval
$(-a(t),a(t))$ with $a(t)< \pi$ for any value of $t$. Hence the 
configurations in which a fraction of the eigenvalues wind around the 
circle do not contribute in the large $N$
limit and  all the integers $l_i$ in eq. (\ref{poisson}) can be
set to zero. The topologically non trivial configurations\footnote{These 
configurations are instantons in the interpretation of the 
eigenvalues as fermions
and of the area as time evolution parameter.} are relevant only if at 
some value of $t$ the distribution covers the whole unit circle.
The role of instantons in inducing the Douglas--Kazakov phase transition
on the sphere was fully investigated by Gross and Matytsin 
\cite{grmat1,grmat2}.

A pictorial view of the DK phase transition is given in Fig. 5.

In order to apply these results to the effective action for the 
Polyakov loop all we have to do is to remember that the role of 
the area is played by $1/J$ and that the radius $r(J)$ of the Wigner 
distribution of the eigenvalues of the Polyakov loop is given by 
(\ref{erretheta}).
Eq. (\ref{4n4}) turns then into an equation for the critical value of $J$:
\eq
\left(\frac{1}{J}\right)^2 = \pi^4 - \pi^2 r(J)^2~.
\label{DKpoint}
\en
This equation can be solved for different values of the number $d$ of
space dimensions and the results can be compared with the ones obtained
in the previous section by requiring the reality of $r(J)$.
The results are summarized in Tab. I.
\begin{table}[t]
\mycaptionl{Table I}{Value of $J$ below which the Wigner distribution
becomes unstable, from the Douglas--Kazakov phase transition
(third column) and from Ref. \cite{zar}.
The radius of the distribution at the critical point is given
in the second column.} 
\begin{center}
\begin{tabular}{c c c c }
\hline\hline
$d$ & $r_c$ & $J_{c}^{\rm (w.c.)}~{\sl[Doug.-Kaz.]}$ &  
$J_{c}^{\rm (w.c.)}~{\sl[Zarembo]}$ \\
\hline
$2$ & $2.96$  & $0.321$ & $0.311$ \\
$3$ & $2.80$ & $0.226$ & $0.218$ \\
$4$ & $2.66$ & $0.192$ & $0.184$ \\
$\infty$ & $0.$ & $1/\pi^2\sim0.101$ & $1/12\sim0.083$ \\
\hline\hline
\end{tabular}
\end{center}
\end{table}
The determination of the critical coupling obtained from 
the DK phase transition is consistently slightly higher
(and hence provides a better lower limit) than the one
by Zarembo. On the other hand the two values are very 
close to each other, thus indicating clearly that we are dealing 
with the same physical phenomenon, as one could argue from the
remark at the end of the previous section, where it was noticed how 
the radius of the distribution developed an imaginary part at small
$J$ as a consequence of the compact and topologically non 
trivial support (the unit circle) of the eigenvalues of unitary matrices.
\iffigs
\begin{figure}[t]
\begin{center}
\null\hskip 1pt
\epsfxsize 12.5cm
\epsffile{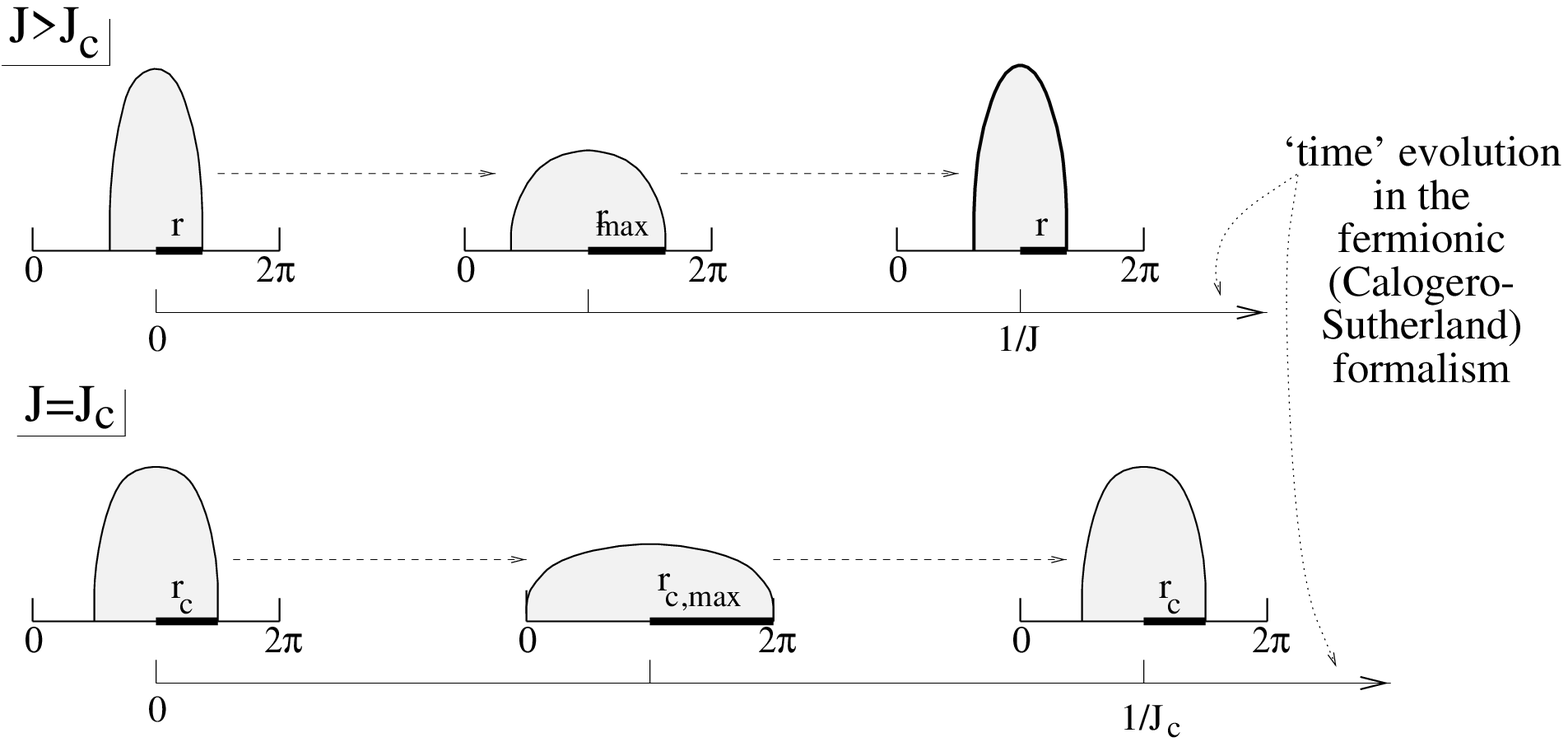}
\end{center}
\vskip 0.2cm\noindent
\mycaptionl{Fig. 5}{Mechanism of the Douglas--Kazakov transition, 
as it applies to our
effective model for the Polyakov loops} 
\label{evol}
\end{figure}
\fi

The  radius of the distribution at the critical DK point,
given in the second column of Tab. I, is consistently  less
than $\pi$ for any number of dimensions, although very close to
it for low dimensions. 
The same calculation, with the critical $J$ obtained by Zarembo,
would instead give a radius larger than $\pi$
for $d=2$ and $d=3$.
In spite of the small value of $J$ we expect the weak coupling solution
to be very reliable above the DK phase transition,
particularly for large $d$, where the small value of $r_c$ ensures that
the quartic terms which  have been neglected in eq. (\ref{zetahk3})
are indeed small.

To sum up, we have established that the Wigner distribution of
eigenvalues becomes unstable at a value of the coupling constant given
by (\ref{DKpoint}), and the transition is driven by the winding modes
in the configuration space of the eigenvalues.
This analogy with the Douglas--Kazakov phase transition on a sphere
gives an almost compelling argument that the corresponding phase
transition is of the third order.
The appearance at the critical point of new classical trajectories
corresponding to winding eigenvalues has presumably the effect of
spreading the distribution, especially at the extremes.
On the other hand, for low dimensions, the transition occurs at a
radius of the eigenvalue distribution very near to $\pi$. This probably
means that the maximum corresponding to the broken, deconfined phase
becomes unstable, and the distribution would collapse into the uniform
one. However 
this is not the case for high dimensions, where the critical radius
is small, approaching zero as $d$ increases. In this case the
Douglas--Kazakov phase transition is a transition from a classical Wigner
distribution to another one, so far unknown, but still presumably
peaked around the origin. This point will be discussed further in sec.
\ref{spdimsec}.
\subsection{Strong coupling expansion}
\label{scesec}
Let us turn now our attention to the  strong
coupling expansion of the effective model
(\ref{zetaheat}).
In this section, which is mainly based on Ref.s \cite{bcdmp,zar,semzar},
we will pursue this expansion  up to the fourth order, obtaining a 
satisfactory picture of the
deconfinement transition as it is seen by approaching the critical value
of the coupling $J_c$ from the strong coupling side and also
obtaining a rather precise determination of the critical value itself.

We will moreover determine the instability of the strong coupling vacuum
(i.e. the point where the uniform eigenvalue distributions turns from
a maximum into a saddle point of the effective action); this instability is
distinct from the first order deconfinement transition, as will be apparent.
\subsubsection{Notations and preliminaries}
\label{notatsec}
In the strong coupling region, the vacuum of the theory is the symmetric 
one: $\langle{\hat P}\rangle = 0$; in terms
of the translationally invariant large $N$ solution, it corresponds  
to a uniform eigenvalue distribution. 
When $J$ grows and reaches a certain critical value, a different non-symmetric 
vacuum, corresponding to a non-uniform eigenvalue distribution, 
becomes eventually energetically favourable and the transition to the
deconfined phase takes place. 

The partition function of the model was given in eq. (\ref{zetahk}),
and is repeated here for commodity:
\eq
Z = \int\, \prod_i {\rm d}\theta_i \, {\cal J}^2(\theta) \,
\left[ {\cal K}_2(P,P^{\dagger};1/{J}) \right]^d~.
\label{zetahk1}
\en
We are interested in the large $N$ limit. We will therefore search for
a ``master field'' solution, described by a certain eigenvalue distribution,
such that it maximizes the free energy that appears in eq. (\ref{zetahk1}).
To proceed, we must find the strong coupling expansion in the large $N$ limit
of this free energy.  

As a preliminary step let us establish some notations. We are
interested in the large $N$ limit, so the fundamental quantity is the
distribution $\rho(\theta)$  of the eigenvalues of the Polyakov loop.
It is convenient to expand $\rho(\theta)$ in its Fourier modes
\eq
\rho(\theta) = \frac{1}{2 \pi} \sum_{n=-\infty}^{\infty}
 \rho_n e^{\ii n \theta}
= \frac{1}{2 \pi} \sum_{n=-\infty}^{\infty} x_n
e^{\ii \alpha_n + \ii n \theta}~,
\label{rhon}
\en
where $\alpha_n \in (- \pi/2,\pi/2)$ is the
argument of $\rho_n$ modulo $\pi$, and $x_n$ coincides with
the modulus of $\rho_n$ up to a sign\footnote{It is convenient 
for the following discussion to restrict
the range for $\alpha_n$ and have $x_n$ taking also negative values.}.
The reality of $\rho(\theta)$ requires $\rho_{-n} =
\rho_n^{*}$, while the normalization  of $\rho(\theta)$ to $1$ in the
interval $(-\pi,\pi)$ fixes $\rho_0 = 1$.

The inverse formula
\eq
\rho_n = \int_{-\pi}^{\pi} \rho(\theta) e^{-\ii n \theta}
\label{rhoinv}
\en
shows that $\rho_n$ corresponds to the large $N$ limit of the
loop winding $n$ times in the time-like direction
with the given eigenvalue distribution.
In particular, $\rho_{\pm 1}$ corresponds
to the large $N$ limit of the Polyakov loop.

The ${\bf Z}_N$ invariance of the effective theory
becomes, in the large $N$ limit, a $U(1)$ invariance under the shift
$\theta \to \theta + \delta$, that is $\alpha_n \to \alpha_n + n
\delta$. If this symmetry is unbroken the eigenvalue
distribution is simply given by $\rho(\theta) = {1 \over 2 \pi}$,
namely $x_n = 0$ for $n \not = 0$. In the broken phase the $U(1)$
symmetry connects the different vacua. If we choose the vacuum
peaked at $\theta = 0$, then the symmetry of the action for $\theta_i
\to - \theta_i$ will force the vacuum
distribution $\rho(\theta)$ to be even in $\theta$, thus fixing all
$\alpha_n$ to zero. In this situation the eigenvalue distribution
takes the form
\eq
\rho(\theta) = {1 \over 2 \pi} \left[ 1 + 2~ \sum_{n=1}^{\infty}
x_n \cos (n \theta) \right]~,
\label{alpha0}
\en
with $x_n$ real.
\subsubsection{The integration measure}
We need to obtain a strong coupling expansion of the  integration measure
${\cal J}^2(\theta)$, which in the large $N$ limit can be expressed
as a function of the loop variables
$\rho_n$, as
\eq
{\cal J}^2(\theta) = \exp \left[\lim_{y \to 1} {N^2 \over 2} \int_{-\pi}^{\pi}
d\theta \int_{-\pi}^\pi d\varphi~ \rho(\theta)~ \rho(\varphi)~ \log(1 -
y\cos (\theta-\varphi)) \right]~,
\label{Jasimp}
\en
where the double integral, which would be
divergent at $\theta=\varphi$, has been regularized by the inclusion of
the parameter $y$, and terms in the exponent suppressed by powers of $N$
have been neglected.
Eq. (\ref{Jasimp}) can be expressed in terms of the modes $x_k$
by expanding in powers of $\theta$
and $\varphi$ and resumming the resulting expression. The result is
\eq
{\cal J}^2(\theta) = \exp \left[\lim_{y \to 1} N^2 \left( C_0(y) +
\sum_{k=1}^{\infty} C_k(y) x_k^2 \right) \right]~,
\label{Jas2}
\en
where $C_0(y)$ is an irrelevant divergent expression and $C_k(y)$ is
given by
\eq
C_k(y) = {1 \over k} \left[ \frac{1 - \sqrt{1 - y^2}}{y} \right]^k~.
\label{ck}
\en
After removing the divergence, the limit $y \to 1$ can be taken,
and gives
\eq
{\cal J}^2(\theta) = \exp \left[ - N^2 \sum_{k=1}^{\infty} {1 \over k} x_k^2
                    \right]~.
\label{Jas3}
\en
\subsubsection{The kernel on the cylinder}
\label{kcscsec}
We have now to calculate the strong coupling expansion for the 
basic building block of the partition function (\ref{zetahk1}), namely
the QCD2 kernel on the cylinder ${\cal K}_2(P,P^\dagger ;1/J)$. 
In the following we shall consider the
more general case in which the holonomies  $g_1,g_2$ at the
boundaries of the cylinder are arbitrary.  

Unlike the weak coupling regime, where we had to use the
expression (\ref{poisson}) for ${\cal K}_2(g_1,g_2^{-1};1/J )$ obtained 
by the Poisson summation formula, we use here  the
the character expansion
\eq
{\cal K}_2(g_1, g_2^{-1} ;1/J ) = \sum_R
e^{- {C_R \over 2N J}} \chi_R (g_1) \chi_R (g_2^{-1})~.
\label{hkcyl1}
\en
which provides directly  a strong coupling
expansion of the free energy\footnote{This is the free energy 
of the QCD2 on the cylinder, of course, and is just an ingredient of the
free energy for the effective model of Polyakov loops that we are managing to
build up.}. The expansion parameter is the following  function of the
coupling $J$: 
\eq
2 \exp \left(-{1\over 2 J}\right) \,\, . 
\label{exppar}
\en 
which vanishes exponentially as $J$ goes to zero.
The free energy $F(g_1,g_2^{-1};\bhk)$ in the large $N$ limit is defined
by:
\eq
{\cal K}_2(g_1,g_2^{-1};{1\over\bhk}) = e^{N^2 F(g_1,g_2^{-1};\bhk)}~,
\label{hkcyl1b}
\en
where, as usual, terms suppressed by powers of $N$ in the exponent will be
neglected.
We have therefore
\eq
F(g_1,g_2^{-1};\bhk) = 2~e^{-1/2\bhk} F_1^H(g_1,g_2^{-1};\bhk)~
 +~4~ e^{-1/\bhk}
F_2^H(g_1,g_2^{-1};\bhk) + \ldots~.
\label{hkcyl1c}
\en
It is found that residual dependence on $\bhk$ in the individual terms is
polynomial in $1/\bhk$, and hence (\ref{hkcyl1c}) is indeed a strong
coupling expansion.

The coefficients $F^H_i(P,P^{\dagger};J)$ of the strong coupling expansion
of the free energy $F(P,P^{\dagger};J)$ can be obtained in principle up 
to any order by using techniques which are summarized in Appendix A of
 \cite{bcdmp}. 
They involve the large $N$ expansion of the Casimirs $C_r$, and the 
expression of the characters, in the large $N$ limit, in terms of the
Fourier coefficients $\rho_k = {1\over N} \trace P^k$ 
of the eigenvalue distribution (see also \cite{Bars,grosstaylor}).
In \cite{bcdmp} this program was carried out up to the $4^{\rm th}$ order.
The result is rather cumbersome and we shall not report it here,
although we will use it in sec. \ref{dectranssec}  
in order to derive the critical 
value of the coupling with the highest possible precision.
We just reproduce here the calculation of
the free energy eq. (\ref{hkcyl1c}) up to the $2^{\rm nd}$ order,
which illustrates all the basic points of the procedure,
without being too involved.
Let us define a representation $r$ of SU$(N)$ to be of order $l$ if 
the large $N$ expansion of its Casimir
is of the form $C_r= lN + O(1)$.
It is clear from eq.s
(\ref{hkcyl1},\ref{hkcyl1b}) that to work out
an expression of the free energy in powers
of $2{\rm e}^{-{1\over 2J}}$ up to $l^{\rm th}$ order, we must take into
account all the representations of order up to $l$. 
The relevant representations up to order two are given 
in Tab. II, together with the corresponding 
Casimirs and the large $N$ expression for their characters.

\begin{table}[t]
\mycaptionl{Table II}{Representations contributing to the strong
coupling expansion of the free energy, up to the second order
in ${\rm e}^{-{1\over J}}$. The fundamental, symmetric and 
antisymmetric rank 2 tensor representations must be considered 
together with their conjugate representations.}
\label{repres}
\begin{center}
\begin{tabular}{c c c c }
\hline\hline
representation & order & Casimir & Character\\
\hline
singlet & $0$ & 1 & 1 \\
fundamental & 1 & ${N^2 - 1 \over N}$& $N \rho_1$\\
symm. rank 2 tensor& 2&${2N^2+2N-4 \over N}$ & $(N^2 \rho_1^2 + N \rho_2)/2$\\
anti-symm. rank 2 tensor& 2&${2N^2-2N-4 \over N}$ & 
$(N^2 \rho_1^2 -N \rho_2)/2$\\
adjoint & 2& $2 N$&$ N^2 |\rho_1|^2 -1$\\
\hline\hline
\end{tabular}
\end{center}
\end{table}
By inserting the expressions in Tab. II into eq. (\ref{hkcyl1}) 
we obtain 
\begin{eqnarray}
\label{k2second}
{\cal K}_2 & \sim & 1 + 2 N^2 |\rho_1|^2 (1 + {1\over 2 N^2 J} + \ldots)
\,{\rm e}^{-{1\over 2J}} + \nonumber\\
&& + 2 \cdot {1\over 4} |N^2\rho_1^2 + N\rho_2|^2 (1 - {1\over N J} -
{2\over N^2 J} + {1\over 2N^2 J^2} + \ldots )\, {\rm e}^{-{1\over J}}
\nonumber\\
&& + 2 \cdot {1\over 4} |N^2\rho_1^2 - N\rho_2|^2 (1 + {1\over N J} -
{2\over N^2 J} + {1\over 2N^2 J^2} + \ldots )\, {\rm e}^{-{1\over J}}
\nonumber\\
&& + (N^2 |\rho_1|^2 - 1)^2 \, {\rm e}^{-{1\over J}} \, + \, O({\rm
e}^{-{3\over 2J}}).
\end{eqnarray}
Eq. (\ref{k2second}) can be exponentiated, as it is easy to check, as follows:
\begin{equation}
\label{fenergysecond}
{\cal K}_2 \sim \exp \biggl\{ N^2 \biggl[ 2 {\rm e}^{-{1\over 2J}} x_1^2
+ 4 {\rm e}^{-{1\over J}} \biggl( -{x_1^2\over 2} + {x_2^2\over 4} 
- {x_1^2 x_2 \cos(2\alpha_1 -\alpha_2)\over 2J} + {x_1^4\over 8J^2}
\biggr) + O({1\over N^2}) \biggr] \biggl\},
\end{equation}
where we have used the notation $\rho_k = x_k {\rm e}^{\ii\alpha_k}$
introduced in the previous section.
Eq. (\ref{fenergysecond}) provides the strong coupling expansion 
of the free energy  $F(P,P^\dagger;J)$ up to the second order in
$2 {\rm e}^{-{1\over 2J}} $.
Notice that  $F_k^H(P,P^\dagger;J)$, obtained from (\ref{fenergysecond})
for $k={1,2}$,  depends only on $x_i$ and 
$\alpha_i$ with $i\leq k$; this property is
general and valid at all orders in the expansion.
\subsubsection{The deconfinement transition}
\label{dectranssec}
By using eq.s (\ref{Jas3}) and (\ref{hkcyl1b},\ref{hkcyl1c}) we can 
now express the large $N$ limit of the partition function 
$Z$ of eq. (\ref{zetahk1}) as
\eq
{1 \over N^2} \log Z =   \sum_{k=1}^{\infty} \left( d~ 
(2 {\rm e}^{-{1\over 2 J}})^k \, \, F_k(x,\alpha) - {x_k^2 \over k} \right)~.
\label{zetaWas}
\en
It is already obvious from (\ref{zetaWas}) that, for small enough
$J$, the free energy has a maximum for $x_k = 0$, that is for
a uniform distribution of eigenvalues.
The first order approximation of (\ref{zetaWas}) can be obtained from
(\ref{fenergysecond}) by neglecting the second order terms, and reads
\eq
{1 \over N^2} \log Z \approx  (2{\rm e}^{-{1\over 2 J}}\, d - 1) x_1^2~,
\label{firstord}
\en
with all other $x_k$ set to zero. The phase structure of
(\ref{firstord}) is very simple: for $J < {1 \over 2\log 2d}$ the
maximum of the free energy occurs at $x_1=0$ and we are in the
unbroken phase. Above the critical point $J ={1 \over 2\log 2d}$ we
have instead $x_1 = 1/2$, which is the maximum
value allowed by the positivity condition on $\rho(\theta)$.

A more accurate analysis of the phase diagram  can be achieved by 
inserting in eq. (\ref{zetaWas}) the available higher orders.     
As we are interested in determining the broken vacuum, we can set 
all $\alpha_n$ to zero, as discussed at the beginning of this section,
and look at the maximum of the free energy regarded as a function of 
the $x_n$'s only.

In looking for the maximum, it is crucial to remain within the domain
where $\rho(\theta)$ is positive or zero for any value of $\theta$.
This is far from trivial when we use the Fourier coefficients $x_k$
as dynamical variables. 
The equations for such a domain can be obtained in principle 
by requiring that, for some $\theta_0$,
\eq  
\rho(\theta_0) = {{\rm d} \over {\rm d} \theta}\rho(\theta) |_{\theta=
\theta_0} = 0
\label{physreg}
\en
 and 
by eliminating $\theta_0$ from the equations. The resulting equations 
for the $x_n$'s give the boundaries of the physical region. 

Let us consider the second order approximation, whose corresponding
free energy can be obtained by substituting (\ref{fenergysecond}) into
(\ref{zetaWas}). The contour plot of the free 
energy as a function of $x_1$ and $x_2$ at various values of the 
coupling $J$ is shown in Fig. 6 for $d=2$.
\iffigs
\begin{figure}
\epsfxsize = 6.7truecm
\epsffile{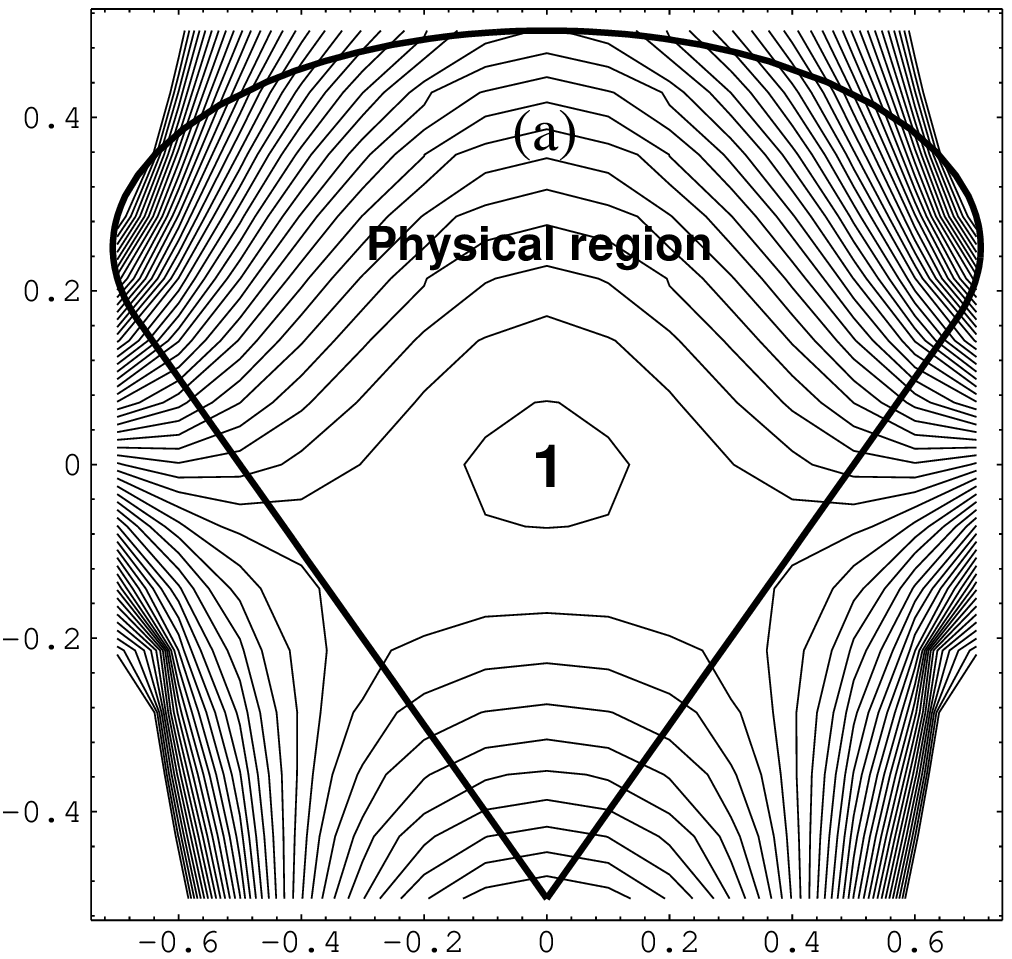}
\vskip -6.4truecm
\hskip 6.7truecm
\epsfxsize = 6.7truecm
\epsffile{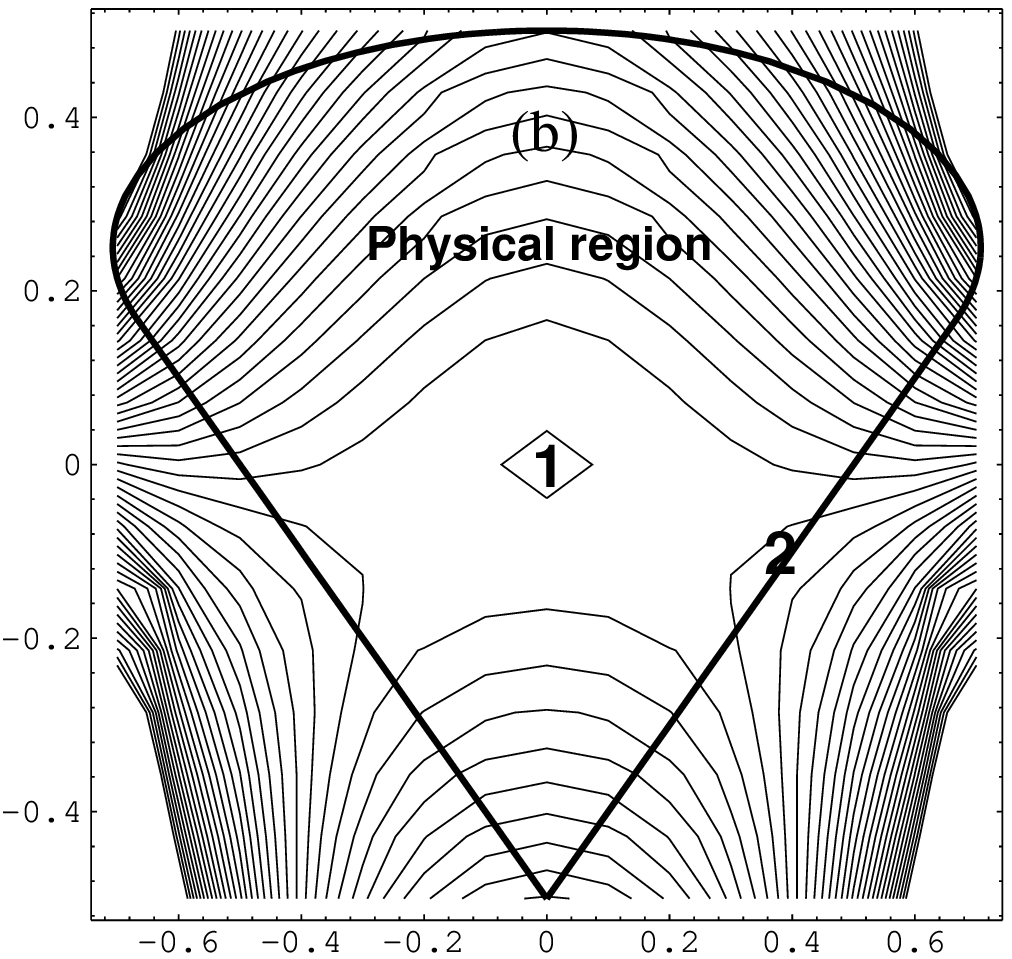}
\newline
\null
\hskip 3.25truecm 
\epsfxsize = 6.7truecm
\epsffile{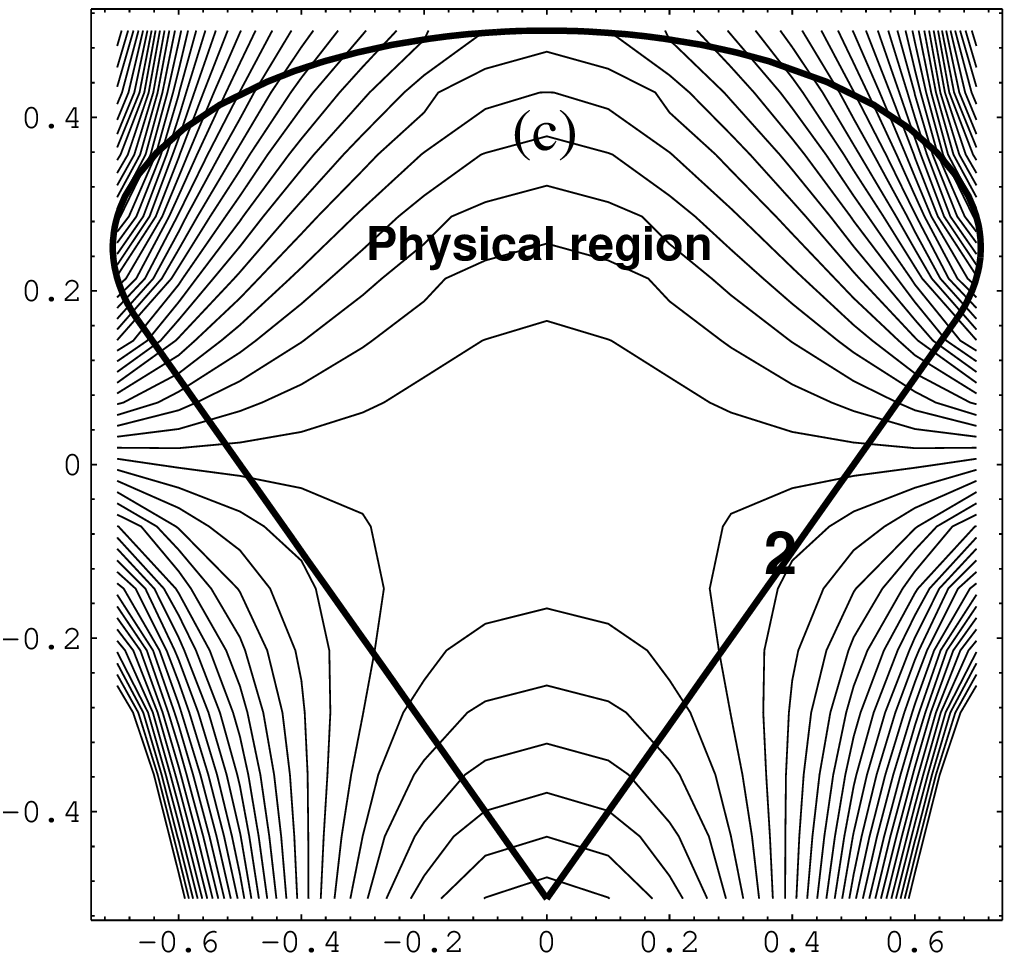}
\vskip 0.2cm\noindent
\mycaptionl{Fig. 6}{Contour plots of the free energy for: {\rm\bf
(a)}, $J = 0.39$; {\rm\bf (b)},$J = 0.41$; {\rm\bf (c)},
$J = 0.44$. The numbers {\rm\bf 1} and {\rm\bf 2} mark respectively
the presence of the symmetric maximum ($x_1 = x_2 = 0$) and the ``broken''
one. The thick line encloses the ``physical'' region} 
\label{contourplots}
\end{figure}
\fi
The physical region in the $(x_1,x_2)$ plane can be easily determined in 
this case, and it is represented by the region inside the thick line. 
The straight edge on the right is given by the equation $x_1 - x_2 = 1/2$ and 
corresponds to density distributions vanishing at $\theta=\pi$, whereas 
the the straight edge on the opposite side corresponds to distributions 
vanishing at $\theta=0$\footnote{The curved section of the boundary is part of
the ellipse of equation $8x_2^2+x_1^2-4x_2=0$ and corresponds to eigenvalue
distributions that vanish in two different points.}.
The plot at $J= 0.39$ clearly shows the maximum at 
$x_1=x_2=0$: the system is in the unbroken phase dominated by a constant 
distribution of eigenvalues. In the next plot, at $J = 0.41$, 
a local maximum has appeared at the edge of the physical region, and it 
is becoming competitive with the unbroken maximum. Notice that there 
is a symmetry $x_1 \to -x_1$, so that a symmetric maximum appears on 
the other edge of the physical region. This symmetry is accidental, and
it is removed when higher order terms are taken into account.
In the last plot, at $J = 0.44$, the maximum at $x_1=x_2=0$ 
has disappeared and the system is clearly in the broken phase.
It is clear from this picture that there are three distinct regions of 
$J$: the first ranging from $J=0$ to a $J_c^{\rm (w.c.)}$ 
where only the symmetric maximum exist, 
the second from $J_c^{\rm (w.c.)}$ to some value $J_c^{\rm (s.c.)}$ 
where both the symmetric and the broken vacuum are present and the third 
for $J>J_c^{\rm (s.c.)}$, where only
the broken vacuum survives. The first order deconfinement transition
occurs in the middle region at the critical  $J_c$ where the value of the
free energy in correspondence of the two maxima is the same.
The point $J_c^{\rm (w.c.)}$ can likely be identified, 
when all orders are taken into
account, with the critical coupling corresponding to the Douglas--Kazakov
phase transition\footnote{This might not be the case, as discussed in 
sec. \ref{spdimsec}, for high values of $d$, the number of space dimensions.}, 
where the weak coupling solution becomes unstable.
The instability of the symmetric solution will be investigated in the next 
subsection, where $J_c^{\rm (s.c.)}$ is determined exactly, namely at
all orders in the strong coupling expansion.

As more orders are taken into account the explicit calculation of the
free energy as a function of the $x_k$'s becomes more and more involved
and the determination of the physical region more complicated.
In ~\cite{bcdmp} the calculation is pushed up to the fourth order,
where the free energy is a function of $x_1, x_2, x_3$ and $x_4$
and the boundaries of the physical region are given by fourth order 
algebraic equations, together with the hyperplanes 
$x_1-x_2+x_3-x_4=1/2$ and $x_1+x_2+x_3+x_4=-1/2$.
 
The critical point can be determined numerically, and it is given  
for different dimensions by the values listed in Table III,
where the values of the $x_k$'s corresponding to the broken vacuum at
the critical point are also reported.

\begin{table}
\mycaptionl{Table III}{Values of the critical coupling $J_c$ at the fourth 
order in the strong coupling expansion of the heat kernel action. 
In the last four columns the 
corresponding values of $x_i,~~(i=1-4)$ are reported.} 
\label{sctable}
\begin{center}
\begin{tabular}{c c c c c c}
\hline\hline
$d$ & $J$ & $x_1$ & $x_2$ & $x_3$ & $x_4$ \\
\hline
$2$ & $0.416$ & $0.41$ & $-0.13$ & $0.03$ & $0.07$ \\ 
$3$ & $0.282$ & $0.45$ & $-0.10$ & $-0.01$ & $0.05$ \\
$4$ & $0.238$ & $0.50$ & $-0.07$ & $-.05$ & $0.02$ \\
$100$ & $0.094$ & $0.64$ & $0.07$ & $-0.10$ & $-0.03$ \\
\hline\hline
\end{tabular}
\end{center}
\end{table}
\subsubsection{Instability of the symmetric vacuum}
\label{instsec}
In the previous section,  based on \cite{bcdmp},
we analysed the $1^{\rm st}$ order phase transition
separating the confined and deconfined phase.
It turns out that the two phases can coexist as meta-stable phases, at least
for not too high values of $d$ (see sec. \ref{spdimsec}). 
Indeed the weak coupling solution is still valid for couplings $J$ lower than
$J_c$, down to the value $J_c^{\rm (w.c.)}$ determined in sec. \ref{dksec} 
from the Douglas--Kazakov transition. 
In this section we shall show that it is possible to determine
exactly the higher limit $J_c^{\rm (s.c.)}$ for the stability of the 
strong coupling symmetric vacuum \cite{zar,boul}, 
i.e. the value for which the configuration
$\rho = 1/2\pi$ ceases to be a local maximum; this value is found to be
higher than $J_c$.
We shall give here the main points of the calculation, a more detailed 
account of it can be found in \cite{zar,boul,semzar}.

The goal is achieved by calculating 
the spectrum of excitations around the strong
coupling symmetric, translationally invariant vacuum $\rho(\theta,{\vec
x})={1\over 2\pi}$. The instability  appears when the lowest-lying
excitation becomes massless. 

As remarked several times, the configuration that dominates in the large $N$
limit is the solution of the classical equations of motion for
the model eq. (\ref{zhk0}).  These equations can be derived from
(\ref{zhk0}) by varying with respect to the eigenvalue distribution
$\rho(\theta,{\vec x})$ of the Polyakov loop and by taking into account the 
large $N$ expression of the integration measure given in (\ref{Jasimp}). 
After taking the derivative with respect to $\theta$ , 
the resulting equations read:
\begin{equation}
\label{sci1}
- \wp\int_{-\pi}^\pi {\rm d}\theta^\prime\, 
\rho(\theta^\prime,{\vec x})\,\cot{\theta - \theta^\prime\over 2} = 
{1\over N^2}\sum_{i=-d}^d
{\partial\over\partial\theta}\,{\delta\over\delta\rho(\theta,{\vec x})}
\log {\cal K}_2(P({\vec x}),P^\dagger({\vec x} + \hat\i); 1/J)~,
\end{equation}
where the left hand side  comes from the integration measure.
The dependence of the kernel 
${\cal K}_2(P({\vec x}),P^\dagger({\vec x} + \hat\i); 1/J)~$
from $\rho(\theta,{\vec x})$ can be obtained in principle from the 
observation, discussed in sec. \ref{dksec}, that in the large $N$ limit
${\cal K}_2(P({\vec x}),P^\dagger({\vec x} + \hat\i); 1/J)~$ is the transition 
amplitude, governed by the Das--Jevicki Hamiltonian (\ref{dasjev}),
from an initial distribution $\rho(\theta,{\vec x})$ to a final distribution
$\rho(\theta,{\vec x}+i)$ in a ``time'' ${1 \over J}$.
Hence the logarithm of the kernel ${\cal K}_2$ is the classical action
corresponding to the Das--Jevicki Hamiltonian, more precisely 
\begin{eqnarray}
\label{classaction}
&&{1 \over N^2} \log {\cal K}_2(P({\vec x}),P^\dagger({\vec x} + \hat\i); 1/J)
\nonumber\\
&&= - \int_0^{{1 \over J}} {\rm d}t 
\int {\rm d} \theta \left[ \Pi(\theta,t) {\partial 
\rho(\theta,t) \over \partial t} - {1 \over 2} 
\rho(\theta,t) \left\{
\left( {\partial \Pi(\theta,t) \over \partial \theta} \right)^2 -
{\pi^2 \over 3} \rho^2(\theta,t) \right\} \right]~,\nonumber
\end{eqnarray}
where the field configurations are the {\it classical} trajectories, namely the
solutions of the Hopf equation (\ref{hopf}) with the boundary conditions
\begin{eqnarray}
\label{sci3}
\rho(\theta,0) & = & \rho(\theta,{\vec x})~,\nonumber\\
\rho(\theta,{1\over J}) & = & \rho(\theta,{\vec x} + \hat\i)~.
\end{eqnarray}
The explicit solution of the Hopf equation with arbitrary boundary condition is
in general not known, but we are interested in small fluctuations around
the symmetric vacuum configuration $\rho(\theta,t)= {1 \over 2 \pi}$, 
and we only need to retain in (\ref{classaction}) 
quadratic terms in such fluctuations.
In this approximation the Hopf equation becomes linear, and hence solvable.
We shall skip here the details of the calculation, which can be found in 
\cite{zar}.
Eq. (\ref{sci1}) eventually reduces to a relation
between the Fourier modes $\rho_n({\vec x})$ of the eigenvalue distribution 
$\rho(\theta,{\vec x})$ on the sites of the lattice:
\begin{equation}
\label{sci6}
\sum_{i = 1}^d \left(\rho_n({\vec x} + \hat\i) - 2 \left[\cosh {n\over 2J} -
{d-1\over d} \sinh {n\over 2J}\right] \rho_n({\vec x}) + 
\rho_n({\vec x} -\hat\i) \right) = 0~,
\end{equation}
which is the free field equation for the scalar excitations $\rho_n({\vec
x})$ propagating on the lattice, provided their masses $M_n$ are identified
as follows:
\begin{equation}
\label{sci7}
M_n^2 = 2d\left(\cosh {n\over 2J} - 1\right) - 2(d-1) \sinh {n\over 2J}~.
\end{equation}
The strong coupling vacuum becomes unstable when the first excitation becomes
massless; setting $M_1^2(J_c^{\rm (s.c.)}) = 0$ we obtain
\begin{equation}
\label{sci8}
{1\over J_c^{\rm(s.c.)}} = 2 \log (2d -1)~.
\end{equation}
The determination of the spectrum of excitations can be quite simply
phrased also in the language of the previous sections 
\ref{notatsec}-\ref{dectranssec}.
Indeed it corresponds to the determination of the free energy (\ref{zetaWas})
{\em exactly} in $J$, but only up to quadratic terms in the $x_n$'s.
In particular, the determination of the strong coupling vacuum instability
$J_c^{\rm (s.c.)}$ requires the determination of all the contributions to
eq. (\ref{zetaWas}) proportional to $x_1^2$. 
This is within reach of the techniques
developed in sec. (\ref{kcscsec}), and the result is:
\begin{eqnarray}
\label{sci9}
{1 \over N^2} \ln Z & \stackrel{x_1^2\,{\rm terms}}{\longrightarrow}&
\biggl[ 2d \biggl({\rm e}^{-{1\over 2 J}} 
- {\rm e}^{-{1\over J}}+ {\rm e}^{-{3\over 2 J}} - 
{\rm e}^{-{2\over J}} + \ldots \biggr)\, d - 1\biggr] x_1^2 \nonumber\\
\null & = & \left({2d\,{\rm e}^{-{1\over 2J}} \over 1 + {\rm e}^{-{1\over
2J}}} - 1\right) x_1^2~.
\end{eqnarray}
By requiring the above expression to vanish we get $\exp (-1/2J_c^{\rm
(s.c.)})$ $= 1/(2d-1)$, namely eq. (\ref{sci8}).

\section{Extended phase diagrams}
\label{extsec}
In the previous section we studied the phase diagram of the effective
model for the Polyakov loop, as a function of the only existing parameter,
namely the coupling $J$. In this section we consider some extensions
of the model involving two parameters, and investigate the phase diagram
in their plane, which reveals some new interesting features.
The first example we shall consider is the same model studied in the previous
section, but regarded as a function of $J$ and of the number $d$ of 
space dimensions. Secondly, following \cite{semzar}, we include the 
effect of external static sources in the adjoint representation
(with coupling parameter $\lambda$) and
consider the phase diagram in the $(J,\lambda)$ plane.
Finally we shall consider 
the effective action with the inclusion of an external
``magnetic'' field, as in the simplified effective action already studied in
sec. 3.1.4, and describe how the phase diagram of Fig. 2 is modified as we
go to higher orders in the strong coupling expansion.
  
\bm
\subsection{Phase diagram in the $(J,d)$ plane}
\ubm
\label{spdimsec}
\iffigs
\begin{figure}
\begin{center}
\null\hskip 1pt
\epsfxsize = 10cm
\epsffile{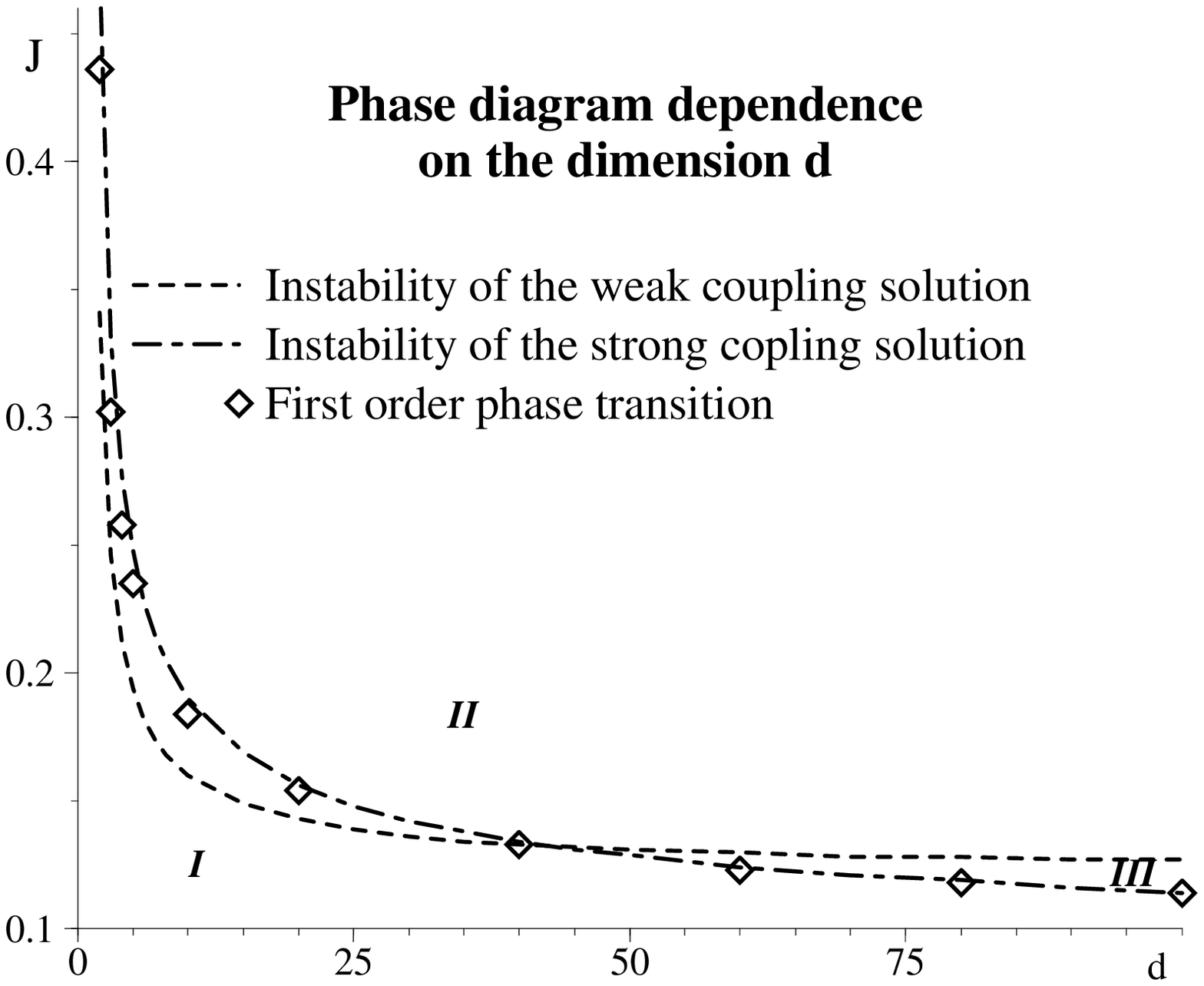}
\end{center}
\vskip -1cm\noindent
\mycaptionl{Fig. 7}{We plot the point of instability $J_c^{\rm (w.k.)}$
of the weak coupling solution, the point  $J_c^{\rm (s.k.)}$
where the strong coupling symmetric vacuum ceases to be a local maximum
and the first-order transition point $J_c$ in dependence of the 
dimensionality $d$ of the lattice. It emerges that for $d\geq 40$ the weak
coupling solution of eq. (\ref{erretheta}) loses 
its validity for higher coupling than
those at which the symmetric vacuum becomes a maximum. This points to
the existence of a new phase (III), beside the usual strongly coupled,
confining one (I) and the weakly coupled, deconfined one described by the
solution (\ref{erretheta}) (II). 
The phases II and III are separated by a third-order phase transition.}  
\label{dimfig}
\end{figure}
\fi
The results obtained in the previous section, both in the weak and strong
coupling regime, hold for any value of the space dimensionality $d$.
So it is possible to plot both  $J_c^{\rm (w.c.)}$ and $J_c^{\rm (s.c.)}$,
marking respectively the DK phase transition and the instability of the
symmetric vacuum, as a function of $d$. 
The plot is shown in Fig. 7.

For low values of $d$ (up to $d\sim 40$) we have 
\begin{equation}
\label{dim1}
J_c^{\rm (w.c.)} < J_c < J_c^{\rm (s.c.)}~.
\end{equation}
This is the situation already described in the previous section: in the 
interval between $J_c^{\rm (w.c.)} $ and $J_c^{\rm (s.c.)}$ both the 
weak coupling solution and the symmetric vacuum are present and a first
order phase transition occurs somewhere in between.
For $d > 40$ the situation is quite different. We have: 
\begin{equation}
\label{dim2}
J_c < J_c^{\rm (s.c.)} < J_c^{\rm (w.c.)}~.
\end{equation}
Eq. (\ref{dim2}) implies that if we approach the phase transition from the 
weak coupling (high $J$) region the semicircular solution becomes unstable
{\em before} the appearing of the symmetric vacuum as a maximum of the
effective action. This means a new phase, or even more than one new phase,
is present for $d > 40$ in the region {\it III} between the two lines of 
Fig. 7. The features of this new, and presumably deconfined, phase(s) are not
known, although a likely possibility is that the system goes from the weak
coupling solution, where the density of eigenvalues vanish in an interval,
to a distribution where it vanishes in one or in a discrete set of points. 
This phase is separated from the semicircular one by
the third order transition due to the  DK transition in QCD2. 
The existence of such intermediate phase(s) is also revealed by the 
circumstance that the radius of the semicircular distribution at the 
Douglas--Kazakov critical point goes to zero as $d$ goes to infinity (see 
Table I), making the possibility of a sudden transition to a uniform 
distribution an extremely unlikely one.

\subsection{Phase diagram in presence of adjoint quarks}
\label{adjsec}
Up to this point we have always been considering {\em pure} SU$(N)$ gauge
theories. Recently in \cite{semzar} an extension of
the effective model considered in section \ref{hksec} has been considered, 
by taking into account the effect of a gas
of non-dynamical, static ``quarks'' (that is, a gas of ``electric''
non-abelian external sources)
transforming in the adjoint representation of the gauge group.
The phase diagram shows a rich structure, similar to that of the diagram 
in the $J,d$ plane discussed in sec. \ref{spdimsec} above.

Static sources couple only to the
``electric'' link matrices $V({\vec x},t)$ pointing in the time direction. 
The minimal introduction of such sources is therefore that of a
quark--antiquark pair inserted both at the same spatial location ${\vec x}$.
In presence of such a pair transforming in the adjoint representation,
it is easy to see, by using gauge invariance and  the periodicity of the 
boundary conditions in the time direction, that the result 
of the integration in eq. (\ref{zetaheat})
over the space-like links of the heat kernel action is multiplied by a
factor of
\begin{equation}
\label{qq1}
{\rm Tr}_{\rm adj} P({\vec x}) \equiv \left|{\hat P}({\vec x})\right|^2 -
1~.
\end{equation}
The insertion of the sources is supposed to be governed by a fugacity
$\lambda = {\rm e}^{-{\mu\over T}}$ ($\mu$ being the chemical potential).
Therefore, in standard fashion, one inserts $k$ such pairs at locations 
${\vec x}_1,\ldots {\vec x}_k$ weighed with a factor of $\lambda^k/k!$, 
and sums over the positions and over $k$. The contribution of (\ref{qq1})
is thus exponentiated and the effective model in the presence of a gas of
adjoint static quarks becomes the following:  
\begin{equation}
\label{ad1}
Z(\lambda) = \int\prod_{\vec x} DP({\vec x})\, \prod_{{\vec x},i} 
{\cal K}_2\bigl(P({\vec x}),P({\vec x} + \hat\i);1/J\bigr)\,
{\rm e}^{\lambda\sum_{\vec x} \left(|{\hat P}({\vec x})|^2 -1\right)}~,
\end{equation}
which obviously reduces to the model (\ref{zhk0}) for $\lambda = 0$.

It is possible \cite{semzar} to describe the phase diagram in the
$(J,\lambda)$ plane of the above model. To do so, in analogy to what we did
in the case $\lambda=0$, one determines the lines $J_c^{\rm (s.c)}(\lambda)$ 
and $J_c^{\rm (w.c)}(\lambda)$ at which the strong coupling and weak
coupling solutions respectively lose their validity, and the line
$J_c(\lambda)$ where the deconfinement transition takes place.
It turns out that all these curves can be determined by the same techniques
already exploited in the previous sections in the $\lambda=0$ case. We will
therefore presently discuss only the changes due to the new 
term, proportional to $\lambda$, in the model (\ref{ad1}).
\iffigs
\begin{figure}
\vskip 0.1cm\noindent
\epsfxsize = 9.2truecm
\epsffile{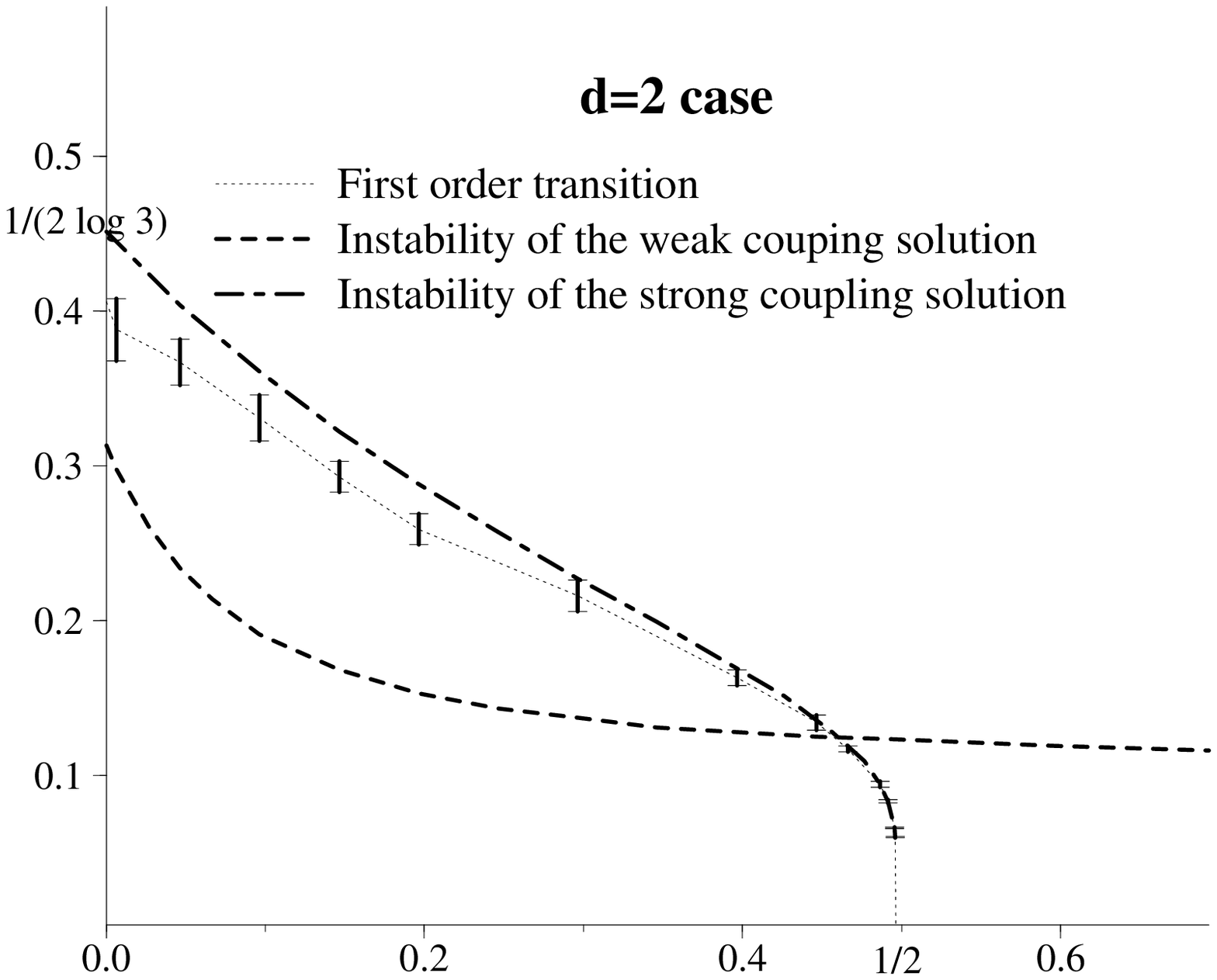}
\vskip -7.2truecm
\hskip 8.2truecm 
\epsfxsize = 5.7truecm
\epsffile{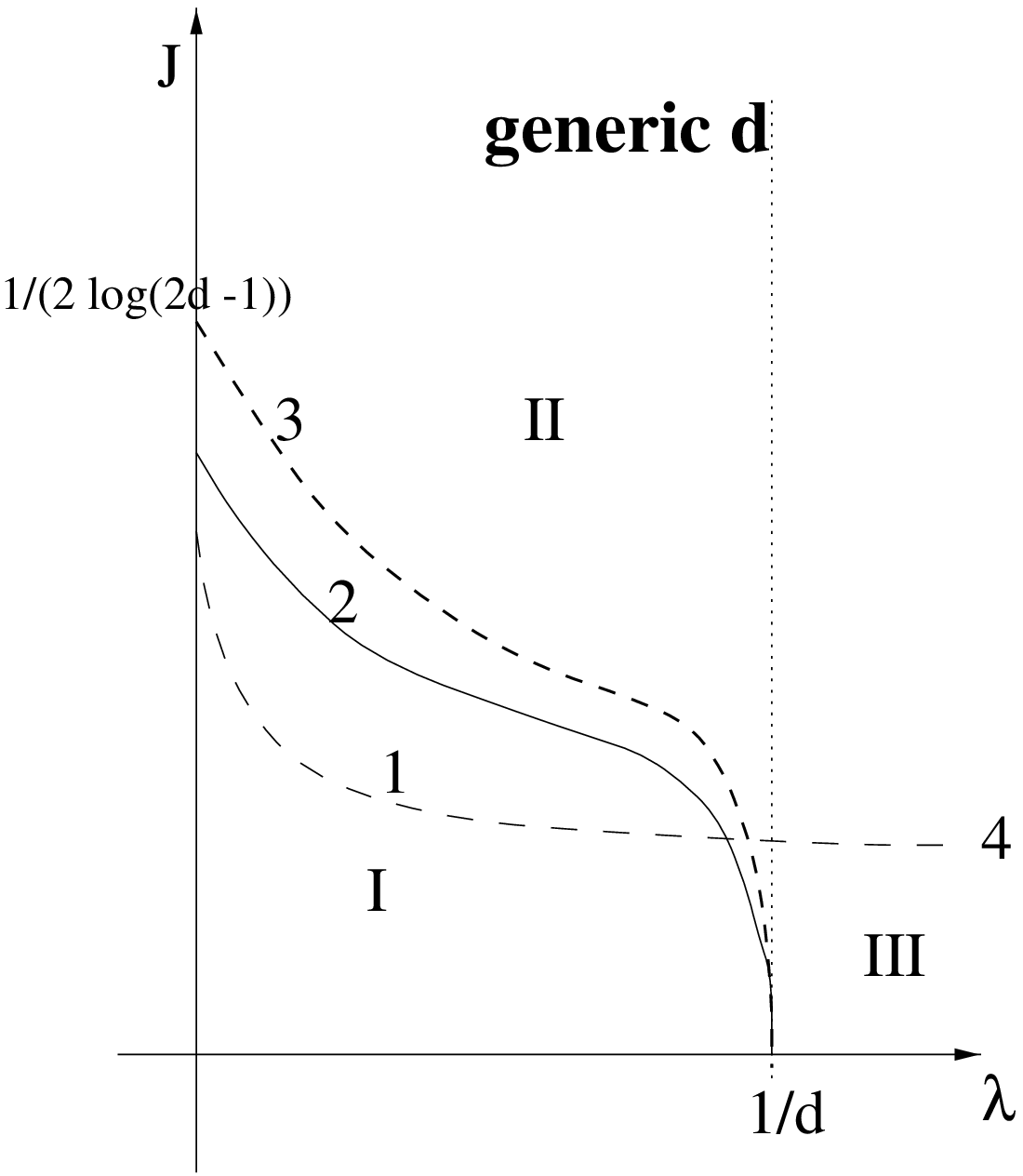}
\vskip 0.2cm\noindent
\mycaptionl{Fig. 8}{The phase diagram 
in the $\lambda,J$ plane for generic dimension
$d$ is determined by the line $J = J_c^{\rm(w.k.)}(\lambda)$ where 
the weak coupling solution becomes unstable (1); the line $J =
J_c^{\rm(s.k.)}(\lambda)$ where the strong coupling symmetric vacuum turns
into a minimum (3); the line of $1^{\rm st}$ order transition $J = J_c(\lambda)$
where the symmetric vacuum ceases to be energetically favourable (2).
As remarked in Ref. \cite{semzar}, besides the strong coupling 
confining phase (I) and
the weak coupling, deconfined one (II), a new region appears (III),
separated from (II) by a line of $3^{\rm rd}$-order transition (4). In the
specific case $d=2$ the actual numerical results for the $1^{\rm st}$ order
transition, obtained as discussed in the text, are reported. The error bars
estimate the uncertainty in the numerical procedure 
utilized for the calculation.} 
\label{adjplots}
\end{figure}
\fi

Let us consider first the instability of the strong coupling vacuum.
It is possible to repeat the same discussion as in sec. \ref{instsec}; the
extra $\lambda$-term gives obviously a contribution, namely
\begin{equation}
\label{ad2}
-2\lambda {\rm Im}\left(\rho_1({\vec x}){\rm e}^{-\ii\theta}\right)~,
\end{equation}
to the eq.s of motion (\ref{sci1}). It is clear from (\ref{ad2}) that 
the new term affects uniquely
the value of the mass of the $1^{\rm st}$ excitation, that becomes:
\begin{equation}
\label{ad3}
m_1^2 = 2d\left(\cosh {1\over 2J} -1\right) - 2(d -1 + \lambda) \sinh
{1\over 2J}
\end{equation}
and vanishes for
\begin{equation}
\label{ad4}
{1\over J_c^{\rm (s.c.)}(\lambda)} = 2 \log {2d -1 + \lambda d\over 1 -
\lambda d}~.
\end{equation}
It is possible determine numerically  the value of $J$ at which the
first order deconfinement transition occurs at fixed values of
$\lambda$, by adding  a term $d\,\lambda (x_1^2-1)$ to the free energy and
using the method described in sec. \ref{dectranssec}. 
The actual values obtained in the case $d=2$ are reported in Fig. 8.
In any case, the curve of instability $J_c^{\rm (s.c.)}$ of the strong 
coupling vacuum represents an upper bound for the curve $J_c(\lambda)$ of
deconfinement transition, as it is indicated in Fig. 8. 

In the weak coupling regime, the model (\ref{ad1}) can again be approximated
by a Kazakov--Migdal model, as in the case (see sec. \ref{kmsec}) 
when $\lambda=0$.
Indeed if we expand the additional term $\lambda (|{\rm Tr}P|^2 -1)$ 
up to the quadratic order in the invariant angles $\theta_i$ of the Polyakov
loop we find that eq.(\ref{zetahk3}) is replaced by:
\begin{equation}
\label{ad6}
\int \prod_i {\rm d}\theta_i \left[ \Delta^2(\theta) \right]^{(1 - d)}
e^{-N [d J - \frac{d-1}{12} + \lambda] \sum_i \theta_i^2 }
\left[\sum_P (-1)^{\sigma(P)} e^{N J
\theta_i \theta_{P(i)}} \right]^d~.
\end{equation} 
Eq. (\ref{ad6}) is a quadratic Kazakov--Migdal model, with $\lambda$-dependent
mass
\begin{equation}
\label{ad7}
m^2(\lambda) = 2 d + \left( 2\lambda - {d-1\over 6}\right){1\over J}~.
\end{equation}
The Gross solution of this model, that describes the vacuum in the weak
coupling, deconfined phase, is a semicircular Wigner distribution with
radius $r(\lambda)$ determined by eq. (\ref{erretheta}), 
but with $m$ replaced with $m(\lambda)$. As described 
in sec. \ref{dksec}, the solution becomes unstable, as a consequence of the
Douglas--Kazakov $3^{\rm rd}$ order transition in QCD2, at the value 
$J_c^{\rm (w.c)}(\lambda)$ that satisfies the equation
\begin{equation}
\label{ad8}
\left({1\over J(\lambda)}\right)^2 = \pi^4 - \pi^2 r^2(\lambda)~.
\end{equation}
The resulting curve is plotted in the case $d=2$ and sketched in the
generic $d$ case in Fig. 8.
In analogy with the $(J,d)$ plane, the $(J,\lambda)$ phase diagram
reveals the presence of a region where neither the symmetric vacuum nor
the weak coupling solution are stable and one or more new phases must
exist. As a matter of fact, for $\lambda> {1 \over d}$, the symmetric vacuum 
is unstable for any value of $J$.
\subsection{Phase diagram in presence of a ``magnetic field''}
\label{magfield}
In sec. \ref{simp} we discussed a simplified exactly solvable model 
for the Polyakov loop, obtained essentially by truncating the strong
coupling expansion to the first order and adding to the free energy
a ``magnetic field'' term  $ hN\sum_{\vec x}\left[
{\rm Tr}(P(\vec x) + P^\dagger(\vec{x}))\right]$. The phase diagram
of this model consists of a line of $3^{\rm rd}$ order phase transition 
ending into a point of first order phase transition (see Fig. 2).   
The inclusion of the ``magnetic field'' to the full ``zeroth order 
approximation'' (\ref{zetaheat}) can be studied in both the weak and 
strong coupling regime.

In the strong coupling expansion it corresponds to adding a term
$ -2 h x_1$ to the free energy, namely to the l.h.s. of eq. (\ref{zetaWas}).
The resulting action can be investigated, in the spirit of sec.
\ref{dectranssec}, by truncating for instance the expansion at the second
order (instead of the first order as in the simplified model of \ref{simp}).
With reference to Fig. 6 of sec. \ref{dectranssec}, we find that the inclusion
of the magnetic term shifts the central maximum towards 
the edge of the physical region even for $J=0$: 
the symmetry is explicitly broken by the magnetic term.
We can distinguish three different cases: 
$${\em a)}~~  h <\: \sim0.15~,\hskip 2cm 
{\em b})~~ \sim 0.15 <  h < 1/2~,\hskip 2cm {\em c})~~  h > 1/2~.$$ 
For $h <\: \sim 0.15$ the phase transition is described 
by the same pattern shown in
Fig. 6, except that the maximum ``1'' is not anymore located at the origin and
it actually moves towards the edge of the physical region as $J$ increases.
In the simplified model corresponding to the first order truncation of the
strong coupling expansion, this interval ($h < \:\sim0.15$) 
shrinks to the point $h=0$ where the first order phase transition occurs.
For $ \: \sim0.15 < h < 1/2$ the maximum ``1'' reaches 
the edge of the physical region
as $J$ increases {\it before} maximum ``2'' develops. The transition is
therefore of the third order. This interval corresponds to the  line of third 
order phase transition in the simplified model.
Finally, for $ h > 1/2$ ``1'' is already outside the physical region at $J=0$
and, as in the simplified model,  there is no obvious phase transition at this
order of the strong coupling expansion\footnote{However, due to the non
analyticity of the boundary of the physical region, a phase transition occurs
if the maximum moves from the straight to the curved section of the boundary.
Numerical calculations indicate that this is indeed the case. Besides, the
weak coupling analysis, as mentioned below, shows that a DK phase transition
is bound to occur even for $h>1/2$.}.

In the weak coupling limit, one can reduce the model with magnetic
term to a Kazakov--Migdal model, by expanding it in powers of $\theta_i$ up to
quadratic terms. The result is identical to eq. (\ref{ad6}) with $\lambda$ 
replaced by $h$. This shows that even for $h>1/2$ a phase transition 
of the Douglas--Kazakov type occurs, separating the weak coupling solution
from some unknown phase.  

\section{Results in the framework of the Wilson Action}
\label{wilssec}
\subsection{Eguchi--Kawai Model}
\label{egusec}
In sec. 3.2.2  a complete dimensional reduction was obtained
by using techniques of the type used in the derivation
of the twisted Eguchi--Kawai model.
The resulting action (\ref{ttekaction}) has $N_t=1$ and it is completely 
equivalent in the large $N$ limit to the original unreduced action. 
By inserting the correct value of $m$ determined in eq. (\ref{emme}) into 
eq. (\ref{ttekaction}) we obtain the following expression for the dimensionally
reduced effective action:
\begin{eqnarray}
\label{5.1}
S_{\rm red} &=& \beta_t N \sum_{\vec x} \sum_{i=1}^d
\real~ \ee{\ii{2\pi \over N_t}}\trace
\bigl[U_i(\vec x) V(\vec x + \hat\i)
\Udag_i(\vec x) \Vdag(\vec x)\bigr] +
\nonumber\\
&&\beta_s N \sum_{\vec x} \sum_{i > j} \real\trace \bigl[
U_{i}(\vec x) U_{j}(\vec x+ \hat\i) \Udag_{i}(\vec x+ \hat\j)
\Udag_{j}(\vec x)\bigr]~.
\end{eqnarray}
In spite of the complete dimensional reduction, this model is still
too complex to be solved exactly, and a solution has been obtained
so far only in zeroth order approximation ($\beta_s=0$)~\cite{bilda}. 
However, even within this approximation, the solution will retain 
a non trivial dependence from $N_t$, resulting from the exact
dimensional reduction. 
In this section we shall derive and discuss such solution.

In order to obtain from (\ref{5.1}) a solvable matrix model we 
assume that an $\vec x$ independent master field dominates the 
functional integral so that the field variables
in (\ref{ttekaction}) can be replaced by constants.
Then, setting $\beta_s = 0$, we get the following partition function:
\begin{equation}
\label{model}
Z = \int \hskip -2pt DV \int\hskip -1pt \prod_{i = 1}^d DU_{i}\,\, 
\ee{\beta_t N \real 
\ee{\ii{2 \pi \over N_t}} \sum_{i = 1}^d \trace 
(V U_{i} \Vdag \Udag_{i} )}~.
\end{equation}
For $d=3$, and more generally for odd values of $d$ (that is even space-time 
dimensions), this result coincides with the one we would have obtained
from a hot twisted Eguchi--Kawai model; the derivation shown here however 
holds for any $d$.
The integrations over the unitary matrices $U_i$ in eq. (\ref{model})
are all independent and the partition function can then be written
as
\begin{eqnarray}
\label{wctekpartfunc}
Z &=& \int \hskip -2pt DV \left[\int \hskip -2pt DU\,\, 
\ee{\beta_t N \real \ee{\ii{2\pi\over N_t}}
\trace (V U \Vdag \Udag)}\right]^d\nonumber\\
&=& \int\bigl(\prod_{i=1}^N \dop \theta_i\bigr)
\hskip 3pt {\cal J}^2(\theta)
\left[\int\hskip -2pt DU\,\,
\ee{\beta_t N \sum_{i,j} |U_{ij}|^2 \cos (\theta_i -
\theta_j + {2\pi\over L})}\right]^d~,
\end{eqnarray}
where in the last line we have gauge-rotated the matrix $U$ to
diagonalize $V$: 
\begin{equation}
\label{nn1}
V\rightarrow \diag (\ee{\ii\theta_0},\ldots,\ee{\ii\theta_{N-1}})~.
\end{equation}
${\cal J}(\theta)$ is the
Haar measure of SU$(N)$ expressed in terms of integration over the
eigenvalues $\ee{\ii\theta_i}$ and defined in (\ref{vandermonde}).
We want to solve the model (\ref{wctekpartfunc}) in the weak
coupling limit. In the extreme weak coupling region the functional integral
is dominated by the vacuum configuration given in eq. (\ref{ttekvacuum}), 
namely
\begin{equation}
\label{conto1}
V=Q_{N_t} \otimes {\bf 1}_{N/N_t}~.
\end{equation}
In this configuration the eigenvalues of $V$ are organised in $N_t$ 
bunches\footnote{Notice however that the open Polyakov loop is given 
by $V^{N_t}={\bf 1}$.}
each composed of ${N \over N_t}$ identical values $\ee{2\pi\ii a\over N_t}$,
$a =1,\ldots N_t$.
\iffigs
\begin{figure}
\epsfxsize 12cm
\begin{center}
\null\hskip 1pt\epsffile{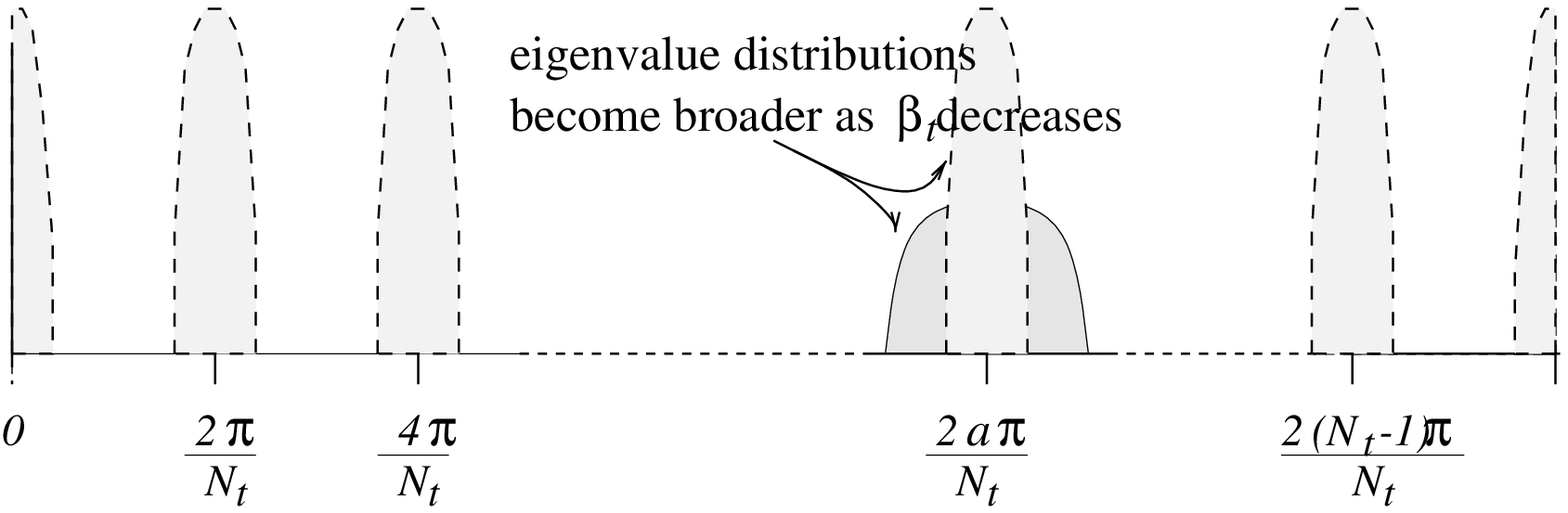}
\vskip 3pt\noindent
\end{center}
\vskip -1.1cm
\mycaptionl{Fig. 9}{The weak coupling solution in the large ${N\over N_t}$
consists of $N_t$
distributions, centered at the multiples of ${2\pi\over N_t}$.
The width of the distributions grows as
$\beta_t$ decreases, until finally an instability is attained at a certain
critical value $\beta_{t,c}$.}
\label{solut}
\end{figure}
\fi
Quantum fluctuations around this vacuum configurations can be considered
by parametrizing the invariant angles $\theta_i$ according to the 
equation:
\begin{equation}
\label{conto3}
\theta_i \equiv
\theta_{a,\alpha} =  {2\pi a\over N_t} +\varphi_{a,\alpha}~,
\end{equation}
with $a =1,\ldots N_t$ , $\alpha=1,\ldots {N\over N_t}$. 
A solution of the model given in eq. (\ref{model}) can be obtained through
the following steps: {\em a)}\, insert eq. (\ref{conto3}) 
into (\ref{wctekpartfunc})
and evaluate the integral in $DU$ by using a saddle point method; 
{\em b)}\, in the resulting effective action keep only 
the terms quadratic in the fluctuations $\varphi_{a,\alpha}$; 
{\em c)}\, solve the model obtained in this way, which turns
out to be a Kazakov--Migdal model, and find the the eigenvalue distribution.
The details of the calculation can be found in Ref. ~\cite{bilda}. 
Here we just remark about some of its most delicate points.
As in the Itzykson--Zuber integral, the extrema of the integrand in the 
integral over $DU$ are given by $U$ coinciding with any element of the Weyl 
group, namely with any permutation of the eigenvalues $\theta_i$.
However in computing the integral with the saddle point method we
only consider permutations that map each bunch of eigenvalues into the 
next one, namely:
\begin{equation}
\label{joe4}
P:\hskip 0.2cm \theta_{a,\alpha} \rightarrow \theta_{a+1,P_a(\alpha)}~.
\end{equation}
where $P_a(\alpha)$ corresponds to a permutation of the ${N \over N_t}$ 
eigenvalues of the bunch at ${ 2 \pi a \over N_t}$.
It is easy to see that the other permutations are exponentially depressed 
compared to the ones above, typically by a factor
\begin{equation}
\label{conto5}
\exp \left(\beta_t N(1-\cos {2\pi s\over N_t})\right) 
\sim \exp\left(-\beta_t N {2 \pi^2 s^2\over N_{t}^2}\right)~.
\end{equation}
for each eigenvalue mapped from a bunch $a$ into $a+1+s$.
One can assume that neglecting these non perturbative contributions is 
justified in the weak coupling (large $\beta_t$) regime, 
while below a critical
value of $\beta_t$ one expects a phase transition to occur as a
result of their condensation.
As we shall see, this is indeed the case, and we shall be able to conclude 
that the non perturbative contributions corresponding to permutations different
from the ones in (\ref{joe4}) play in the present formulation the same
role as the instantons in the Douglas--Kazakov phase transition discussed in
the previous sections.

The result of the calculation, up to quadratic terms in $\varphi_{a,\alpha}$,
is the following (see~\cite{bilda}):
\begin{eqnarray}
\label{nuova33}
&& \int \Bigl[ \prod_{a\alpha} {\rm d}\varphi_{a,\alpha} \Bigr] 
\Bigl[\prod_{a}
\Delta^2(\varphi_a)\Bigr]^{(1-d)} \,\,\exp\Bigl\{ -{N\over N_t} \Bigl[ 
(\beta_t N_t- {N_t\over 4}) d - {d-1\over 12}N_{t}^2 \Bigr]
\sum_{a,\alpha} \varphi_{a,\alpha}^2\Bigr\}
\nonumber\\
&& \hskip 1.2cm\times \Biggl( \sum_{\{P_a\}} (-1)^{\sigma(P_a)} \, 
\exp\Bigl\{ {N\over N_t} (\beta_t N_t - {N_t\over 4}) \sum_{a,\alpha} 
\varphi_{a,\alpha} \varphi_{a+1,P_a(\alpha)} \Bigr\} \,\Biggr)^d~.
\end{eqnarray}
The solution of this model can be obtained by assuming 
that the master field is translational invariant, namely invariant under
the ${\bf Z}_{N_t}$ symmetry of the vacuum . With this assumption, which is 
proved {\it a posteriori} to be correct in ~\cite{bilda},
all bunches of eigenvalues have the same distribution and we can set:
\begin{equation}
\label{conto21}
\varphi_{a,\alpha} \rightarrow {\varphi_\alpha\over N_t}~,
\end{equation}
where $\varphi_\alpha$ is now of order $1$ in the large $N_t$ limit, whereas 
the fluctuations $\varphi_{a,\alpha}$are  of order $1\over N_t$.
Notice that since the Polyakov line is $P =V^{N_t}$, the angles $\varphi_\alpha$
are the invariant angles \footnote{Each eigenvalue $\varphi_\alpha$ 
has obviously a degeneracy $N_t$, as it appears in each bunch.} of $P$, and the 
corresponding eigenvalue distribution is the eigenvalue distribution of $P$.
With the position (\ref{conto21}) the partition function (\ref{nuova33})
reduces to the product of the partition functions of $N_t$ identical models:
\begin{eqnarray}
\label{conto22}
Z & = & \Biggl[\int\prod_{\alpha=1}^n {\rm d}\varphi_\alpha\, 
\Bigl[\Delta^2(\varphi)\Bigr]^{1-d}
\,\,\exp\Bigl\{-n \Bigl[J d - {d-1\over 12}  
\Bigr] \sum_\alpha(\varphi_\alpha)^2
\Bigr\}\nonumber\\
&&\times\Biggl(\sum_P (-1)^{\sigma(P)} \exp\Bigl\{
n J \sum_\alpha 
\varphi_\alpha\varphi_{P\alpha}\Bigr\}\Biggr)^d \Biggr]^{N_t} \,\, ,
\end{eqnarray}
where we have defined $n = N/N_t$ and  done the replacement  
\begin{equation}
\label{conto23}
 {1\over N_t}(\beta_t -{1\over 4})~~=  ~~J~,
\end{equation}
where $J$ is just the ``normalized'' coupling of the heat kernel action
(see eq. (\ref{331})). In fact, if we consider eq. (\ref{conto23})
at $\rho=1$ we find the linear rescaling 
\eq
\label{conto30bis}
\beta(n_t)= n_t~J + \frac{1}{4}~,
\en
in agreement with eq. (\ref{331bis}). The comparison between 
eq.s (\ref{conto30bis}) and (\ref{331bis}) also determines the
value of $\alpha_{\tau}^0$. A more complete analysis of the
$\rho$ dependence however can be obtained by writing (\ref{conto23})
 with $N_t=\rho n_t$  and with $N_t= n_t$, 
and by eliminating $J$ from the two equations. We find
\begin{equation}
\label{conto30ter}
\beta_t(\rho n_t) = \rho \biggl(\beta(n_t) - {1 \over 4}\biggr) + {1 \over
4}~.
\end{equation}
This equation gives the rescaling of the coupling constant induced by
varying the asymmetry parameter $\rho$, and it should be compared with
eq. (\ref{rel3}), where $\beta_t\to\beta_t(\rho n_t)$ and $\beta\to\beta(n_t)$.
{}From the rescaling (\ref{conto30ter}) we can read off immediately the  
function $c_\tau(\rho)$:
\begin{equation}
\label{ctau1}
c_\tau(\rho) = - {1\over 4} + {1\over 4\rho} + \ldots~.
\end{equation}
This has to be compared with the expression (\ref{rhoex}) 
obtained by Karsch~\cite{k81}:
\begin{equation}
\label{ctau2}
c_\tau(\rho) = - 0.2609 + {1\over 4\rho} + \ldots~.
\end{equation}
The agreement between the two results is quite remarkable. In fact, although
both calculations are based on evaluating one loop effects around the classical
vacuum, this is done  in the approach described here {\it after} a complete
dimensional reduction and {\it neglecting} the effect of the space-like 
plaquettes, which were instead taken into account in  Karsch's calculation.
This  confirms a feature, that
already emerged in the case of SU(2)~\cite{ourpaper}, 
namely that the corrections 
to $c_{\tau}(\rho)$ due to the space-like plaquettes are relatively small.
As a consequence, and to the extent to which the space-like plaquettes
can be safely neglected, the result (\ref{ctau1}) is independent from
the number $d$ of space dimensions, whereas Karsch's result was limited to
$d=3$. 
The agreement between (\ref{ctau1}) and (\ref{ctau2}) also represents a 
check of the consistency of the calculation, and in 
particular that only the fluctuations around the classical vacuum are
relevant in the deconfined phase.
 
Apart from the replacement of $N$ with $n$ the expression in
(\ref{conto22}) raised to the $N_t$th power coincides with the
Kazakov--Migdal partition function (\ref{zetahk3}) obtained from 
the heat kernel action. 
Its solution is therefore given by the Wigner semicircular 
distribution of eigenvalues described in section 4.3, and the
instability of the solution occurs exactly for the same value of
$J$ as described there.
The complete agreement between the results of the heat kernel action
and the ones obtained from the Wilson action in the present section
confirms that for large $n_t$, where the coupling constants coincide,
the two models are equivalent. 
This was expected, as discussed in sec. 3.1, and this result justifies
{\it a posteriori} the assumption that the non perturbative 
contributions which have been neglected in the saddle point method, 
i.e. those involving permutations not of the form (\ref{conto3}), 
are indeed irrelevant in the weak coupling phase.
It is shown in ~\cite{bilda} that these contributions can be put in one to one 
correspondence with the winding configurations (instantons) of the heat kernel
model. So their condensation  is expected to be responsible of the phase 
transition at the critical value of $\beta_t$, in the same way as the instantons 
are responsible for the Douglas--Kazakov phase transition.

In terms of the original invariant angles $\theta_i$ the 
eigenvalue distribution of our solution consists, as shown in Fig. 9, of a
sequence of bunches, each made of $n=N/N_t$ eigenvalues, at intervals of 
${2 \pi \over N_t}$. Each bunch is described by a Wigner semicircular 
distribution of radius $\frac{r}{N_t}$ with $r^2$ given by (\ref{erretheta}).
Above the phase transition  different bunches do not communicate
with each other, except for the canonical shift of ${2 \pi \over N_t}$ 
due to the twist, and the partition function is just the one corresponding 
to a single bunch raised to  the power $N_t$ (see eq. (\ref{conto22})).
The non-perturbative contributions, corresponding to interactions between
other bunches, are suppressed by the exponential factor given in 
eq. (\ref{conto5}).  

The picture, discussed in sec \ref{dksec}, for the eigenvalues of the Polyakov 
loop in the heat kernel model is quite similar. 
Their distribution can be represented on an
infinite line as a periodic distribution of period $2 \pi$, that before
the phase transition consists of separate bunches of $N$ eigenvalues.
The contribution of a one-instanton configuration connecting  a bunch to
its neighbour is weighted by an exponential factor of
$\exp((- 2 \pi^2 \beta_t N/ N_t)$, as can be seen  
from eq. (\ref{poisson}), with ${\cal A}= 1/J \sim {N_t\over\beta_t}$.  
This contribution coincides with the one-instanton contribution in the
one-plaquette model, that is obtained setting $s=1$ in eq. (\ref{conto5}),
if one takes into account that the
number of eigenvalues in the latter case is $n=N/N_t$ instead of $N$.
The contribution of the instantons to  the classical action is
then the same in the two models. This  is
a strong hint that they describe the same physics and that their
condensation play the same role in the phase transition.    

\subsection{Strong coupling expansion}
\label{scwsec}
Let us now consider the strong coupling expansion of the 
effective model for the Polyakov loops obtained from the Wilson action.
We proceed by mimicking the treatment of the heat kernel 
case done in sec. \ref{scesec}, and we will just sketch the main points, 
referring to \cite{bcdmp} for further details. 

By assuming  translational invariance, 
we can rewrite the Wilson partition function (\ref{wilson}) as
\eq
Z_{\rm W} = \int \prod_i {\rm d}\theta_i\, {\cal J}^2(\theta)\,
\left( {\cal I}(\theta,J_{\rm W}) \right)^d~,
\label{zetaW}
\en
whence it is evident that the integral
\begin{equation}
\label{ical}
{\cal I}(\theta,J_{\rm W}) = \int DU \exp \biggl(N\, J_{\rm W} \, 
\real\trace U{\rm e}^{\ii\theta}
U^\dagger{\rm e}^{-\ii\theta}\biggr)
\end{equation}
plays the same role as ${\cal K}_2(P,P^\dagger;1/J)$ in the heat kernel case
(compare with eq. (\ref{zetahk})).
It turns out \cite{bcdmp} that the integral (\ref{zetaW}) admits a large $N$ 
strong coupling expansion of the form:
\eq
{\cal I}(\theta,J) = \exp\biggl(N^2 \sum_{k=1}^{\infty} 
J_{\rm W}^k\, {\tilde F}_k(x,\alpha)\biggr)~,
\label{calIas}
\en
where the functions ${\tilde F}_k(x,\alpha)$ can be shown to depend only on 
$x_i$ and $\alpha_i$ with $i \leq k$, and can be determined in 
principle by using Schwinger--Dyson equations. 
Correspondingly, the large $N$ limit of the partition function $Z_{\rm W}$ 
becomes
\eq
{1 \over N^2} \ln Z_{\rm W} =   \sum_{k=1}^{\infty} \left( d~J_{\rm W}^k\, 
{\tilde F}_k(x,\alpha) - {x_k^2 \over k} \right)~.
\label{zetaWas1}
\en
This is exactly the same type of expansion occurring in the heat kernel case,
see eq. (\ref{zetaWas}). The expansion parameter is now the Wilson coupling 
$J_{\rm W}$ instead of the function $2\exp[-1/(2J)]$ 
of the heat kernel coupling.

In particular, one finds that at the first order
\begin{equation} 
{\tilde F}_1(x,\alpha) = x_1^2 = F_1(x,\alpha;J) ~,
\label{f1}
\end{equation}
while at higher orders  ${\tilde F}_k$ and $F_k$ begin to 
differ\footnote{The expression of ${\tilde F}_k(x,\alpha)$ 
up to the $4^{\rm th}$ order is given in \cite{bcdmp}, eq.s (65)--(68).}.

At first order therefore the phase diagram resulting from the Wilson strong
coupling expansion is {\em exactly} the same obtained in the heat kernel 
case (see eq. (\ref{firstord}) and following text), expressed in 
terms of $J_{\rm W}$ instead of $2\exp [-1/(2J)]$; thus the critical Wilson 
coupling is indeed given at this order by $J_{{\rm W},c}=1/d$. 
Notice that this is, as it was
to be expected, the result that is found within the simple model reviewed in
sec. 3.5, a model that was indeed obtained from the truncation 
to the first order of the Wilson action. 

Of course the deconfinement transitions observed in the Wilson
and the heat kernel regularization schemes correspond to the same physical 
phenomenon. 
{}From the above discussion it follows that this is the case if
\begin{equation}
J_{\rm W} = 2\exp\left(-{1\over 2J}\right) + \ldots~,
\label{whk}
\end{equation}
the extra terms in the r.h.s. being the effect of higher order corrections.
This is just the relation between the two couplings given in (\ref{335}) 
and valid in the strong coupling regime. 
This relation was derived in sec. 3.3 by comparing the strong coupling 
behaviour of the coefficients in the character expansions of the heat 
kernel and Wilson actions. 
The analysis of the deconfinement transition, by using the
strong coupling expansion of the Wilson action, can be carried on
in exactly the same way as in sec. 4.2.4, and the results up to the fourth 
order are summarized in Table IV.
\begin{table}
\mycaptionl{Table IV}{Values of the critical coupling 
$J_{{\rm W},c}$ at the fourth 
order in the strong coupling expansion of the Wilson action. 
In the last four columns the 
corresponding values of $x_i,~~(i=1-4)$ are reported.} 
\label{scwtable}
\begin{center}
\begin{tabular}{c c c c c c}
\hline\hline
$d$ & $J_{{\rm W},c}$ & $x_1$ & $x_2$ & $x_3$ & $x_4$ \\
\hline
$2$ & $0.601$ & $0.50$ & $0.06$ & $-0.03$ & $-0.03$ \\ 
$3$ & $0.379$ & $0.52$ & $0.03$ & $-0.03$ & $-0.02$ \\
$4$ & $0.275$ & $0.53$ & $0.02$ & $-0.03$ & $-0.02$ \\
$100$ & $0.010$ & $0.49$ & $0.004$ & $0.004$ & $0.005$ \\
\hline\hline
\end{tabular}
\end{center}
\end{table}
The results of Table IV all refer to $n_t=1$. It is possible
in principle to apply the same method to extract from the 
dimensionally reduced twisted action of eq. (\ref{model}) with
$\rho=1$ a strong coupling expansion for any value of $n_t$. However, as
the Polyakov loop coincides in (\ref{model}) with $\trace V^{n_t}$,
namely with $x_{n_t}$, higher and higher orders in the strong coupling 
expansion are needed to proceed to higher values of $n_t$.
It is more convenient to use the values of Table IV as an input and use
the scaling law (\ref{conto30bis}) to extract the dependence from
$n_t$.

\section{Comparison with results from  Montecarlo simulations}
\label{mcsec}
When comparing our results with those of  Montecarlo simulations we must
expect two kinds of systematic deviations. 
\begin{itemize}
\item[{\em a)}]
The first one is due to the large $N$ approximation. 
This is however a rather small deviation. In fact,
the lesson that we learn by looking at the
Montecarlo simulations is that the large $N$ limit 
results are reached for rather small values of $N$. For instance
recent calculations on the glue-ball spectrum in (2+1) dimensions for low values
of $N$ have shown that already for $N\geq 3$ some mass ratios are well
described with the leading term in the large $N$ limit~\cite{teplat96}. 
Thus we can consider our large $N$ results as rather good approximations of the
$N=3$ case in which we are interested.  
At the same time rather large corrections are certainly present in the
$N=2$ case. Obviously, it would be highly desirable to have results from
simulations for larger values of $N$, so as to study in detail the approach to
the large $N$ limit and  test the analytic calculations reviewed here in a 
more stringent way. But this is not an easy task.
Large $N$ simulations have been performed only in the 
framework of
the Eguchi--Kawai reduction scheme and are plagued by the presence of a bulk
phase transition which shadows the deconfinement one. 
\item[{\em b)}]
The second, more serious, problem is that, in order to solve the model,
we had to neglect the  space-like  plaquettes in
the original action. We have already discussed the consequences that this
approximation has on the scaling laws in sec.s 3.1.3 and \ref{egusec}:
only for $d=2$ we find the correct scaling behaviour. 
The experience gained with the SU(2) \cite{ourpaper} model however suggests 
that in the $n_t=1$ case
the corrections due to the space-like plaquettes 
are very small and that 
their effect is essentially negligible. Even if this case is not
interesting from a phenomenological point of view, since $n_t=1$ is 
too small to give informations on the continuum limit, 
it becomes very interesting as a test
of our method. For this reason we shall devote sec. \ref{trepiuuno} 
to a discussion of this comparison.
For larger values of $n_t$ 
these corrections are not any more negligible and their 
importance increases as $n_t$ increases. Notice  that, as mentioned above,
the corrections due to the inclusion of the space-like plaquettes
have a different relevance in (2+1) and (3+1) dimensions. In
$(2+1)$ dimensions they do not affect the scaling behaviour, and in fact
we shall be able to successfully compare not just the  leading, but the next to
leading term in the scaling law (\ref{conto30bis}) with the results of
Montecarlo simulations.
On the contrary in (3+1) dimensions they completely change the scaling law. 
For this reason we shall discuss separately the $(2 + 1 )$ and the $(3 + 1)$
dimensional cases in the following subsections.
\end{itemize}
\bm
\subsection{$n_t=1$ in (3+1) dimensions}
\ubm
\label{trepiuuno}
In the $n_t=1$ case we have only two results from Montecarlo simulations; they
are however very interesting and carry a lot of information. The first one 
is for the SU(2) model~\cite{bems}. It is very precise since it was obtained
by using an original and very powerful non local algorithm (see~\cite{bems} for
the details). The value of the critical temperature in our units is
 $J_{{\rm W},c}\equiv\beta_c/N^2=0.2185(1)$. The second results is for $N=30$, hence in a
 situation in which the large $N$ limit is a very good approximation.
The critical coupling is $J_{{\rm W},c}\sim 0.38$. 
It was obtained in~\cite{fhklink}
within the framework of the Eguchi--Kawai reduction scheme.
These results must be compared with what we have found in the case of the
Wilson action. In this case
our best estimate for the critical coupling in $(3 + 1)$ dimensions,
obtained by means of the strong coupling expansion,
is $J_{{\rm W},c}=0.378$ as shown in Table IV. 
The agreement with the $N=30$ value is indeed
impressive. This is an interesting result and,
compared with the naive results $J_{{\rm W},c}=1/d=0.333$ which one
would obtain by using the simplified action of sec. 3.5, it shows the relevance of
the higher order terms in the character expansion of our action.
The rather large gap between the value of $J_{{\rm W},c}$ for $N=2$ and $N=30$
can be understood already at the level of the lowest order in the strong
coupling  expansion where, for $N=2$ only, the reality of the fundamental
representation leads to a critical coupling $J_{{\rm W},,c} = {1 \over 2 d}$.
It would be  interesting to have the Montecarlo simulations for 
some intermediate value of $N$, to see how the large $N$ limit is approached.
Analytically this result could  be obtained within the strong coupling 
expansion scheme developed for the large $N$ limit. 
A different technique has been applied both for $N=2$~\cite{mean} and for
higher values of $N$ ~\cite{tesi} consisting in an improved mean field Bethe
approximation that determines the asymptotic expansion of $J_{{\rm W},,c}$ 
in powers of $1/d$. The results obtained so far can be summarized as follows:
\begin{eqnarray}
\label{larged}
N=2~~~~~~~~ & J_{{\rm W},c} & = {1 \over 2 d}  + {1 \over 4 d^2} + 
{ 13 \over 48 d^3} + O({1 \over d^4})~, \nonumber \\
N=3~~~~~~~~ & J_{{\rm W},c} & = {1 \over d} - {1 \over 4 d^2} + 
O({1 \over d^3})~,  \nonumber \\
N>3~~~~~~~~ & J_{{\rm W},c} & = {1 \over d} +{1 \over 2  d^2} + 
O({1 \over d^3})~. 
\end{eqnarray} 
Notice that for $N$ larger than 3 the expansion is the same up to 
terms of order $1 \over d^3$, which means that already at $N=4$ we are 
quite close to the large $N$ limit.
If one substitutes   $d=3$ in (\ref{larged})
keeping all known terms one obtains the values 0.2045, 0.306 and 
0.389 for $N=2$, $N=3$ and $N>3$ respectively. They significantly 
improve the lowest order results and they agree, within
the approximation one would expect from the truncation, with the 
Montecarlo results.
\subsection{2+1 dimensions}
Very precise Montecarlo estimates of the
deconfinement temperature in (2+1) dimensions  exist for the $N=2$ and
$N=3$ models in the range $2\leq n_t\leq 6$. These can be found 
in~\cite{t2,ctdw,eklllps} and are reported in Tab. V.
All these simulations were made with the standard Wilson
action and at $\rho=1$. 

Looking at the data one can see that the expected linear dependence on $n_t$ is
very well fulfilled.  This allows us to use the 
variable $\frac{T_c}{N g^2}\equiv \frac{\beta_c}{n_t}$, which is the natural
one to take the continuum limit (see Tab. VI).
Once the linear term is factored out  we  can  see in Tab. VI a 
residual dependence on $n_t$ which is due to the higher order terms of 
eq. (\ref{251}). 
These corrections are so small that one can safely use a simple
one parameter fit, keeping only the leading $O(1/\beta)$
correction,  to extract  reliable estimates for the continuum limit values 
of the critical temperature. These are reported in the last row of Tab. VI

As discussed in the previous sections, in the framework of the EK reduction
scheme we can predict both the leading and sub-leading scaling behaviour.
Remarkably enough, not only the functional form of these two terms, but also
the {\em numerical estimates} of the correction that we find are in 
very good agreement with the simulations. They are reported in the last column
of Tab. VI. They have been obtained by taking for 
$\beta_c(n_t=1)=J_{{\rm W},c}$ our best 
strong coupling estimate given in Tab. IV and by using the scaling law
(\ref{conto30bis}) to extrapolate the result to arbitrary $n_t$.
The result is:
\eq
\label{anal}
\frac{T_c}{N g^2}\equiv \frac{\beta_c(n_t)}{n_t}=0.351 + \frac{1}{4 n_t}~.
\en
This analytical prediction has to be compared with the following
best fits, for the $N=2$ and $N=3$ cases, obtained from the Montecarlo 
data of Tab. VI by assuming a
linear dependence of the critical coupling from $1/n_t$.
Both fits have a high confidence level and give:
\begin{eqnarray}
\label{bestfit}
N=2 & ~~~~~~~~~ \frac{\textstyle \beta_c(n_t)}{\textstyle n_t}& = 
0.380(3)+ {0.106(11) \over n_t}~, \nonumber \\
N=3 & ~~~~~~~~~ \frac{\textstyle \beta_c(n_t)}{\textstyle n_t}& = 
0.366(2)+ {0.174(6) \over n_t}~. 
\end{eqnarray}

The agreement of the analytic results (\ref{anal}) with the best
fit (\ref{bestfit}) of the Montecarlo simulations is quite
remarkable. In fact the trend shown in the $N=2$ and $N=3$
simulations is perfectly consistent with the theoretical 
large $N$ limit not only in the leading term but also in
the sub-leading one. Although it is premature to draw any
conclusion, one can say that the results are at least 
consistent so far with the space-like plaquettes being
negligible in $2+1$ dimensions as far as the determination 
of the deconfinement transition is concerned. 
Notice that in $d=2$ there is no
simulation for $n_t=1$ to make a direct comparison as we did in the (3+1) case
above.  In~\cite{ourpaper,varenna} however we obtained 
a rather reliable  estimate of $\beta_c$ in the SU(2) case. 
This value is reported in the first line of Tab. VI. 

\begin{table}
\mycaptionl{\bf Table V}{The critical coupling $\beta_c$
as  a function of the lattice size in the t direction, $n_t$, 
in the (2+1) dimensional SU(2) and SU(3) LGT, taken from~\cite{t2},
\cite{eklllps} and~\cite{ctdw}.}
\label{d2atable}
\begin{center}
\begin{tabular}{c c c }
\hline\hline 
$n_t$ & $N=2$ & $N=3$ \\  
\hline
$2$ & $\sim 0.866$ & $0.906(2)$  \\
$3$ & $\sim 1.251$ & $$           \\
$4$ & $1.630(8)$ & $1.638(6)$ \\
$6$ & $2.388(10)$ & $2.371(17)$ \\
\hline\hline
\end{tabular}
\end{center} 
\end{table}
\begin{table}
\mycaptionl{\bf Table VI}{The critical temperature
$\frac{T_c}{N g^2}\equiv \frac{\beta_c}{n_t}$, in (2 + 1) dimensions,
 as a function
of $n_t$ and $N$.  In the first line  of the $N=2$ column 
is reported an analytic prediction 
obtained in~\cite{varenna} for $n_t=1$, and in the last line the extrapolations
to the continuum limit. The prediction of
the analysis reviewed in this work is displayed in the last column.}
\label{d2btable}
\begin{center}
\begin{tabular}{c c c c }
\hline\hline 
$n_t$ & $N=2$ & $N=3$ & $N=\infty$\\  
\hline
$1$ & $\sim 0.459$ & $$           & $0.601$\\
$2$ & $\sim 0.433$ & $0.4531(8)$  & $0.476$     \\
$3$ & $\sim 0.417$ & $$           & $0.434$     \\
$4$ & $0.4075(19)$ & $0.4094(14)$ & $0.413$     \\
$6$ & $0.3979(16)$ & $0.3952(27)$ & $0.393$     \\
$\infty$ & $0.380(3)$ & $0.366(2)$ & $0.351$     \\
\hline\hline
\end{tabular}
\end{center} 
\end{table}
\begin{table}
\mycaptionl{\bf Table VII}{The critical coupling $\beta_c/N^2$,
in (3 + 1) dimensions, as a function
of $n_t$ and $N$. In the last column our predictions, according to the
scaling law eq.(\ref{332}) and $\beta_c/N^2=0.378$ for $n_t=1$.}
\label{d3table}
{\footnotesize
\begin{center}
\begin{tabular}{c c c c c c c  c}
\hline\hline 
$n_t$ & $N=2$ & $N=3$ & $N=24$  & $N=54$& $N=81$& $N=96$ & $N=\infty$\\  
\hline
$2$ & $0.4700(8)$ & $\sim 0.568$ & $\sim 0.70$ &  & $$ & $$ & $0.506$\\
$3$ & $0.5442(8)$ & $0.6167(11)$  & $$ & $\sim 0.70$ & $$ & $$ & $0.634$\\
$4$ & $0.5746(2)$ & $0.63250(2)$  & $$ & $$ & $$ & $\sim 0.70$ & $0.762$\\
$5$ & $0.5932(11)$ & $$ & $$ & $$ & $$ & $$ & $0.890$\\
$6$ & $0.6066(8)$ & $0.65490(6)$  & $$ & $$ & $0.705(5)$ & & $1.018$\\
\hline\hline
\end{tabular}
\end{center} }
\end{table}
\bm
\subsection{$n_t>1$ in (3+1) dimensions}
\ubm
Very precise data on SU(2) and SU(3) in the range $n_t=2-16$ can be found
in~\cite{fhk}. Some results for higher values of $N$, 
obtained within the context of the EK model can be found 
in~\cite{daskogut,fhklink}. 
We have collected in Tab. VII these results for the lowest values of $n_t$.

As mentioned above in this case we do not obtain in our approach the correct
scaling law for the critical temperature as a function of $n_t$ for which the
inclusion of the space-like plaquette contribution is mandatory. Hence we can
trust  our results only for low values of $n_t$. We have already seen before
that in the $n_t=1$ case our prediction is indeed successful, we shall now
extend the comparison to higher values of $n_t$. To obtain our best estimate
for the critical temperature in this case we use again the 
scaling relation eq. (\ref{conto30bis})  and take our
$n_t=1$  value  $(J_c=0.378)$  as input.
 These estimates are reported in the last
column of Tab. VII.

A few comments are in order at this point:
\begin{itemize}
\item[{\em a)}]
It is interesting to notice that the critical couplings for small values
of $n_t$ have a dependence on $n_t$ which is not too different from
ours.
In fact, even if the scaling law eq. (\ref{conto30bis}) is definitely different 
from the scaling behaviour eq. (\ref{tccont}), for  low values of $n_t$ the 
Montecarlo estimates of the critical couplings
are still far from the asymptotic
scaling region and do not follow the  scaling behaviour of
eq. (\ref{tccont}).
\item[{\em b)}]
At fixed $n_t$, as  $N$ increases also the critical temperature
systematically increases,
and our large $N$ estimates seem to be the upper limit of this behaviour.
\item[{\em c)}]
For large values of  $N$ all the Montecarlo result cluster around the
value $\sim 0.70$. This is an artifact of the EK approximation. This value
is due to the presence of a bulk transition which shadows the real 
deconfinement transition. 
\end{itemize}
As we have seen, our results, obtained analytically in the large $N$
limit, show a more than qualitative agreement with Montecarlo
simulations performed on the full theory. 
We think that a careful analysis of higher order contributions in $\beta_s$,
may allow in future further improvements in the analytic
determination of the deconfinement critical temperature.

\setcounter{totalnumber}{0}
\vskip 1cm
\centerline{\bf Acknowledgements}
\vskip 0.5cm\noindent
We would like to thank for discussions F. Gliozzi 
and particularly L. Magnea, with
whom the work contained in Ref. \cite{bcdmp} was carried out.\\
This work was partially supported by the 
European Commission TMR programme ERBFMRX-CT96-0045.
\vfill\eject

\end{document}